%% file: emulate_2.tex
\documentclass{new_tlp}
\usepackage{times}
\usepackage{helvet}
\usepackage{courier}
\usepackage{amsmath}
\usepackage{graphicx}
\usepackage{xspace}
\usepackage[defblank]{paralist}
\usepackage{latexsym}

\newcommand{\set}[1]{\ensuremath{\{#1\}}}

\title{Relative Expressiveness of Defeasible Logics II}

\author[M.J. Maher]
{Michael J. Maher \\
School of Engineering and Information Technology \\
University of New South Wales, Canberra \\
ACT 2600, Australia    \\
E-mail: m.maher@adfa.edu.au
}

\newcommand{\ignore}[1]{}
\newcommand{\finish}[1]{}
\newcommand{\skipit}[1]{}

\newcommand{\DL}{{\bf DL}}
\newcommand{\WFDL}{{\bf WFDL}}

\newcommand{\ARROW}{\hookrightarrow}

\newcommand{\supp}{\sigma}   

\newcommand{\f}{{\it undefeated}}

\newcommand{\T}{{\cal T}}

\newcommand{\non}{{\sim}}

\newcommand{\seq}[2][n]{\ensuremath{#2_{1},\dots,#2_{#1}}\xspace}

\newcounter{clause}
\def\theclause{$c$\arabic{clause}}

{\end{tabbing}}

{\end{tabbing}}

\newtheorem{theorem}{Theorem}
\newtheorem{lemma}[theorem]{Lemma}
\newtheorem{definition}[theorem]{Definition}

\newtheorem{corollary}[theorem]{Corollary}
\newtheorem{example}[theorem]{Example}

\begin{document}

\ignore{
\pagerange{\pageref{firstpage}--\pageref{lastpage}}
\volume{\textbf{10} (3):}
\jdate{March 2012}
\setcounter{page}{1}
\pubyear{2012}
}

\maketitle

\label{firstpage}

\begin{abstract}
\cite{Maher12} introduced an approach for relative expressiveness of defeasible logics,
and two notions of relative expressiveness were investigated.
Using the first of these definitions of relative expressiveness,
we show that all the defeasible logics in the $\DL$ framework are equally expressive
under this formulation of relative expressiveness.
The second formulation of relative expressiveness is stronger than the first.
However, we show that logics incorporating individual defeat are equally expressive as
the corresponding logics with team defeat.
Thus the only differences in expressiveness of logics in $\DL$
arise from differences in how ambiguity is handled.
This completes the study of relative expressiveness in $\DL$ begun in \cite{Maher12}.
\end{abstract}

\finish{12 page limit + biblio}

\finish{ remove finish}

\section{Introduction}


Defeasible logics provide several linguistic features to support the expression of defeasible knowledge.
There are also a variety of such logics, supporting different intuitions about reasoning in a defeasible setting.
The $\DL$ framework \cite{flexf,TOCL10} provides logics that allow ambiguity in the ``truth'' status of a literal to propagate,
and logics that block ambiguity; it has logics that require an individual rule to defeat all competitors,
and logics that allow a ``team'' of rules to defeat competitors.
Given the different inferences supported by the different logics,
it is interesting to determine whether these logics are equally powerful
or whether, perhaps, some are more powerful than the others.

In terms of inference strength, \cite{TOCL10} established the relationship between
the different logics of $\DL$.
In terms of computational complexity,
the logics of $\DL$ are equivalent: all have linear complexity \cite{Maher2001,TOCL10}.
Relative expressiveness of the different logics was first investigated in \cite{Maher12},
which developed a framework, based on simulation in the presence of additional elements.
Two notions of relative expressiveness within this framework were investigated:
polynomial simulation wrt the addition of facts,
and simulation wrt the addition of rules.

In this paper we continue this investigation.
We will see that all the logics of $\DL$ are equally expressive,
using the first notion of relative expressiveness.
Thus we cannot distinguish the logics based on this notion.
We also establish that individual defeat has equal expressiveness to team defeat
in the logics of $\DL$ wrt addition of rules.
This is somewhat surprising, given the apparent greater sophistication of the
team defeat inference rules.
Given results in \cite{Maher12}, this completes the study of relative expressiveness for $\DL$.

\finish{
There is an analogy between ambiguity propagation in defeasible reasoning
and the propagation the undefined value $\bot$ in strict programming languages.
Given that call-by-value and call-by-name lambda calculus have incomparable expressiveness \cite{Felleisen}
the equal expressiveness of ambiguity propagating and blocking defeasible logics
is somewhat surprising.
In particular, the ability of an ambiguity propagating logic to express ambiguity blocking behaviour
is unexpected because ambiguity propagating defeasible logics are weaker in inference strength
than the corresponding ambiguity blocking logics \cite{TOCL10}.
}
\ignore{
The next two sections summarize the $\DL$ framework of defeasible logics and the notions of relative expressiveness
introduced in \cite{Maher12}.
Then the following two sections together provide the proof that the logics of $\DL$ are of equal expressivity
(in terms of simulation wrt addition of facts).
The first shows the simulation of $\DL(\delta^*)$ by $\DL(\partial^*)$.
In this simulation it is necessary for $\partial^*$ to simulate both $\delta^*$ and $\supp^*$.
The second completes the proof by simulating  $\DL(\partial^*)$ in $\DL(\delta^*)$.
The proof in this case is complicated by the fact that
inference rules for $\delta^*$ and $\supp^*$ are defined mutually recursively,
while the inference rules for $\partial^*$ are directly recursive.
This difference in structure makes a direct inductive proof difficult.
The problem is resolved by a ``tight'' simulating transformation that is able to simulate $\DL(\partial^*)$
(wrt addition of facts) in any of the $\DL$ logics.

The following sections investigate relative expressiveness via the second notion.
We first show that $\DL(\partial)$ can simulate $\DL(\partial^*)$
and $\DL(\delta)$ can simulate $\DL(\delta^*)$ wrt addition of rules.
We then show the reverse: that $\DL(\partial^*)$ can simulate $\DL(\partial)$
and $\DL(\delta^*)$ can simulate $\DL(\delta)$ wrt addition of rules.
This establishes that individual defeat has equal expressiveness to team defeat
in the logics of $\DL$.
Proofs of the results are detailed and lengthy.  They appear in an appendix.
}

The next two sections summarize the $\DL$ framework of defeasible logics and the notions of relative expressiveness
introduced in \cite{Maher12}.
Then the following two sections together provide the proof that the logics of $\DL$ are of equal expressivity
(in terms of simulation wrt addition of facts).
The first shows the simulation of an ambiguity propagating logic by an ambiguity blocking logic,
while the second shows a simulation in the reverse direction.
Combined with results of \cite{Maher12}, this establishes that the logics in $\DL$ all
have the same expressiveness in this formulation.

\ignore{
In this simulation it is necessary for $\partial^*$ to simulate both $\delta^*$ and $\supp^*$.
The second completes the proof by simulating  $\DL(\partial^*)$ in $\DL(\delta^*)$.
The proof in this case is complicated by the fact that
inference rules for $\delta^*$ and $\supp^*$ are defined mutually recursively,
while the inference rules for $\partial^*$ are directly recursive.
This difference in structure makes a direct inductive proof difficult.
The problem is resolved by a ``tight'' simulating transformation that is able to simulate $\DL(\partial^*)$
(wrt addition of facts) in any of the $\DL$ logics.

The following sections investigate relative expressiveness via the second, stronger formulation.
We first show that $\DL(\partial)$ can simulate $\DL(\partial^*)$
and $\DL(\delta)$ can simulate $\DL(\delta^*)$ wrt addition of rules.
We then show the reverse: that $\DL(\partial^*)$ can simulate $\DL(\partial)$
and $\DL(\delta^*)$ can simulate $\DL(\delta)$ wrt addition of rules.
This establishes that individual defeat has equal expressiveness to team defeat
in the logics of $\DL$.
Proofs of the results are detailed and lengthy.  They appear in an appendix.
}
The following sections investigate relative expressiveness via the second, stronger formulation.
Adapting simulations of \cite{Maher12} to the stronger formulation,
we establish that individual defeat has equal expressiveness to team defeat
in the logics of $\DL$.
Proofs of the results in this paper are detailed and lengthy.  They appear in an appendix.

\section{Defeasible Logic}

In this section we can only present an outline of the defeasible logics we investigate.
Further details can be obtained from \cite{TOCL10} and the references therein.
We address propositional defeasible logics,
but the results should extend to a first-order language.

A defeasible theory is built from a language $\Sigma$ of literals (which we assume is closed under negation)
and a language $\Lambda$ of labels.
A \emph{defeasible theory} $D = (F, R, >)$ consists of a set of facts $F$, a finite set of rules $R$,
each rule with a distinct label from $\Lambda$,
and an acyclic relation $>$ on $\Lambda$ called the \emph{superiority relation}.
This syntax is uniform for all the logics considered here.
Facts are individual literals
expressing indisputable truths.
Rules relate a set of literals (the body), via an arrow, to a literal (the head), and are one of three types:
a strict rule, with arrow $\rightarrow$;
a defeasible rule, with arrow $\Rightarrow$;
or
a defeater,  with arrow $\leadsto$.
Strict rules represent inferences that are unequivocally sound if based on definite knowledge;
defeasible rules represent inferences that are generally sound.
Inferences suggested by a defeasible rule may fail, due to the presence in the theory
of other rules.
Defeaters do not support inferences, but may impede inferences suggested by other rules.
The superiority relation provides a local priority on rules.
Strict or defeasible rules whose bodies are established defeasibly represent claims
for the head of the rule to be concluded.
The superiority relation contributes to the adjudication of these claims by an inference rule,
leading (possibly) to a conclusion.
Given a theory $D$, the corresponding languages are expressed by $\Sigma(D)$ and $\Lambda(D)$.

Defeasible logics derive conclusions that are outside the syntax of the theories.
Conclusions may have the form 
${+}d q$, which denotes that under the inference rule $d$ the literal $q$ can be concluded,
or
$-d q$, which denotes that the logic can establish that under the inference rule $d$ the literal $q$ cannot be concluded.
The syntactic element $d$ is called a tag.
In general, neither conclusion may be derivable:
$q$ cannot be concluded under $d$, but the logic is unable to establish that.
Tags ${+}\Delta$ and $-\Delta$ represent monotonic provability (and unprovability)
where inference is based on facts, strict rules, and modus ponens.
We assume these tags and their inference rules are present in every defeasible logic.
What distinguishes a logic is the inference rule for defeasible reasoning.
The four logics discussed in the Introduction correspond to four different pairs of inference rules,
labelled $\partial$, $\delta$, $\partial^*$, and $\delta^*$;
they produce conclusions of the form (respectively) ${+}\partial q$, $-\partial q$, ${+}\delta q$, $-\delta q$, etc. 
The inference rules $\delta$ and $\delta^*$ require auxiliary tags and inference rules,
denoted by $\supp$ and $\supp^*$, respectively.
For each of the four principal defeasible tags $d$, the corresponding logic is denoted by $\DL(d)$.

The four principal tags and corresponding inference rules represent different intuitions about defeasible reasoning:
in $\partial$ and $\partial^*$ ambiguity is blocked, while in $\delta$ and $\delta^*$ ambiguity is propagated;
in $\partial$ and $\delta$ rules for a literal act as a team to overcome competing rules, 
while in $\partial^*$ and $\delta^*$ a single rule must overcome all competing rules.
A more detailed discussion of ambiguity and team defeat in the $\DL$ framework is given in \cite{TOCL10} and \cite{Maher12}.

The inference rules are presented 
in the appendix in the form of the definition of a function $\T_D$ for a given theory $D$.
Given a defeasible theory $D$, for any set of conclusions $E$,
$\T_D(E)$ denotes the set of conclusions inferred from $E$ using $D$
and one application of an inference rule.
$\T_D$ is a monotonic function on the complete lattice of sets of conclusions ordered by containment.
The least fixedpoint of $\T_D$ is the set of all conclusions that can be drawn from $D$.
We follow standard notation in that $\T_D \uparrow 0 = \emptyset$ and $\T_D \uparrow (n+1) = \T_D(\T_D \uparrow n)$.

The relative inference strength of the different logics in $\DL$ was established in
the inclusion theorem of \cite{TOCL10}.
For any tag $d$, $+d(D)$ denotes the set of conclusions of $D$ of the form $+d q$
and similarly for $-d$.
\begin{theorem}[Inclusion Theorem \cite{TOCL10}]  \label{thm:inc} 
Let $D$ be a defeasible theory.
\begin{itemize}
\item[(a)] ${+}\Delta(D) \subseteq {+}\delta^*(D) \subseteq {+}\delta(D) \subseteq {+}\partial(D)
\subseteq {+}\supp(D) \subseteq {+}\supp^*(D)$.

\item[(b)] $-\supp^*(D) \subseteq -\supp(D) \subseteq -\partial(D) \subseteq -\delta(D)
\subseteq -\delta^*(D) \subseteq -\Delta(D)$.

\item[(c)] ${+}\delta^*(D) \subseteq {+}\partial^*(D) \subseteq {+}\supp^*(D)$

\item[(d)] $-\supp^*(D) \subseteq {-}\partial^*(D) \subseteq -\delta^*(D)$
\end{itemize}
\end{theorem}
Parts (a) and (b) are proved in \cite{TOCL10}.
Parts (c) and (d) can be established by similar methods.

\section{Simulating Defeasible Logics}   \label{sect:sim}

\cite{Maher12}  introduced a framework for addressing the relative expressiveness
of defeasible logics.
The framework identifies 
the greater (or equal) expressiveness of $L_2$ compared to $L_1$ with
the ability to simulate any theory $D$ in a logic $L_1$
by a theory $T(D)$ in the logic $L_2$.
Simple simulation was shown not to be sufficiently discriminating,
so simulation was required to hold in the presence of an addition to the theory.

The \emph{addition} of a theory $A$ to a theory $D$ is denoted by $D + A$.
Addition is essentially the union of the theories,
but we require $\Lambda(D) \cap \Lambda(A) = \emptyset$,
so that the addition of theories preserves the property that distinct rules have distinct labels.
This requirement also has the effect that a superiority statement in $D$ cannot affect a rule in $A$,
and vice versa.
Let $D = (F, R, >)$ and $A = (F', R', >')$.
Then $D + A = (F  \cup F', R  \cup R', >  \cup >')$.
$\Lambda(D{+}A) = \Lambda(D) \cup \Lambda(A)$ and
$\Sigma(D{+}A) = \Sigma(D) \cup \Sigma(A)$.

A simulating theory $T(D)$ in general will involve additional literals, rules and labels beyond those of $D$.
If additions $A$ were permitted to affect these, the notion of simulation would become trivial,
so we restrict additions to have only an indirect effect on $T(D)$, via $\Sigma(D)$.
Given a theory $D$ and a possible simulating theory $T(D)$,
we say an addition $A$ is \emph{modular} if
$\Sigma(A) \cap \Sigma(T(D)) \subseteq \Sigma(D)$,
$\Lambda(D) \cap \Lambda(A) = \emptyset$, and
$\Lambda(T(D)) \cap \Lambda(A) = \emptyset$.
In general, we will consider a class of additions
but for any $D$ and $T(D)$ only the modular additions in the class will be considered.

Since different logics involve different tags, conclusions from theories in different logics
cannot be identical.
For simulation it suffices that conclusions are equal modulo tags.
Given logics $L_1$ and $L_2$, with principal tags $d_1$ and $d_2$, respectively,
we say two conclusions $\alpha$ in $L_1$ and $\beta$ in $L_2$ are \emph{equal modulo tags}
if $\alpha$ is $+d_1 q$ and $\beta$ is $+d_2 q$ or
$\alpha$ is $-d_1 q$ and $\beta$ is $-d_2 q$.

Thus we have the following definition of simulation and relative expressiveness.
For more discussion on the motivations for the definitions, see \cite{Maher12}.

\begin{definition}  \label{defn:simC}
Let $C$ be a class of defeasible theories.

We say $D_1$ in logic $L_1$ is \emph{simulated} by $D_2$ in $L_2$ with respect to a class $C$
if, for every modular addition $A$ in $C$,
$D_1 + A$ and $D_2 + A$ have the same conclusions in $\Sigma(D_1 + A)$, modulo tags.

We say a logic $L_1$ can be simulated by a logic $L_2$ with respect to a class $C$
if every theory in $L_1$ can be simulated by some theory in $L_2$ with respect to additions from $C$.

We say $L_2$ is \emph{more (or equal) expressive than} $L_1$
if $L_1$ can be simulated by $L_2$ with respect 
$C$.
\end{definition}

Different notions of relative expressiveness arise from different choices for $C$.
There were two classes of additions investigated in \cite{Maher12}:
the addition of \emph{facts} (that is, $A$ has the form $(F, \emptyset, \emptyset)$),
and the addition of \emph{rules} (that is, $A$ has the form $(\emptyset, R, \emptyset)$).
Simulation with respect to addition of rules is stronger than simulation with respect to addition of facts
because any fact can equally be expressed as a strict rule with an empty body.
We might also consider arbitrary additions, where $A$ can be any defeasible theory.


The main results of \cite{Maher12} are that:
\begin{itemize}
\item
$DL(\partial)$ and $DL(\partial^*)$ have equal expressiveness, with respect to addition of \emph{facts},
as do $DL(\delta)$ and $DL(\delta^*)$
\item
neither $DL(\partial)$ nor $DL(\partial^*)$ is more expressive than $DL(\delta)$ or $DL(\delta^*)$,
and vice versa, with respect to addition of \emph{rules}
\item
when arbitrary additions are permitted, of the four defeasible logics under consideration,
none is more expressive than any other
\end{itemize}
\ignore{
In the following we will:
\begin{itemize}
\item
show that the four logics in the $\DL$ framework are equally expressive with respect to addition of facts,
thus extending the first point above
\item
show that, wrt addition of rules, $DL(\partial)$ is more expressive than $DL(\partial^*)$ and 
$DL(\delta)$ is more expressive than $DL(\delta^*)$,
in contrast to the second point above
\end{itemize}

\finish{
This settles all the relative expressiveness relations under additions of facts and rules.
??????????????
}
}
\ignore{
It was shown in \cite{Maher12} that simply using the conclusions that a logic defines
makes for an ineffective notion of expressiveness:
all defeasible logics in the framework $\DL$ are equally expressive in a trivial way,
because the conclusions themselves are easily encoded in a theory.
Instead it is necessary to require that the structure of a theory is preserved
.......
}

\ignore{
Consider addition limited to a set of facts, that is $A = (F, \emptyset, \emptyset)$.
Allowing arbitrary addition of facts corresponds to treating each theory $D_1$ under logic $L_1$
as defining a non-monotonic inference relation from facts to consequences.
This is similar to Dix's treatment of logic programs in \cite{Dix1} where
a logic program is viewed as defining a non-monotonic inference relation from the input atoms to the output atoms.
It also reflects a common practice of keeping the rules static while facts vary.
Simulation then requires that any inference relation expressed by $D_1$ under $L_1$ 
can be expressed by some $D_2$ under $L_2$.
}

\section{Blocked Ambiguity Simulates Propagated Ambiguity}

We now show that every theory over an ambiguity propagating logic 
can be simulated by a theory over the corresponding ambiguity blocking logic.
To begin, we show that $\DL(\partial^*)$ can simulate $\DL(\delta^*)$.
Any defeasible theory $D$ is transformed into a new theory.
The new theory employs new propositions 
$strict(q)$ and $supp(q)$, for each literal $q$,
and $supp\_body(r)$, $comp(r)$, and $o(r)$, for each rule $r$.
The new theory also introduces labels $p_d(r)$, $n_d(r, s)$, $p_s(r)$, $n_s(r, s)$,
for each pair $r, s$ of opposing rules in $D$.
These are families of propositions and labels, not predicates, despite the notation.

\begin{definition}  \label{defn:ABsimAP}
Let $D = (F, R, >)$ be a defeasible theory with language $\Sigma$.
We define the transformation $T$ of $D$ to $T(D) = (F', R', >')$ as follows:

\begin{enumerate}
\item \label{t1:facts}
The facts of $T(D)$ are the facts of $D$.
That is, $F' = F$.
\item \label{t1:strict}
Every strict rule of $R$ is included in $R'$.

\item \label{t1:strictq}  
For every literal $q$, $R'$ contains

\[
\begin{array}{lrll}
str(q):     &              q           & \rightarrow & \phantom{\neg} strict(q) \\ 
nstr(q):   &                           & \Rightarrow &  \neg strict(q) \\
\end{array}
\]

and the superiority relation contains
$nstr(q) >' str(q)$, for every $q$.

\item \label{t1:supp1}
For each literal $q$ in $\Sigma$, $R'$ contains
\[
\begin{array}{lrll}
& q & \Rightarrow & supp(q) \\
\end{array}
\]

\item  \label{t1:supp2} 
For each strict or defeasible rule $r$ of the form $b_1, \ldots, b_n \ARROW_r q$ in $R$, $R'$ contains
\[
\begin{array}{lrll}
& supp(b_1), \ldots, supp(b_n) & \Rightarrow & supp\_body(r) \\
& supp\_body(r), \neg o(r)         & \Rightarrow & supp(q) \\   
\end{array}
\]

\noindent
and, further, for each rule $s = B_s \ARROW_{s} \non q$ for $\non q$ in $R$, where $s > r$, $R'$ contains

\[
\begin{array}{lrll}
n_s(r, s):  & B_s & \Rightarrow & \phantom{\neg} o(r) \\
p_s(r):        &      & \Rightarrow  & \neg o(r) \\ 
\end{array}
\]
and the superiority relation contains
$n_s(s, r) >' p_s(s)$.


\item \label{t1:def}  
For each strict or defeasible rule $r = B_r \ARROW_r q$ in $R$,  $R'$ contains
\[
\begin{array}{lrll}
inf(r): & B_r, \neg comp(r), \neg strict(\non q) & \Rightarrow & q \\

\end{array}
\]

\noindent
and, further, for each rule $s = B_s \ARROW_{s} \non q$ for $\non q$ in $R$, where $s \not< r$, $R'$ contains

\[
\begin{array}{lrll}
n_d(r, s): & supp\_body(s) & \Rightarrow &  \phantom{\neg} comp(r) \\
p_d(r):     &                           & \Rightarrow &  \neg comp(r) \\
\end{array}
\]

and the superiority relation contains
$n_d(r, s) >' p_d(r)$.

\end{enumerate}
\end{definition}

\finish{Need to define "fails"}

Parts \ref{t2:facts} and \ref{t2:strict} of the transformation preserve all the strict inferences from $D$.
Part \ref{t2:strictq} allows us to distinguish strict conclusions from defeasible conclusions.
The structure of these rules -- where $str(q)$ is strict,  $nstr(q)$ is defeasible, and $nstr(q) > str(q)$ --
ensures that $strict(q)$ is inferred defeasibly iff $q$ is inferred strictly,
and $strict(q)$ fails iff strict inference of $q$ fails.
A similar structure of rules was previously used in \cite{Maher12} 
in showing that $\DL(\partial^*)$ can simulate $\DL(\partial)$ wrt addition of facts.

We use the proposition $supp(q)$ to indicate that the literal $q$ is supported (i.e. $+\supp^* q$ can be inferred),
while the literal $q$ refers to defeasible provability (wrt $\delta^*$).
Part \ref{t1:supp1} ensures that every literal that holds defeasibly is also supported.
This property is justified by the inclusion theorem of  \cite{TOCL10}.
Part \ref{t1:supp2} encodes the inference rules for support (i.e. $\supp^*$).
$supp\_body(r)$ indicates that all literals in the body of rule $r$ are supported.
The head $q$ of a rule $r$ is supported if the body of $r$ is supported
and $r$ is not overruled
(i.e. all rules $s$ that are superior to $r$ fail).
The overruling of $r$ is indicated by $o(r)$.
The rules $n_s(r,s)$ and $p_s(r)$ and the superiority relation
ensure that $\neg o(r)$ is derived defeasibly iff there is no overruling rule $s$.

Rules $inf(r)$ in part \ref{t1:def} encode the inference rules for $\delta^*$.
$q$ holds defeasibly iff the body of a rule $r$ for $q$ holds defeasibly and $r$ has no competing rules
(i.e. all rules for $\non q$ not inferior to $r$ have a body that fails wrt $\supp^*$).
The rules $n_d(r,s)$ and $p_d(r)$ and the superiority relation
ensure that $\neg comp(r)$ is derived defeasibly iff there is no competing rule.

In this translation, 
the superiority relation in $D$ is not directly represented by the superiority relation in $T(D)$.
Instead, the superiority relation in $D$ is used to restrict the instantiation of rules in the transformation,
while the superiority relation in $T(D)$ is used to ensure that $o(r)$ and $\neg o(r)$ do not both fail,
and similarly for $comp(r)$.


\begin{example}
To see the operation of this transformation, consider the following theory $D$,
which demonstrates the difference between ambiguity propagation and blocking logics.
\begin{tabbing}
1234123412341234\=123412341234\=12341234\=1234\kill

\>$r_1: \  \ \Rightarrow \phantom{\neg} p$
\>\> $r_3: \ \neg p \Rightarrow \neg q$ \\

\> $r_2: \  \ \Rightarrow \neg p$ 
\>\>$r_4: \ \ \hspace{15pt} \Rightarrow \phantom{\neg} q$ \\

\end{tabbing}
In $\DL(\partial^*)$ from $D$ we conclude $-\partial^* p$ and $-\partial^* \neg p$, 
$+\partial^* q$ and $-\partial^* \neg q$.
In $\DL(\delta^*)$ from $D$ we conclude $-\delta^* p$ and $-\delta^* \neg p$, 
$-\delta^* q$ and $-\delta^* \neg q$.
We also conclude
$+\supp^* p$ and $+\supp^* \neg p$, 
$+\supp^* q$ and $+\supp^* \neg q$.

$T(D)$ contains the following rules.
\begin{tabbing}
1234\=12341234123412\=341234123412341234\=1234123412341234\kill

\ignore{
\>$str(p): \hspace{27pt}  p \rightarrow strict(p)$
\>\> $nstr(p): \hspace{18pt}  \Rightarrow  \neg strict(p)$ \\
\>$str(\neg p): \hspace{13pt}  \neg p \rightarrow strict(\neg p)$
\>\> $nstr(\neg p): \hspace{12pt}   \Rightarrow \neg strict(\neg p)$ \\

\>$str(q): \hspace{27pt}  q \rightarrow strict(q)$
\>\> $nstr(q): \hspace{18pt}  \Rightarrow  \neg strict(q)$ \\
\>$str(\neg q): \hspace{13pt}  \neg q \rightarrow strict(\neg q)$
\>\> $nstr(\neg q): \hspace{12pt}   \Rightarrow \neg strict(\neg q)$ \\

\ \\
}

\>$  \hspace{43pt} \Rightarrow supp\_body(r_1)$ 
\>\>$supp\_body(r_1), \neg o(r_1) \Rightarrow supp(p)$ \\
\>$  \hspace{43pt} \Rightarrow  supp\_body(r_2)$ 
\>\>$supp\_body(r_2), \neg o(r_2) \Rightarrow supp(\neg p)$ \\
\>$    supp(\neg p) \Rightarrow  supp\_body(r_3)$ 
\>\>$supp\_body(r_3), \neg o(r_3) \Rightarrow supp(\neg q)$ \\
\>$  \hspace{43pt} \Rightarrow  supp\_body(r_4)$ 
\>\>$supp\_body(r_4), \neg o(r_4) \Rightarrow supp(q)$ \\
\\
\>$\hspace{22pt} p \Rightarrow supp(p)$ 
\>\>$p_s(r_1): \hspace{10pt} \Rightarrow \neg o(r_1)$ \\
\>$\hspace{15pt} \neg p \Rightarrow supp(\neg p)$ 
\>\>$p_s(r_2): \hspace{10pt} \Rightarrow \neg o(r_2)$ \\
\>$\hspace{22pt} q \Rightarrow supp(q)$ 
\>\>$p_s(r_3): \hspace{10pt} \Rightarrow \neg o(r_3)$ \\
\>$\hspace{15pt} \neg q \Rightarrow supp(\neg q)$ 
\>\>$p_s(r_4): \hspace{10pt} \Rightarrow \neg o(r_4)$ \\

\\
\>$inf(r_1): \hspace{19pt} \neg comp(r_1), \neg strict(\neg p) \Rightarrow \phantom{\neg} p$ \\
\>$inf(r_2):  \hspace{26pt} \neg comp(r_2), \neg strict(p) \Rightarrow \neg p$ \\
\>$inf(r_3):  \hspace{10pt} \neg p,  \neg comp(r_3), \neg strict(q) \Rightarrow \neg q$ \\
\>$inf(r_4):  \hspace{20pt} \neg comp(r_4), \neg strict(\neg q) \Rightarrow \phantom{\neg} q$ \\

\\
\>$p_d(r_1):\hspace{10pt} \Rightarrow \neg comp(r_1)$ 
\>\>$n_d(r_1, r_2): \hspace{10pt} supp\_body(r_2) \Rightarrow  comp(r_1)$ \\
\>$p_d(r_2):\hspace{10pt} \Rightarrow \neg comp(r_2)$ 
\>\>$n_d(r_2, r_1): \hspace{10pt} supp\_body(r_1) \Rightarrow  comp(r_2)$ \\
\>$p_d(r_3):\hspace{10pt} \Rightarrow \neg comp(r_3)$ 
\>\>$n_d(r_3, r_4): \hspace{10pt} supp\_body(r_4) \Rightarrow  comp(r_3)$ \\
\>$p_d(r_4):\hspace{10pt} \Rightarrow \neg comp(r_4)$ 
\>\>$n_d(r_4, r_3): \hspace{10pt} supp\_body(r_3) \Rightarrow  comp(r_4)$ \\
\\
\end{tabbing}

$T(D)$ also contains the following superiority statements.
\begin{tabbing}
1234\=12341234123412\=341234123412341234\=1234123412341234\kill

\>$nstr(p) >' str(p)$ 
\>\> $n_d(r_1, r_2) >' p_d(r_1)$ \\
\> $nstr(\neg p) >' str(\neg p)$ 
\>\> $n_d(r_2, r_1) >' p_d(r_2)$ \\
\> $nstr(q) >' str(q)$ 
\>\> $n_d(r_3, r_4) >' p_d(r_3)$ \\
\> $nstr(\neg q) >' str(\neg q)$ 
\>\> $n_d(r_4, r_3) >' p_d(r_4)$ \\

\end{tabbing}

There are some points to highlight in this example.
Rules for $strict$ and $\neg strict$ are omitted from the listing above because they are not of interest 
($D$ has no strict rules or facts);
we will have conclusions ${+}\partial^* \neg strict(l)$ and $-\partial^* strict(l)$, for every literal $l$. 
There are no rules $n_s(r, s)$ in $T(D)$ because they only occur when $s > r$,
and the superiority relation  in $D$ is empty.
Consequently, there are no superiority statements of the form $n_s(r, s) >' p_s(r)$.
It also follows that  ${+}\partial^* \neg o(r)$ is concluded, for each rule $r$,
and hence we can infer ${+}\partial^* supp(l)$, for each literal $l$ except $\neg q$,
reflecting the fact that these literals are supported in $D$,
and ${+}\partial^* supp\_body(r)$, for each rule $r$.
We can then infer also ${+}\partial^* supp(\neg q)$.
It then follows that $-\partial^* \neg comp(r)$ is concluded, for each $r$,
using the superiority relation.
Then, as a consequence of the rules $inf(r)$, we find that all literals $l$ fail to be inferred
(i.e. we conclude $-\partial^* l$, for each literal $l$).
This expresses the ambiguity propagating behaviour of $\DL(\delta^*)$
from within $\DL(\partial^*)$.
\ignore{
The conclusions involving $strict$ (see part \ref{t1:strictq} of the translation)
will have the form ${+}\partial^* \neg strict(l)$ and $-\partial^* strict(l)$, for every literal $l$, 
since there are no facts or strict rules in $D$.
See Lemma \ref {lemma:strictAB} below.
}
\end{example}

\ignore{
The facts and strict rules of $D{+}A$ and $T(D){+}A$ are the same, except for rules for $strict(q)$ in $T(D){+}A$.
However $strict(q)$ is not used in any other strict rule.
Consequently, for any addition $A$, $D{+}A$ and $T(D){+}A$ draw the same strict conclusions in $\Sigma(D{+}A)$.
Furthermore, these conclusions are reflected in the defeasible conclusions of $strict(q)$.

\begin{lemma}  \label{lemma:strictAB}
Let $A$ be any defeasible theory, and let $\Sigma$ be the language of $D{+}A$.
Then, for every $q \in \Sigma$,
\begin{itemize}
\item
$D{+}A \vdash {+}\Delta q$ iff $T(D){+}A \vdash {+}\Delta q$ 

\hspace{67pt}
iff $T(D){+}A \vdash {+}\partial^* strict(q)$ iff $T(D){+}A \vdash -\partial^* \neg strict(q)$
\item
$D{+}A \vdash -\Delta q$ iff $T(D){+}A \vdash -\Delta q$  

\hspace{67pt}
iff $T(D){+}A \vdash -\partial^* strict(q)$  iff $T(D){+}A \vdash {+}\partial^* \neg strict(q)$
\end{itemize}
\end{lemma}
\skipit{
\begin{proof}
The proof of
$D{+}A \vdash \pm\Delta q$ iff $T(D){+}A \vdash \pm\Delta q$ 
is straightforward, 
by induction on length of proofs.

In the inference rule for ${+}\partial^* strict(q)$,
clause $.2.3$ must be false, by the structure of the rules in part \ref{t2:strictq} of the transformation.
Consequently, we infer ${+}\partial^* strict(q)$ iff we infer ${+}\Delta strict(q)$,
which happens iff we infer ${+}\Delta q$ since there is only the one rule for $strict(q)$.
Similarly,
clause $.2.3$ of the inference rule for $-\partial^* strict(q)$ is true, so
we infer $-\partial^* strict(q)$ iff we infer $-\Delta strict(q)$,
which happens iff  we infer $-\Delta q$ since there is only the one rule for $strict(q)$.

In the inference rule for $-\partial^* \neg strict(q)$,
clause $.2.1$ is false because the body of $nstr(q)$ is empty,
and clause $.2.3$ is false because $nstr(q) >' str(q)$.
Thus we infer $-\partial^* \neg strict(q)$ iff we infer ${+}\Delta strict(q)$.
Finally,
in the inference rule for ${+}\partial^* \neg strict(q)$,
clause $.1$ is false, because there is no fact or strict rule for $\neg strict(q)$.
and clauses $.2.1$ and $.2.3$ are true (the latter because $nstr(q) >' str(q)$).
Thus, we can infer ${+}\partial^* \neg strict(q)$ iff we can infer $-\Delta strict(q)$.
\end{proof}
}

This lemma establishes that strict provability ($\pm \Delta$) from $D{+}A$ in $\DL(\delta^*)$
is captured in $\DL(\partial^*)$ by the transformation defined above,
no matter what the addition $A$.
We now show that $\DL(\partial^*)$ can simulate 
the behaviour of
both $\delta^*$ and $\supp^*$
with respect to addition of facts.

\begin{lemma}
Let $D$ be a defeasible theory, 
$T(D)$ be the transformed defeasible theory as described in Definition \ref{defn:ABsimAP},
and let $A$ be a modular set of facts.
Let $\Sigma$ be the language of $D{+}A$ and let $q \in \Sigma$.
Then
\begin{itemize}
\item
$D{+}A \vdash {+}\delta^* q$ iff $T(D){+}A \vdash {+}\partial^* q$
\item
$D{+}A \vdash -\delta^* q$ iff $T(D){+}A \vdash -\partial^* q$ 
\item
$D{+}A \vdash {+}\supp^* q$ iff $T(D){+}A \vdash {+}\partial^* supp(q)$
\item
$D{+}A \vdash -\supp^* q$ iff $T(D){+}A \vdash -\partial^* supp(q)$
\end{itemize}
\end{lemma}

This result concerns only addition of facts.
It was established in \cite{Maher12} that
it cannot be extended to addition of rules.

Given that the ambiguity blocking logics can simulate each other, as can the ambiguity propagating logics
(see \cite{Maher12}) we have
}

\begin{theorem}  \label{thm:ABsimAP}
The ambiguity blocking logics ($\DL(\partial)$ and $\DL(\partial^*)$)
can simulate the ambiguity propagating logics ($\DL(\delta)$ and $\DL(\delta^*)$)
with respect to addition of facts.
\end{theorem}

\finish{
Consequently, under this formulation of relative expressiveness
the ambiguity blocking logics are of equal or greater expressiveness
than the ambiguity propagating logics.
In the next section we show that, in fact, the logics have equal expressiveness.
}

\section{Propagated Ambiguity Simulates Blocked Ambiguity}


We now show that every theory over an ambiguity blocking logic 
can be simulated by a theory over the corresponding ambiguity propagating logic.
To begin,
we simulate $\DL(\partial^*)$ by $\DL(\delta^*)$.
Any defeasible theory $D$ is transformed into a new theory $T(D)$.
The new theory employs new propositions 
$strict(q)$ and $\f(q)$ for each literal $q$ in $\Sigma$,
and employs labels 
$str(q)$ and $nstr(q)$ for each literal $q$ in $\Sigma$,
and
$n_d(r, s)$ and $p_d(r)$ for each pair of opposing rules $r, s$ in $R$.

\begin{definition}   \label{defn:APsimAB}
Let $D = (F, R, >)$ be a defeasible theory with language $\Sigma$.
We define the transformation $T$ of $D$ to $T(D) = (F', R', >')$ as follows:

\begin{enumerate}
\item \label{t2:facts}
The facts of $T(D)$ are the facts of $D$.
That is, $F' = F$.
\item \label{t2:strict}
Every strict rule of $R$ is included in $R'$.

\item \label{t2:strictq}  
For every literal $q$, $R'$ contains

\[
\begin{array}{lrll}
str(q):     &              q           & \rightarrow & \phantom{\neg} strict(q) \\ 
nstr(q):   &                           & \Rightarrow &  \neg strict(q) \\
t(q):        &    strict(q)          & \Rightarrow & \phantom{\neg} true(q) \\ 
nt(q):     &                           & \Rightarrow &  \neg true(q) \\
\end{array}
\]

and the superiority relation contains
$nstr(q) >' str(q)$
and $t(q) >' nt(q)$, for every $q$.

\item \label{t2:def}  
For each literal $q$,  $R'$ contains
\[
\begin{array}{lrll}
              &              \f(q)           & \Rightarrow & q \\ 
\end{array}
\]

For each strict or defeasible rule $r = B_r \ARROW_r q$ for $q$ in $R$,  $R'$ contains
\[
\begin{array}{lrll}
p_d(r): & B_r,  \neg true(\non q) & \Rightarrow & \f(q) \\
\end{array}
\]

\ignore{
$q$ holds defeasibly if the body of a rule $r$ for $q$ holds defeasibly, $r$ is not defeated
(i.e. all rules not inferior to $r$ have a body that fails wrt $\supp^*$),
and $\non q$ is not established strictly.
}

\noindent
and, further, for each rule $s = B_s \ARROW_{s} \non q$ for $\non q$ in $R$, where $r \not> s$, $R'$ contains

\[
\begin{array}{lrll}
n_d(r, s): & B_s & \Rightarrow & \neg \f(q) \\
\end{array}
\]

and  the superiority relation contains
$n_d(r, s) >' p_d(r)$.
\end{enumerate}
\end{definition}

Parts \ref{t2:facts} and \ref{t2:strict} preserve all the strict inferences from $D$.
Part \ref{t2:strictq} allows us to distinguish strict conclusions from defeasible conclusions.
For this transformation -- compared to the transformation in the previous section --
extra rules $t$ and $nt$ are needed.
These rules ensure that $\delta^*$ and $\supp^*$ agree on the literals $true(q)$,
that is, from $T(D){+}A$ we conclude $+\delta^* true(q)$ iff we conclude $+\supp^* true(q)$ iff
$D{+}A \vdash +\Delta q$.
(See Lemma 18 in the appendix.)
In comparison,
we never infer $-\supp^* \neg strict(q)$ and always infer $+\supp^* \neg strict(q)$,
independent of $D$.\footnote{
This also demonstrates a flaw in \cite{Maher12}.
In that paper, the transformation used to simulate $\DL(\delta)$ with $\DL(\delta^*)$
fails to use these extra rules, and thus is incorrect.  
Definition \ref{defn:IDsimTD} has a corrected transformation.
}

Part \ref{t2:def} encodes the inference rules for $\partial^*$:
$q$ holds defeasibly if the body of a rule $r$ for $q$ holds defeasibly, 
$\non q$ is not established strictly,
and $r$ is not defeated
(i.e. all rules not inferior to $r$ have a body that fails wrt $\partial^*$).
The requirement that $r$ is not defeated is expressed through
the use of rules $n_d(r, s)$ opposing $p_d(r)$ for each rule $s$ in $D$ not inferior to $r$.
The rules $n_d(r, s)$ are superior to $p_d(r)$ in $T(D)$,
thus ensuring that 
$\f(q)$ is inferred iff $r$ is not defeated.


\begin{example}
To see the operation of this transformation, consider (again) the following theory $D$,
which demonstrates the difference between ambiguity propagation and blocking logics.
\begin{tabbing}
1234123412341234\=123412341234\=12341234\=1234\kill

\>$r_1: \  \ \Rightarrow p$
\>\> $r_3: \ \neg p \Rightarrow \neg q$ \\

\> $r_2: \  \ \Rightarrow \neg p$ 
\>\>$r_4: \ \ \hspace{15pt} \Rightarrow q$ \\

\end{tabbing}
In $\DL(\partial^*)$ from $D$ we conclude $-\partial^* p$ and $-\partial^* \neg p$, 
${+}\partial^* q$ and $-\partial^* \neg q$.
In $\DL(\delta^*)$ from $D$ we conclude $-\delta^* p$ and $-\delta^* \neg p$, 
$-\delta^* q$ and $-\delta^* \neg q$.
We also conclude
${+}\supp^* p$ and ${+}\supp^* \neg p$, 
${+}\supp^* q$ and ${+}\supp^* \neg q$.

$T(D)$ contains the following rules and superiority relation.
\begin{tabbing}
1234\=12341234123412\=34123412341234123412\=34123412341234\kill
\ignore{
\>$str(p): \hspace{24pt}  p \rightarrow strict(p)$
\>\> $nstr(p): \hspace{15pt}  \Rightarrow  \neg strict(p)$ \\
\>$str(\neg p): \hspace{10pt}  \neg p \rightarrow strict(\neg p)$
\>\> $nstr(\neg p): \hspace{09pt}   \Rightarrow \neg strict(\neg p)$ \\

\>$str(q): \hspace{24pt}  q \rightarrow strict(q)$
\>\> $nstr(q): \hspace{15pt}  \Rightarrow  \neg strict(q)$ \\
\>$str(\neg q): \hspace{10pt}  \neg q \rightarrow strict(\neg q)$
\>\> $nstr(\neg q): \hspace{09pt}   \Rightarrow \neg strict(\neg q)$ \\
\ \\
\>$t(p): \hspace{24pt}  strict(p) \Rightarrow true(p)$
\>\> $nt(p): \hspace{15pt}  \Rightarrow  \neg true(p)$ \\
\>$t(\neg p): \hspace{10pt}  strict(\neg p) \Rightarrow true(\neg p)$
\>\> $nt(\neg p): \hspace{09pt}   \Rightarrow \neg true(\neg p)$ \\

\>$t(q): \hspace{24pt}  strict(q) \Rightarrow true(q)$
\>\> $nt(q): \hspace{15pt}  \Rightarrow  \neg true(q)$ \\
\>$t(\neg q): \hspace{10pt}  strict(\neg q) \Rightarrow true(\neg q)$
\>\> $nt(\neg q): \hspace{09pt}   \Rightarrow \neg true(\neg q)$ \\

\ \\
}
\> $n_d(r_1, r_2):\hspace{27pt} \Rightarrow \neg \f(p)$
\>\>$p_d(r_1): \hspace{17pt} \neg true(\neg p) \Rightarrow \f(p)$
\\
\> $n_d(r_2, r_1):\hspace{27pt} \Rightarrow \neg \f(\neg p)$
\>\>$p_d(r_2): \hspace{24pt}  \neg true(p) \Rightarrow \f(\neg p)$
\\
\> $n_d(r_3, r_4):\hspace{27pt} \Rightarrow \neg \f(\neg q)$
\>\>$p_d(r_3): \hspace{1pt}  \neg p, \neg true(\neg q) \Rightarrow \f(\neg q)$
\\
\> $n_d(r_4, r_3):\hspace{10pt}\neg p \Rightarrow \neg \f(q)$
\>\>$p_d(r_4): \hspace{24pt}  \neg true(q) \Rightarrow \f(q)$
\\
\\
\>$\hspace{7pt} \f(p) \Rightarrow p$ \\
\>$\f(\neg p) \Rightarrow \neg p$ \\
\>$\hspace{7pt} \f(q) \Rightarrow q$ \\
\>$\f(\neg q) \Rightarrow \neg q$ \\

\\
\> $n_d(r_1, r_2) >' p_d(r_1)$
\>\> $nstr(p) >' str(p)$ \\
\> $n_d(r_2, r_1) >' p_d(r_2)$
\>\> $nstr(\neg p) >' str(\neg p)$ \\
\> $n_d(r_3, r_4) >' p_d(r_3)$
\>\> $nstr(q) >' str(q)$ \\
\> $n_d(r_4, r_3) >' p_d(r_4)$
\>\> $nstr(\neg q) >' str(\neg q)$ \\

\end{tabbing}

The rules concerning $strict$ and $true$ have been omitted.
Because there are no facts or strict rules in $D$ we will infer ${-}\delta^* strict(s)$,
and hence ${+}\delta^* \neg true(s)$ and ${+}\supp^* \neg true(s)$
for each literal $s \in \Sigma$.
However, because of the superiority of $n_d$ over $p_d$,
we infer ${-}\supp^* \f(\neg p)$ and ${-}\delta^* \f(\neg p)$ and hence ${-}\supp^* \neg p$ and ${-}\delta^* \neg p$
(and similarly for $p$).
Hence, the body of $p_d(r_3)$ fails, and we infer ${-}\delta^* \neg q$.
Similarly, the body of $n_d(r_4, r_3)$ fails, and hence we infer ${+}\delta^* q$.
This reflects the ambiguity blocking behaviour of $\DL(\partial^*)$ from within
the ambiguity propagating logic $\DL(\delta^*)$.
\end{example}

\ignore{
As with the previous simulation,
the facts and strict rules of $D$ and $T(D)$ are the same, except for rules for $strict(q)$ in $T(D)$.
Thus, again, for any addition $A$, 
$D{+}A$ and $T(D){+}A$ draw the same strict conclusions in $\Sigma(D{+}A)$.
Furthermore, these conclusions are reflected in the defeasible conclusions of $strict(q)$ and $\neg true(q)$,
and also in support conclusions.

\begin{lemma}  \label{lemma:strictAP}
Let $D$ be a defeasible theory, 
$T(D)$ be the transformed defeasible theory as described in Definition \ref{defn:APsimAB},
and let $A$ be a modular defeasible theory.
Let $\Sigma$ be the language of $D{+}A$ and let $q \in \Sigma$.
Then
\begin{itemize}
\item
$D{+}A \vdash {+}\Delta q$ iff $T(D){+}A \vdash {+}\Delta q$ \hspace{32.5pt} iff $T(D){+}A \vdash {+}\delta^* strict(q)$ 

\hspace{67pt}
iff $T(D){+}A \vdash {+}\delta^* true(q)$  \hspace{5pt} iff $T(D){+}A \vdash {+}\supp^* true(q)$

\hspace{67pt}
iff $T(D){+}A \vdash -\delta^* \neg true(q)$ iff $T(D){+}A \vdash -\supp^* \neg true(q)$ \\

\item
$D{+}A \vdash -\Delta q$ iff $T(D){+}A \vdash -\Delta q$ \hspace{32.5pt} iff $T(D){+}A \vdash -\delta^* strict(q)$  

\hspace{67pt}
iff $T(D){+}A \vdash -\delta^* true(q)$  \hspace{5pt} iff $T(D){+}A \vdash -\supp^* true(q)$

\hspace{67pt}
iff $T(D){+}A \vdash {+}\delta^* \neg true(q)$ iff $T(D){+}A \vdash {+}\supp^* \neg true(q)$

\end{itemize}
\end{lemma}

We need this more detailed characterization of strict consequence,
compared to Lemma \ref{lemma:strictAB},
because both $\delta^*$ and $\supp^*$ intermediate conclusions influence $\delta^*$ conclusions.

The next lemma is a key part of the proof.
It shows that the structure of $T(D){+}A$ tightly constrains the inferences that can be made
in the sense that, for the literals of interest,
the inference rules $\delta^*$ and $\supp^*$ draw the same conclusions.

\begin{lemma}  \label{lemma:tight}
Let $D$ be a defeasible theory, 
$T(D)$ be the transformed defeasible theory as described in Definition \ref{defn:APsimAB},
and let $A$ be a modular set of facts.
Let $\Sigma$ be the language of $D{+}A$ extended with 
literals of the forms $\f(p)$, $\neg \f(p)$ and $\neg true(p)$, for $p \in \Sigma(D)$.
 
Then, for any $q \in \Sigma$,
\begin{itemize}
\item
$T(D){+}A \vdash {+}\delta^* q$  iff $T(D){+}A \vdash {+}\supp^* q$ 
\item
$T(D){+}A \vdash {-}\delta^* q$  iff $T(D){+}A \vdash {-}\supp^* q$ 
\end{itemize}
\end{lemma}

As a consequence of the inclusion theorem and the previous lemma,
any inference rule between $\supp^*$ and $\delta^*$ 
(that is, any inference rule except for $\Delta$ and $\delta$)
behaves the same way on $\Sigma$-literals
in $T(D){+}A$.
In particular, it applies to $\partial^*$.

\ignore{
\begin{corollary}  \label{cor:tight}
Let $\Sigma$ be the language of $D$, $\Sigma'$ be as defined in the previous lemma.
Let $A$ be any set of facts.
Then if $q \in \Sigma'$

\begin{itemize}
\item
$T(D){+}A \vdash {+}\delta^* q$  iff $T(D){+}A \vdash {+}\partial^* q$ 
\item
$T(D){+}A \vdash {-}\delta^* q$  iff $T(D){+}A \vdash {-}\partial^* q$ 
\end{itemize}
\end{corollary}
}
Now we show that the transformation preserves the $\partial^*$ consequences of $D{+}A$.

\begin{theorem}   \label{thm:partial*}
Let $D$ be a defeasible theory, 
$T(D)$ be the transformed defeasible theory as described in Definition \ref{defn:APsimAB},
and let $A$ be a modular set of facts.
Let $\Sigma$ be the language of $D{+}A$ and let $q \in \Sigma$.
Then
\begin{itemize}
\item
$D{+}A \vdash +\partial^* q$ iff
$T(D){+}A \vdash +\partial^* q$
\item
$D{+}A \vdash -\partial^* q$ iff
$T(D){+}A \vdash -\partial^* q$
\end{itemize}
\end{theorem}

Combining Theorem \ref{thm:partial*} with Lemma \ref{lemma:tight} and the inclusion theorem,
we see that $\DL(\partial^*)$ can be simulated by $\DL(\delta^*)$ and $\DL(\delta)$.
\ignore{
\begin{theorem}   \label{thm:APsimAB}
Let $D$ be a defeasible theory, 
$T(D)$ be the transformed defeasible theory as described in Definition \ref{defn:APsimAB},
and let $A$ be a modular set of facts.
Let $\Sigma$ be the language of $D{+}A$, let $q \in \Sigma$ and let $d \in \{\delta, \delta^*, \partial \}$.
Then
\begin{itemize}
\item
$D{+}A \vdash +\partial^* q$ iff
$T(D){+}A \vdash +d q$
\item
$D{+}A \vdash -\partial^* q$ iff
$T(D){+}A \vdash -d q$
\end{itemize}
\end{theorem}
}

}
The proof of correctness of this simulation is complicated by the fact that
inference rules for $\delta^*$ and $\supp^*$ are defined mutually recursively,
while the inference rules for $\partial^*$ are directly recursive.
This difference in structure makes a direct inductive proof difficult.
The problem is resolved by a ``tight'' simulating transformation that is able to simulate $\DL(\partial^*)$
(wrt addition of facts) in any of the $\DL$ logics.

\begin{theorem}   \label{thm:APsimAB}
For $d \in \{\delta, \delta^*, \partial \}$,
$\DL(d)$ can simulate $\DL(\partial^*)$ with respect to addition of facts
\end{theorem}
\skipit{
\begin{proof}
$D{+}A \vdash +\partial^* q$ iff
$T(D){+}A \vdash +\partial^* q$ (by Theorem \ref{thm:partial*})
iff
$T(D){+}A \vdash +\delta^* q$ (by Corollary \ref{cor:tight})
iff
$T(D){+}A \vdash +d q$ (by Lemma \ref{lemma:tight} and the inclusion theorem).
The proof is similar for $-\partial^* q$.
\end{proof}
}

Combining Theorems \ref{thm:ABsimAP} and \ref{thm:APsimAB} with results from \cite{Maher12},
we see that all logics of the $\DL$ framework are equally expressive
in terms of simulation wrt addition of facts.

\section{Simulation of Individual Defeat wrt Addition of Rules}

\finish{define competitor}

The following definition defining $D' = (F', R', <')$ from $D$ is repeated from \cite{Maher12}.
\begin{definition}   \label{defn:TDsimID}
We add the following rules
\begin{enumerate}
\item
The facts of $D'$ are the facts of $D$.
That is, $F' = F$.
\item
For each rule $r = B \ARROW_r q$ in $R$,  $R'$ contains
\[
\begin{array}{lrll}
p(r): & B & \ARROW_r & h(r) \\
s(r):  & h(r) &  \rightarrow & q
\end{array}
\]

\noindent
and, further, for each rule $r' = B' \ARROW_{r'} \non q$ for $\non q$ in $R$, $R'$ contains

\[
\begin{array}{lrll}
n(r, r'):  & B' & \ARROW_{r'} & \neg h(r)
\end{array}
\]

\item
For every $r > r'$ in $D$, where $r$ and $r'$ are rules for opposite literals,
$D'$ contains $p(r) >' n(r, r')$ and $n(r', r) >' p(r')$.

\end{enumerate}
\end{definition}

It was shown in \cite{Maher12} that, using this transformation, 
$\DL(\partial)$ simulates $\DL(\partial^*)$ and
$\DL(\delta)$ simulates $\DL(\delta^*)$, wrt addition of facts.

On the surface, it might appear that this result extends readily to addition wrt rules:
since the added rules do not participate in the superiority relation of the combined theory,
it might be expected that the difference between team defeat and individual defeat is irrelevant.
However, that expectation is misleading.
The following example shows that
this transformation
\emph{does not} provide a simulation of $\DL(\partial^*)$ by $\DL(\partial)$ wrt addition of \emph{rules}.

\begin{example}   \label{ex:failAB}
Let $D$ consist of the rules

\[
\begin{array}{lrll}
r_1:   &                           & \Rightarrow & \phantom{\neg} p \\ 
r_2:   &                           & \Rightarrow &  \neg p \\
\end{array}
\]

Then $T(D)$ consists of the following rules

\[
\begin{array}{lrll}
p(r_1):          &                           & \Rightarrow & \phantom{\neg} h(r_1) \\ 
n(r_1, r_2):   &                           & \Rightarrow & \neg h(r_1) \\ 
p(r_2):           &                           & \Rightarrow & \phantom{\neg} h(r_2) \\ 
n(r_2, r_1):   &                           & \Rightarrow & \neg h(r_2) \\ 
s(r_1):           &       h(r_1)          & \Rightarrow & \phantom{\neg} p \\ 
s(r_2):            &       h(r_2)        & \Rightarrow &  \neg p \\
\end{array}
\]

Now, let $A$ be the rule
\[
\begin{array}{lrll}
   &                           & \Rightarrow & p \\ 
   \ \\
\end{array}
\]

\noindent
Clearly, $D+A \vdash -\partial^* p$ (and $D+A \vdash -\partial^* \neg p$),
since $r_2$ cannot be overruled.
However, $T(D)+A \vdash -\partial h(r_2)$, 
since $n(r_2, r_1)$ cannot be overruled,
and hence $s(r_2)$ fails.
This leaves the rule for $p$ in $A$ without competition, and so
$T(D)+A \vdash +\partial p$.
\end{example}

\ignore{
To fix the rules problem
$o(r):  B_r => one(q)$
$o(q): one(q) => q$
$t(q): => q$
$o(q) > t(\non q)$

if $h(r)$ then also $one(q)$ and normal rules for  $\non q$ fail, $t(\non q)$ is overruled.
BUT $o(\non q)$ is not overruled.
So, add $s(r) > o(\non q)$.

if $one(q)$ and $one(\non q)$ and all $h(r)$'s fail (for $q$ and $\non q$)
then there is only the $o$ and $t$ rules for $q$ and $\non q$.
both o's are undefeated.

If $B_s$ is infinite and none succeed
then $one(\non q)$ is infinite and so $\non q$ doesnt fail
so $q$ doesnt succeed ... \finish{how to fix?} ... its ok b/c $s(r) > o(\non q)$

What about:
$h(r) => succ(q)$
$s(q): succ(q) => q$
$o(q): one(q) => q$
$o(q) > t(\non q)$
$s(q) > o(\non q)$

Does this work for $\delta$?????????????
$t(q)$ doesnt work when there are no rules for $q$ and $\non q$.
In that case $D + =>q \vdash +d q$ but $T(D) + =>q \vdash -d q$.

Just $o(q)$ might not work when $q$ neither succeeds nor fails (~= infinite).
Then either 
(1) all rules for $q$ fail or are infinite, with >=1 infinite, or
(2) some rule for $q$ succeeds and some rule for $\non q$ is infinite, or
(3) some rules for $q$ and $\non q$ succeed but rule for $q$ overrules rule(s) for $\non q$,
and  $\non q$ has an infinite rule that is not overruled.

NEED to check what i mean by succeeds/fail .. should be $D{+}A \vdash +d rule body$.
Similarly failure.

.....
}

A similar but more complex example (given in the appendix) shows the transformation
also does not provide  a simulation of $\DL(\delta^*)$ by $\DL(\delta)$ wrt addition of rules.
These problems arise because if 
a rule body succeeds in $D$ for each of $q$ and $\non q$, and the rules are not overruled,
the simulation $D'$ has all bodies for $q$ and $\non q$ failing.
An applicable rule for $q$ (or $\non q$) in $A$ thus has a competitor in $D$, but not in $D'$.
In this way $D'$ differs from $D$, and the examples show that
addition of rules can make this difference observable.

To avoid these problems, we add extra rules to those in Definition \ref{defn:TDsimID}.

\begin{definition}   \label{defn:newTDsimID}
We define $T(D)$ as the theory $(F', R', >')$
consisting of the facts, rules and superiority statements
from $D'$ in Definition \ref{defn:TDsimID},
and the following.
\begin{enumerate}
\setcounter{enumi}{3}
\item
For each literal $q$, $R'$ contains
\[
\begin{array}{lrll}
o(q): & one(q) & \leadsto &  q
\end{array}
\]
\item
For each rule $r = B_r \ARROW_r q$ in $R$,  $R'$ contains
\[
\begin{array}{lrll}
 & B_r & \Rightarrow & one(q) \\
\end{array}
\]
\item
For every rule $r$ for $q$,
$>'$ contains $s(r) >' o(\non q)$.
\end{enumerate}
\end{definition}

Parts 4 and 5 of this definition introduce an additional rule for each literal $\non q$ which, however,
is subordinate to the methods to derive $q$ in the original transformation
in the sense that a derivation of $q$ in the original transformation
will overrule (part 6) a derivation of $\non q$ using part 4.
The rules in part 4 are defeaters, so they cannot be used to derive any conclusions.

The effect of the extended definition on Example \ref{ex:failAB} is to add the following
to the transformed theory:

\[
\begin{array}{llrllllll}
 \Rightarrow & one(p) &
o(p)~~~: & one(p) & \leadsto & \phantom{\neg} p \\
 \Rightarrow & one(\neg p) &
o(\neg p): & one(\neg p) & \leadsto & \neg p \\
 \\
 & s(r_1) > o(\neg p) & & & & s(r_2) > o(p) \\
\end{array}
\]
\smallskip

We now have $T(D)+A \vdash -\partial p$,
since the rule $o(\neg p)$ provides a non-failed competitor to the rule in $A$.
More generally, we find that, through the extended transformation,
team defeat logics can simulate the corresponding individual defeat logics
with respect to addition of rules.

\begin{theorem}  \label{thm:TDsimAD}
The logic $\DL(\partial)$ can simulate $\DL(\partial^*)$
with respect to addition of rules.

\noindent
The logic $\DL(\delta)$  can simulate $\DL(\delta^*)$
with respect to addition of rules.
\end{theorem}

\section{Simulation of Team Defeat wrt Addition of Rules}

\finish{linking text}

The same theory $D$ and addition $A$ as in Example \ref{ex:failAB} demonstrates
that the simulation of $\DL(\partial)$ by $\DL(\partial^*)$ wrt addition of facts exhibited in \cite{Maher12}
does not extend to addition of rules.

The transformation below modifies the one of \cite{Maher12} by
treating strict rules differently (following Definition \ref{defn:APsimAB}),
adding a competitor for each literal $q$ (following Definition \ref{defn:newTDsimID}),
and employing separate defeasible rules to accommodate differences between the
$\delta$ and $\supp$ inference rules.
We use a construction to restrict one class of defeasible rules to use only in simulating $\supp$ inference;
it is not necessary to restrict the other class because $\delta \subseteq \supp$, by the inclusion theorem.

\finish{
Need example for $\delta$.  Need to use the revised transformation, though strictness should not affect it
}

\begin{definition}   \label{defn:IDsimTD}
We define the transformation $T$ of $D$ to $T(D) = (F', R', >')$ as follows:
\begin{enumerate}
\item  \label{t4:facts}
The facts of $T(D)$ are the facts of $D$.
That is, $F' = F$.
\item \label{t4:strict}
Every strict rule of $R$ is included in $R'$.

\item \label{t4:strictq}  
For every literal $q$, $R'$ contains

\[
\begin{array}{lrll}
str(q):     &              q           & \rightarrow & \phantom{\neg} strict(q) \\ 
nstr(q):   &                           & \Rightarrow &  \neg strict(q) \\
t(q):        &    strict(q)          & \Rightarrow & \phantom{\neg} true(q) \\ 
nt(q):     &                           & \Rightarrow &  \neg true(q) \\
\end{array}
\]

and the superiority relation contains
$nstr(q) >' str(q)$
and $t(q) >' nt(q)$, for every $q$.

\ignore{
\item  \label{pt:facts}
The facts of $T(D)$ are the facts of $D$.
That is, $F' = F$.
\item  \label{t4:strict}
For each literal $q$, and each strict rule $r = (B \rightarrow  q)$ in $R$, $R'$ contains 
\[
\begin{array}{lrlrl}
ns(q): &   & \Rightarrow & \neg & strict(q) \\
s(r):    & B & \rightarrow &          & strict(q) \\
\end{array}
\]
and $ns(q) >' s(r)$.
\item  \label{t4:strict2}
For each literal $q$ defined by at least one strict rule in $R$,  $R'$ contains 
\[
\begin{array}{lrll}
& strict(q) & \rightarrow & q \\
\end{array}
\]
}

\item  \label{t4:dft}
For each ordered pair of opposing rules $r_i = (B_i \ARROW_{i} \non q)$ and $r_j = (B_j \ARROW_{j} q)$ in $R$,  
where 
$r_j$ is not a defeater, $R'$ contains
\[
\begin{array}{lrlrl}
R1_{ij}: & B_i &  \ARROW_i & \neg & d(r_i, r_j) \\
R2_{ij}: & B_j &  \Rightarrow && d(r_i, r_j) \\
R3_{ij}:  & true(q) &  \Rightarrow && d(r_i, r_j) \\
          & d(r_i, r_j) & \Rightarrow && d(r_i) \\
        & fail(r_i) & \Rightarrow & & d(r_i) \\
NF_i:  & B_i & \Rightarrow & \neg & fail(r_i) \\
F_i:  &           & \Rightarrow && fail(r_i) \\
\end{array}
\]
and
$R2_{ij} >' R1_{ij}$ iff $r_j > r_i$, 
$R3_{ij} >' R1_{ij}$ for every $i$ and $j$,
and $NF_i > F_i$ for every $i$. 

If there is no strict or defeasible rule $r_j$ for $q$ in $D$ then only the last three rules appear in $R'$, for each $i$.

\item  \label{t4:one}
For each literal $q$, and each strict or defeasible rule $r = (B_r \ARROW_r  q)$ in $R$, $R'$ contains 
\[
\begin{array}{lrll}
& B_r & \Rightarrow & one(q) \\
\end{array}
\]
\item  \label{t4:oneq}
For each literal $q$,  $R'$ contains 
\[
\begin{array}{lrll}
s(q): & one(q), \neg true(\non q), d(s_1), \ldots, d(s_k) & \Rightarrow & q \\
\end{array}
\]
where $s_1, \ldots, s_k$ are the rules for $\non q$
\item  \label{t4:support}
For each literal $q$ and for each strict or defeasible rule $r$ for $q$,  $R'$ contains 
\[
\begin{array}{lrll}
supp(q): & B_r,  d_\supp(s_1, r), \ldots, d_\supp(s_k, r), g, \neg g & \Rightarrow & q \\
\end{array}
\]
where 
$B_r$ is the body of $r$, 
$s_1, \ldots, s_k$ are the rules for $\non q$,
and for every strict or defeasible rule $r$ and opposing rule $s$,  $R'$ contains 
\[
\begin{array}{lrll}
a(s,r): & B_s & \Rightarrow & \neg d_\supp(s, r) \\
b(s,r): & B_r & \Rightarrow & \phantom{\neg} d_\supp(s, r) \\
\end{array}
\]
The superiority relation contains $a(s,r) > b(s,r)$ iff $s > r$.
$R'$ also contains the rules
\[
\begin{array}{lrll}
 &  & \Rightarrow & \phantom{\neg} g \\
 &  & \Rightarrow & \neg g \\
\end{array}
\]

\item   \label{t4:o}
For each rule $r = B_r \ARROW_r q$ in $R$,  $R'$ contains
\[
\begin{array}{lrll}
 & B_r & \Rightarrow & o(q) \\
\end{array}
\]

\item  \label{t4:oq}
For each literal $q$, $R'$ contains
\[
\begin{array}{lrll}
o(q): & o(q) & \leadsto &  q
\end{array}
\]
and $>'$ contains $s(q) >' o(\non q)$.
\end{enumerate}
\end{definition}

\ignore{ OLD VERSION
\begin{definition}   \label{defn:IDsimTD}
We define the transformation $T$ of $D$ to $T(D) = (F', R', >')$ as follows:
\begin{enumerate}
\item  \label{t4:facts}
The facts of $T(D)$ are the facts of $D$.
That is, $F' = F$.
\item \label{t4:strict}
Every strict rule of $R$ is included in $R'$.

\item \label{t4:strictq}  
For every literal $q$, $R'$ contains

\[
\begin{array}{lrll}
str(q):     &              q           & \rightarrow & \phantom{\neg} strict(q) \\ 
nstr(q):   &                           & \Rightarrow &  \neg strict(q) \\
t(q):        &    strict(q)          & \Rightarrow & \phantom{\neg} true(q) \\ 
nt(q):     &                           & \Rightarrow &  \neg true(q) \\
\end{array}
\]

and the superiority relation contains
$nstr(q) >' str(q)$
and $t(q) >' nt(q)$, for every $q$.

\ignore{
\item  \label{pt:facts}
The facts of $T(D)$ are the facts of $D$.
That is, $F' = F$.
\item  \label{t4:strict}
For each literal $q$, and each strict rule $r = (B \rightarrow  q)$ in $R$, $R'$ contains 
\[
\begin{array}{lrlrl}
ns(q): &   & \Rightarrow & \neg & strict(q) \\
s(r):    & B & \rightarrow &          & strict(q) \\
\end{array}
\]
and $ns(q) >' s(r)$.
\item  \label{t4:strict2}
For each literal $q$ defined by at least one strict rule in $R$,  $R'$ contains 
\[
\begin{array}{lrll}
& strict(q) & \rightarrow & q \\
\end{array}
\]
}

\item  \label{t4:dft}
For each ordered pair of opposing rules $r_i = (B_i \ARROW_{i} \non q)$ and $r_j = (B_j \ARROW_{j} q)$ in $R$,  
where 
$r_j$ is not a defeater, $R'$ contains
\[
\begin{array}{lrlrl}
R1_{ij}: & B_i &  \ARROW_i & \neg & d(r_i, r_j) \\
R2_{ij}: & B_j &  \Rightarrow && d(r_i, r_j) \\
R3_{ij}:  & true(q) &  \Rightarrow && d(r_i, r_j) \\
          & d(r_i, r_j) & \Rightarrow && d(r_i) \\
        & fail(r_i) & \Rightarrow & & d(r_i) \\
NF_i:  & B_i & \Rightarrow & \neg & fail(r_i) \\
F_i:  &           & \Rightarrow && fail(r_i) \\
\end{array}
\]
and
$R2_{ij} >' R1_{ij}$ iff $r_j > r_i$, 
 $R3_{ij} >' R1_{ij}$ for every $i$ and $j$,
and $NF_i > F_i$ for every $i$. 

If there is no strict or defeasible rule $r_j$ for $q$ in $D$ then only the last three rules appear in $R'$, for each $i$.

\item  \label{t4:one}
For each literal $q$, and each strict or defeasible rule $r = (B \ARROW_r  q)$ in $R$, $R'$ contains 
\[
\begin{array}{lrll}
& B & \Rightarrow & one(q) \\
\end{array}
\]
\item  \label{t4:oneq}
For each literal $q$,  $R'$ contains 
\[
\begin{array}{lrll}
s(q): & one(q), not\Delta(\non q), d(r_1), \ldots, d(r_k) & \Rightarrow & q \\
\end{array}
\]
where $r_1, \ldots, r_k$ are the rules for $\non q$
\item  \label{t4:notDELTA}
For each literal $q$,  $R'$ contains 
\[
\begin{array}{lrll}
 & \neg true(\non q) & \Rightarrow & not\Delta(q) \\
 & p & \Rightarrow & not\Delta(q) \\
\end{array}
\]
$R'$ also contains the rules
\[
\begin{array}{lrll}
 &  & \Rightarrow & \phantom{\neg} g \\
 &  & \Rightarrow & \neg g \\
\end{array}
\]

\item  \label{t4:oq}
For each literal $q$, $R'$ contains
\[
\begin{array}{lrll}
o(q): & o(q) & \leadsto &  q
\end{array}
\]
and $>'$ contains $s(q) >' o(\non q)$.
\item   \label{t4:o}
For each rule $r = B_r \ARROW_r q$ in $R$,  $R'$ contains
\[
\begin{array}{lrll}
 & B_r & \Rightarrow & o(q) \\
\end{array}
\]
\end{enumerate}
\end{definition}
}

Parts \ref{t4:facts}--\ref{t4:strictq} allow us to characterize strict conclusions.
Part \ref{t4:dft} expresses whether a rule is defeated or not,
while part \ref{t4:oneq} expresses that $q$ can be concluded if 
there is an applicable strict or defeasible rule for $q$,
all attempts to strictly derive $\non q$ fail finitely,
and all opposing rules are defeated.
While this expresses properly the inference rules for $\partial$ and $\delta$, 
the inference rule for $\supp$ omits the condition on strict derivation of $\non q$ and has a slightly different form of defeat.
We need part \ref{t4:support} to express inference (and defeat) for $\supp$.
$g$ and $\neg g$ are used to restrict the applicability of this rule to $\supp^*$;
we have $T(D){+}A \vdash +\supp^* g$, but
$T(D){+}A \vdash -\partial^* g$ and $T(D){+}A \vdash -\delta^* g$ (and the same for $\neg g$).
Parts \ref{t4:o} and \ref{t4:oq} redress the lack of a competitor in the same way as in Definition \ref{defn:newTDsimID}.

\begin{theorem}   \label{thm:IDsimTD}
The logic $\DL(\partial^*)$ can simulate $\DL(\partial)$
with respect to addition of rules.

\noindent
The logic $\DL(\delta^*)$  can simulate $\DL(\delta)$
with respect to addition of rules.
\end{theorem}

\finish{
This example demonstrates that the simulation of \cite{Maher12} wrt addition of facts
is little more than simple simulation without addition.
The addition of facts only affects defeasible inference via strict inference.
This example shows that even the addition of defeasible ``facts''
is enough to break the transformation.
??????????????????????????????????

IN the conclusion, or earlier:

Example \ref{ex:failAB} only demonstrates that the transformation
in Definition \ref{defn:TDsimID} does not provide a simulation of $\DL(\partial^*)$ by $\DL(\partial)$
wrt addition of rules;
it leaves open the question of whether some other transformation can achieve this.
Similarly the question of whether 
$\DL(\delta^*)$ can be simulated by $\DL(\delta)$ wrt addition of rules
is left open by Example \ref{ex:failAP}.
}
\ignore{
Combined definition, just for convenience: \\
\ \\
\ \\

\begin{enumerate}
\item
The facts of $\T(D)$ are the facts of $D$.
That is, $F' = F$.
\item
For each rule $r = B \ARROW_r q$ in $R$,  $R'$ contains
\[
\begin{array}{lrll}
p(r): & B & \ARROW_r & h(r) \\
s(r):  & h(r) &  \rightarrow & q
\end{array}
\]

\noindent
and, further, for each rule $r' = B' \ARROW_{r'} \non q$ for $\non q$ in $R$, $R'$ contains

\[
\begin{array}{lrll}
n(r, r'):  & B' & \ARROW_{r'} & \neg h(r)
\end{array}
\]

\item
For every $r > r'$ in $D$, where $r$ and $r'$ are rules for opposite literals,
$T(D)$ contains $p(r) >' n(r, r')$ and $n(r', r) >' p(r')$.

\item
For each literal $q$, $R'$ contains
\[
\begin{array}{lrll}
o(q): & one(q) & \Rightarrow &  q
\end{array}
\]
\item
For each rule $r = B_r \ARROW_r q$ in $R$,  $R'$ contains
\[
\begin{array}{lrll}
 & B_r & \Rightarrow & one(q) \\
\end{array}
\]
\item
For every rule $r$ for $q$,
$>'$ contains $s(r) >' o(\non q)$.

\end{enumerate}
}

\finish{
simulating NTD with TD (or vice versa?):

Interestingly, this transformation uses the superiority relation only when there are two single rules opposed.
In this case team defeat and non-team defeat produce the same conclusions.
Thus this transformation is also correct for transforming a theory within the logic $\DL(\partial)$.
}

\section{Conclusions}

\begin{figure}[t]
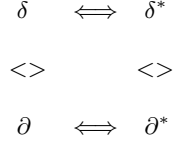
   

\[
\begin{array}{lcl}
        \delta          &  \Longleftrightarrow  & \delta^*     \\
 \\
               \!\!<>             &                                 &  \!\!<>   \\
 \\
        \partial          &  \Longleftrightarrow  & \partial^*     \\
\end{array}
\]
\caption{Relative expressiveness of logics in $\DL$ using simulation wrt addition of rules}
\label{fig:RE}
\end{figure}

We have shown that the logics of the $\DL$ framework are equally expressive
when relative expressiveness is formulated as ability to simulate in the presence of additional facts.
This involved the introduction of two new transformations simulating, respectively,
a logic that blocks ambiguity and a logic that propagates ambiguity.

We also completed the study of relative expressiveness wrt addition of rules.
Figure \ref{fig:RE} shows this relation on the logics in $\DL$,
where an arrow from $d_1$ to $d_2$ expresses that $DL(d_1)$ can be simulated by $DL(d_2)$
with respect to the addition of rules.
$<>$ between tags expresses that the two corresponding logics have incomparable expressiveness.
It is clear that $\DL$ breaks into two classes of logics of different expressiveness.

While the issue of relative expressiveness within the framework $\DL$ is now largely resolved,
this same approach can be applied to relate these logics to other logics.
We can expect the same results for the $\WFDL$ logics \cite{MG99,WFDL},
because of their similarity to $\DL$,
but their relation to the defeasible logics of Nute and Maier \cite{MN06,MN10}
will be of interest.
Even more interesting will be to address other systems of defeasible reasoning, such as argumentation
\cite{Dung95,RahwanSimari}.

\ignore{
\section{Conclusion}

We have shown that the logics of the $\DL$ framework are equally expressive
when relative expressiveness is formulated as ability to simulate in the presence of additional facts.
This involved demonstrating simulations of ambiguity blocking logics in ambiguity propagating logics,
and simulations of ambiguity propagating logics in ambiguity blocking logics.
The existence of such simulations is somewhat surprising, given the different treatments of ambiguity.

We then investigated a stronger notion of relative expressiveness,
where simulation must be achieved in the presence of additional rules.
We showed that each logic with individual defeat has the same expressiveness as the corresponding logic
with team defeat.
reflecting an intuition that inference with team defeat is more sophisticated than with individual defeat.

\finish{what about individual defeat simulating team defeat??????
if we have that then:

We showed that logics with team defeat are no more expressive than logics with individual defeat,
despite an intuition that suggested differently.
Thus we have completed the analysis of the $\DL$ framework of defeasible logics begun in \cite{Maher12}
on these two notions of relative expressiveness.

}

While the issue of relative expressiveness within the framework $\DL$ is now largely resolved,
this same approach can be applied to relate these logics to other logics.
We can expect the same results for the $\WFDL$ logics \cite{MG99,WFDL},
but their relation to the defeasible logics of Nute and Maier \cite{MN06,MN10}
will be of interest.
Even more interesting will be to address other systems of defeasible reasoning, such as argumentation
\cite{Dung95,RahwanSimari}.
}
\ignore{

We have seen that relative expressiveness in defeasible logics is not directly related to
the relative inference strength of the logics.
Logics with or without team defeat are equally expressive while, on the other hand,
the alternative treatments of ambiguity are distinct and independent in terms of expressiveness.
}

{\bf Acknowledgements:}  The author thanks the referees for their comments,
and UNSW, Canberra for a grant supporting this work.


\bibliographystyle{acmtrans}
\bibliography{emulate_2}

\appendix

\input{emulate_2_app_body}

\end{document}

%% file: emulate_2_app_body.tex
\begin{appendix}

\section{Appendix}

This appendix contains the inference rules for the logics in $\DL$,
proofs of results in the body of the paper,
and some examples.
Theorems, Lemmas, or Examples numbered 1--14 refer to items in the body of the paper.
Larger numbers refer to items in this appendix.

\section{ Inference Rules for $\DL$}

For every inference rule ${+}d$ there is a closely related inference rule $-d$
allowing to infer that some literals $q$ cannot be consequences of $D$ via $+d$.
The relationship between $+d$ and $-d$ is described as the Principle of Strong Negation \cite{flexf}.
These inference rules are placed adjacently to emphasize this relationship.

Some notation in the inference rules requires explanation.
Given a literal $q$, its complement $\non q$ is defined as follows:
if $q$ is a proposition then $\non q$ is $\neg q$; if $q$ has form $\neg p$ then $\non q$ is $p$.
We say $q$ and $\non q$ (and the rules with these literals in the head)  \emph{oppose} each other.
$R_s$ ($R_{sd}$) denotes the set of strict rules (strict or defeasible rules) in $R$.
$R[q]$ ($R_s[q]$, etc) denotes the set of rules (respectively, strict rules) of $R$ with head $q$.
Given a rule $r$, $A(r)$ denotes the set of literals in the body (or antecedent) of $r$.
\smallskip

{\small
\begin{minipage}[t]{.45\textwidth}
\begin{tabbing}
$+\Delta)$  $+\Delta q \in \T_D(E)$  iff  either  \\
\hspace{0.2in}  .1)  $q \in F$; or  \\
\hspace{0.2in}  .2)  $\exists r \in R_{s}[q]$  such that \\
\hspace{0.4in}      .1)  $\forall a \in A(r),  +\Delta a  \in  E$ \\
\end{tabbing}
\end{minipage}
\begin{minipage}[t]{.10\textwidth}
\hspace{1.0cm}
\end{minipage}
\begin{minipage}[t]{.45\textwidth}
\begin{tabbing}
$-\Delta)$  $-\Delta q \in \T_D(E)$  iff  \\
\hspace{0.2in}  .1)  $q \notin  F$, and \\
\hspace{0.2in}  .2)  $\forall r \in R_{s}[q]$  \\
\hspace{0.4in}      .1)  $\exists a \in A(r),  -\Delta a  \in  E$ \\
\end{tabbing}
\end{minipage}

\noindent\begin{minipage}[t]{.45\textwidth}
\begin{tabbing}
$+\partial)$   $+\partial q \in \T_D(E)$  iff either \\
\hspace{0.2in}  .1)  $+\Delta q \in E$; or  \\
\hspace{0.2in}  .2)  The following three conditions all hold. \\
\hspace{0.4in}      .1)  $\exists r \in R_{sd}[q] \  \forall a \in A(r),  +\partial a \in E$,  and \\
\hspace{0.4in}      .2)  $-\Delta \non q \in E$,  and \\
\hspace{0.4in}      .3)  $\forall s \in R[\non q]$  either \\
\hspace{0.6in}          .1)  $\exists a \in A(s),  -\partial a \in E$;  or \\
\hspace{0.6in}          .2)  $\exists t \in R_{sd}[q]$  such that \\
\hspace{0.8in}              .1)  $\forall a \in A(t),  +\partial a \in E$,  and \\
\hspace{0.8in}              .2)  $t > s$.
\end{tabbing}
\end{minipage}
\begin{minipage}[t]{.45\textwidth}
\begin{tabbing}
$-\partial)$  $-\partial q \in \T_D(E)$  iff  \\
\hspace{0.2in}  .1)  $-\Delta q \in E$, and \\
\hspace{0.2in}  .2)  either \\
\hspace{0.4in}      .1)  $\forall r \in R_{sd}[q] \  \exists a \in A(r),  -\partial a \in E$; or \\
\hspace{0.4in}      .2)  $+\Delta \non q \in E$; or \\
\hspace{0.4in}      .3)  $\exists s \in R[\non q]$  such that \\
\hspace{0.6in}          .1)  $\forall a \in A(s),  +\partial a \in E$,  and \\
\hspace{0.6in}          .2)  $\forall t \in R_{sd}[q]$  either \\
\hspace{0.8in}              .1)  $\exists a \in A(t),  -\partial a \in E$;  or \\
\hspace{0.8in}              .2)  not$(t > s)$.\\
\end{tabbing}
\end{minipage}
\smallskip

\begin{minipage}[t]{.45\textwidth}
\begin{tabbing}
$+\partial^{*})$  $+\partial^* q \in \T_D(E)$  iff  either  \\
\hspace{0.2in}  .1)  $+\Delta q \in E$; or  \\
\hspace{0.2in}  .2)  $\exists r \in R_{sd}[q]$  such that \\
\hspace{0.4in}      .1)  $\forall a \in A(r),  +\partial^{*} a  \in  E$,  and\\
\hspace{0.4in}      .2)  $-\Delta  \non q  \in  E$,  and \\
\hspace{0.4in}      .3)  $\forall s  \in R[ \non q]$  either \\
\hspace{0.6in}          .1)  $\exists a \in A(s),  -\partial^{*}a  \in  E$;  or \\
\hspace{0.6in}          .2)  $r > s$.
\end{tabbing}
\end{minipage}
\begin{minipage}[t]{.07\textwidth}
\hspace{1.0cm}
\end{minipage}
\begin{minipage}[t]{.45\textwidth}
\begin{tabbing}
$-\partial^{*})$  $-\partial^* q \in \T_D(E)$  iff  \\
\hspace{0.2in}  .1)  $-\Delta q  \in  E$, and \\
\hspace{0.2in}  .2)  $\forall r \in R_{sd}[q]$  either \\
\hspace{0.4in}      .1)  $\exists a \in A(r),  -\partial^{*}a  \in  E$; or \\
\hspace{0.4in}      .2)  $+\Delta  \non q  \in  E$; or \\
\hspace{0.4in}      .3)  $\exists s  \in R[ \non q]$  such that \\
\hspace{0.6in}          .1)  $\forall a \in A(s),  +\partial^{*}a  \in  E$,  and \\
\hspace{0.6in}          .2)  not$(r > s)$. \\
\end{tabbing}
\end{minipage}
}

\smallskip
{\small
\smallskip
\noindent\begin{minipage}[t]{.45\textwidth}
\begin{tabbing}
$+\delta)$   $+\delta q \in \T_D(E)$  iff either \\
\hspace{0.2in}  .1)  $+\Delta q \in E$; or  \\
\hspace{0.2in}  .2)  The following three conditions all hold. \\
\hspace{0.4in}      .1)  $\exists r \in R_{sd}[q] \  \forall a \in A(r),  +\delta a \in E$,  and \\
\hspace{0.4in}      .2)  $-\Delta \non q \in E$,  and \\
\hspace{0.4in}      .3)  $\forall s \in R[\non q]$  either \\
\hspace{0.6in}          .1)  $\exists a \in A(s),  -\sigma a \in E$;  or \\
\hspace{0.6in}          .2)  $\exists t \in R_{sd}[q]$  such that \\
\hspace{0.8in}              .1)  $\forall a \in A(t),  +\delta a \in E$,  and \\
\hspace{0.8in}              .2)  $t > s$.
\end{tabbing}
\end{minipage}
\begin{minipage}[t]{.45\textwidth}
\begin{tabbing}
$-\delta)$   $-\delta q \in \T_D(E)$  iff \\
\hspace{0.2in}  .1)  $-\Delta q \in E$, and \\
\hspace{0.2in}  .2)  either \\
\hspace{0.4in}      .1)  $\forall r \in R_{sd}[q] \  \exists a \in A(r),  -\delta a \in E$; or \\
\hspace{0.4in}      .2)  $+\Delta \non q \in E$; or \\
\hspace{0.4in}      .3)  $\exists s \in R[\non q]$  such that \\
\hspace{0.6in}          .1)  $\forall a \in A(s),  +\sigma a \in E$,  and \\
\hspace{0.6in}          .2)  $\forall t \in R_{sd}[q]$  either \\
\hspace{0.8in}              .1)  $\exists a \in A(t),  -\delta a \in E$;  or \\
\hspace{0.8in}              .2)  not$(t > s)$.\\
\end{tabbing}
\end{minipage}

\begin{minipage}[t]{.45\textwidth}
\begin{tabbing}
$+\sigma)$   $+\sigma q \in \T_D(E)$  iff either  \\
\hspace{0.2in}  .1)  $+\Delta q \in E$; or\\
\hspace{0.2in}  .2)  $\exists r \in R_{sd}[q]$  such that  \\
\hspace{0.4in}      .1)  $\forall a \in A(r),  +\sigma a \in E$,  and \\
\hspace{0.4in}      .2)  $\forall s \in R[\non q]$  either \\
\hspace{0.6in}          .1)  $\exists a \in A(s),  -\delta a \in E$;  or \\
\hspace{0.6in}          .2)  not$(s > r)$.
\end{tabbing}
\end{minipage}
\begin{minipage}[t]{.10\textwidth}
\hspace{1.0cm}
\end{minipage}
\begin{minipage}[t]{.45\textwidth}
\begin{tabbing}
$-\sigma)$    $-\sigma q \in \T_D(E)$  iff \\
\hspace{0.2in}  .1)  $-\Delta q \in E$, and \\
\hspace{0.2in}  .2)  $\forall r \in R_{sd}[q]$  either \\
\hspace{0.4in}      .1)  $\exists a \in A(r),  -\sigma a \in E$;  or \\
\hspace{0.4in}      .2)  $\exists s \in R[\non q]$  such that \\
\hspace{0.6in}          .1)  $\forall a \in A(s),  +\delta a \in E$,  and \\
\hspace{0.6in}          .2)  $s > r$.\\
\end{tabbing}
\end{minipage}

\begin{minipage}[t]{.45\textwidth}
\begin{tabbing}
$+\delta^{*})$    $+\delta^{*}q \in \T_D(E)$  iff either \\
\hspace{0.2in}  .1)  $+\Delta q  \in  E$; or  \\
\hspace{0.2in}  .2)  $\exists r  \in R_{sd}[q]$  such that \\
\hspace{0.4in}      .1)  $\forall a \in A(r),  +\delta^{*}a  \in  E$,  and \\
\hspace{0.4in}      .2)  $-\Delta  \non q  \in  E$,  and \\
\hspace{0.4in}      .3)  $\forall s  \in R[ \non q]$  either \\
\hspace{0.6in}          .1)  $\exists a \in A(s),  -\sigma ^{*}a  \in  E$;  or \\
\hspace{0.6in}          .2)  $r > s$.
\end{tabbing}
\end{minipage}
\begin{minipage}[t]{.10\textwidth}
\hspace{1.0cm}
\end{minipage}
\begin{minipage}[t]{.45\textwidth}
\begin{tabbing}
$-\delta^{*})$    $-\delta^{*}q \in \T_D(E)$  iff \\
\hspace{0.2in}  .1)  $-\Delta q  \in  E$, and \\
\hspace{0.2in}  .2)  $\forall r \in R_{sd}[q]$  either \\
\hspace{0.4in}      .1)  $\exists a \in A(r),  -\delta^{*}a  \in  E$; or \\
\hspace{0.4in}      .2)  $+\Delta  \non q  \in  E$; or \\
\hspace{0.4in}      .3)  $\exists s  \in R[ \non q]$  such that \\
\hspace{0.6in}          .1)  $\forall a \in A(s),  +\sigma ^{*}a  \in  E$,  and \\
\hspace{0.6in}          .2)  not$(r > s)$.\\
\end{tabbing}
\end{minipage}

\begin{minipage}{.45\textwidth}
\begin{tabbing}
$+\sigma ^{*})$    $+\sigma ^{*}q \in \T_D(E)$  iff either  \\
\hspace{0.2in}  .1)  $+\Delta q  \in  E$; or  \\
\hspace{0.2in}  .2)  $\exists r  \in R_{sd}[q]$  such that  \\
\hspace{0.4in}      .1)  $\forall a \in A(r),  +\sigma ^{*}a  \in  E$,  and \\
\hspace{0.4in}      .2)  $\forall s  \in R[ \non q]$  either \\
\hspace{0.6in}          .1)  $\exists a \in A(s),  -\delta^{*}a  \in  E$;  or \\
\hspace{0.6in}          .2)  not$(s > r)$.
\end{tabbing}
\end{minipage}
\begin{minipage}[t]{.10\textwidth}
\hspace{1.0cm}
\end{minipage}
\begin{minipage}{.45\textwidth}
\begin{tabbing}
$-\sigma ^{*})$    $-\sigma ^{*}q \in \T_D(E)$  iff \\
\hspace{0.2in}  .1)  $-\Delta q  \in  E$, and \\
\hspace{0.2in}  .2)  $\forall r \in R_{sd}[q]$  either \\
\hspace{0.4in}      .1)  $\exists a \in A(r),  -\sigma ^{*}a  \in  E$;  or \\
\hspace{0.4in}      .2)  $\exists s  \in R[ \non q]$  such that \\
\hspace{0.6in}          .1)  $\forall a \in A(s),  +\delta^{*}a  \in  E$,  and \\
\hspace{0.6in}          .2)  $s > r$.
\end{tabbing}
\end{minipage}
}
\smallskip

\section{Proofs of results}
We now turn to proofs of results in the body of the paper,
and some examples.
This part of the appendix has the same structure as the paper itself, to make access easier.

All simulation proofs (of $\DL(d_1)$ by $\DL(d_2)$, say) have two parts:
first we show every consequence of $D{+}A$ in $\DL(d_1)$
has a corresponding consequence of $T(D){+}A$ in $\DL(d_2)$,
and then we show that every consequence of $T(D){+}A$ in $\DL(d_2)$ in the language of $D{+}A$
has a corresponding consequence  of $D{+}A$ in $\DL(d_1)$.
In both cases the proof is by induction on the level $n$ of $\T\uparrow n$
where $\T$ combines the functions in the inference rules for $\pm d_1$
and $\pm\Delta$ for $D{+}A$ in the first part, and
combines the functions in the inference rules for $\pm d_2$
and $\pm\Delta$ for $T(D){+}A$ in the second part.
The induction hypothesis for the first part is:
for $k \leq n$, if $\alpha \in \T_{D{+}A}\uparrow n$ then $T(D){+}A \vdash \alpha'$,
where $\alpha'$ is the counterpart, in $\DL(d_2)$, of $\alpha$.
For the second part it is:
for $k \leq n$, if $\alpha \in \Sigma$ and $\alpha \in \T_{T(D){+}A}\uparrow n$ then $D{+}A \vdash \alpha'$,
where $\alpha'$ is the counterpart, in $\DL(d_1)$, of $\alpha$.
Since $\T_{P}\uparrow 0 = \emptyset$ the induction hypothesis is always valid for $n=0$.

Throughout this appendix, if $r$ is a rule then $B_r$ refers to the body of that rule.
For brevity, we write $+d B$, where $B$ is a set of literals, to mean
$\{+d q ~|~ q \in B\}$.


\section{Blocked Ambiguity Simulates Propagated Ambiguity}

The facts and strict rules of $D{+}A$ and $T(D){+}A$ are the same, except for rules for $strict(q)$ in $T(D){+}A$.
However $strict(q)$ is not used in any other strict rule.
Consequently, for any addition $A$, $D{+}A$ and $T(D){+}A$ draw the same strict conclusions in $\Sigma(D{+}A)$.
Furthermore, these conclusions are reflected in the defeasible conclusions of $strict(q)$.

\begin{lemma}  \label{lemma:strictAB}
Let $A$ be any defeasible theory, and let $\Sigma$ be the language of $D{+}A$.
Then, for every $q \in \Sigma$,
\begin{itemize}
\item
$D{+}A \vdash {+}\Delta q$ iff $T(D){+}A \vdash {+}\Delta q$ 

\hspace{57pt}
iff $T(D){+}A \vdash {+}\partial^* strict(q)$ iff $T(D){+}A \vdash -\partial^* \neg strict(q)$
\item
$D{+}A \vdash -\Delta q$ iff $T(D){+}A \vdash -\Delta q$  

\hspace{57pt}
iff $T(D){+}A \vdash -\partial^* strict(q)$  iff $T(D){+}A \vdash {+}\partial^* \neg strict(q)$
\end{itemize}
\end{lemma}
\begin{proof}
The proof of
$D{+}A \vdash \pm\Delta q$ iff $T(D){+}A \vdash \pm\Delta q$ 
is straightforward, 
by induction on length of proofs.

In the inference rule for ${+}\partial^* strict(q)$,
clause $.2.3$ must be false, by the structure of the rules in part \ref{t2:strictq} of the transformation.
Consequently, we infer ${+}\partial^* strict(q)$ iff we infer ${+}\Delta strict(q)$,
which happens iff we infer ${+}\Delta q$ since there is only the one rule for $strict(q)$.
Similarly,
clause $.2.3$ of the inference rule for $-\partial^* strict(q)$ is true, so
we infer $-\partial^* strict(q)$ iff we infer $-\Delta strict(q)$,
which happens iff  we infer $-\Delta q$ since there is only the one rule for $strict(q)$.

In the inference rule for $-\partial^* \neg strict(q)$,
clause $.2.1$ is false because the body of $nstr(q)$ is empty,
and clause $.2.3$ is false because $nstr(q) >' str(q)$.
Thus we infer $-\partial^* \neg strict(q)$ iff we infer ${+}\Delta strict(q)$.
Finally,
in the inference rule for ${+}\partial^* \neg strict(q)$,
clause $.1$ is false, because there is no fact or strict rule for $\neg strict(q)$.
and clauses $.2.1$ and $.2.3$ are true (the latter because $nstr(q) >' str(q)$).
Thus, we can infer ${+}\partial^* \neg strict(q)$ iff we can infer $-\Delta strict(q)$.
\end{proof}

This lemma establishes that strict provability ($\pm \Delta$) from $D{+}A$ in $\DL(\delta^*)$
is captured in $\DL(\partial^*)$ by the transformation defined above,
no matter what the addition $A$.
We now show that $\DL(\partial^*)$ can simulate 
the behaviour of
both $\delta^*$ and $\supp^*$
with respect to addition of facts.

\begin{lemma}
Let $D$ be a defeasible theory, 
$T(D)$ be the transformed defeasible theory as described in Definition \ref{defn:ABsimAP},
and let $A$ be a modular set of facts.
Let $\Sigma$ be the language of $D{+}A$ and let $q \in \Sigma$.
Then
\begin{itemize}
\item
$D{+}A \vdash {+}\supp^* q$ iff $T(D){+}A \vdash {+}\partial^* supp(q)$
\item
$D{+}A \vdash -\supp^* q$ iff $T(D){+}A \vdash -\partial^* supp(q)$
\item
$D{+}A \vdash {+}\delta^* q$ iff $T(D){+}A \vdash {+}\partial^* q$
\item
$D{+}A \vdash -\delta^* q$ iff $T(D){+}A \vdash -\partial^* q$ 
\end{itemize}
\end{lemma}
\begin{proof}

Suppose ${+}\supp^* q \in \T_{D{+}A} \uparrow (n{+}1)$.
Then, by the ${+}\supp^*$ inference rule,
there is a strict or defeasible rule $r$ in $D$ with head $q$ and body $B_r$ such that
${+}\supp^* B_r \subseteq \T_{D{+}A} \uparrow n$,
and for every rule $s$ in $D$ for $\non q$
either there is a literal $b$ in the body of $s$ such that $-\delta^* b \in  \T_{D{+}A} \uparrow n$
or $s \not> r$.
Hence, by the induction hypothesis,
there is a strict or defeasible rule $r$ in $D$ with head $q$ and body $B_r$ such that
$T(D){+}A \vdash {+}\partial^* supp(b)$ for each $b \in B_r$
and for every rule $s$ in $D$ for $\non q$
either there is a literal $b$ in the body of $s$ such that  $T(D){+}A \vdash -\partial^* b$
or $s \not> r$.
Then
$T(D){+}A \vdash {+}\partial^* supp\_body(r)$ and
for for every rule $s$ in $D$ for $\non q$ with $s > r$
$T(D){+}A \vdash -\partial^* B_s$,
and hence
$T(D){+}A \vdash {+}\partial^* \neg o(r)$.
Combining these two conclusions,
and given that there is no rule for $\neg supp(q)$,
we have
$T(D){+}A \vdash {+}\partial^* supp(q)$.

Suppose ${+}\delta^* q \in \T_{D{+}A} \uparrow (n{+}1)$.
Then, by the ${+}\delta^*$ inference rule,
there is a strict or defeasible rule $r$ in $D$ with head $q$ and body $B_r$ such that
${+}\delta^* B_r \subseteq \T_{D{+}A} \uparrow n$,
$-\Delta \non q \in \T_{D{+}A} \uparrow n$,
and for every rule $s$ in $D$ for $\non q$ where $r \not> s$,
there is a literal $b$ in the body of $s$ such that $-\supp^* b \in  \T_{D{+}A} \uparrow n$.
Hence, by the induction hypothesis,
there is a strict or defeasible rule $r$ in $D$ with head $q$ and body $B_r$ such that
$T(D){+}A \vdash {+}\partial^* B_r$,
$T(D){+}A \vdash -\Delta \non q$,
and for every rule $s$ in $D$ for $\non q$  where $r \not> s$,
there is a literal $b$ in the body of $s$ such that  $T(D){+}A \vdash -\partial^* supp(b)$.
By Lemma \ref{lemma:strictAB}, 
$T(D){+}A \vdash -\partial^* \neg strict(\non q)$.
By repeated application of the $-\partial^*$ inference rule
we have $T(D){+}A \vdash -\partial^* supp\_body(s)$ for each $s$,
and then $T(D){+}A \vdash {+}\partial^* \neg comp(r)$.
Thus the body of the rule $inf(r)$ in $T(D)$  holds defeasibly.
On the other hand,
for every rule $s$ for $\non q$ in $D$ where $r \not> s$
there is a literal $b$ in the body of $s$ such that  $T(D){+}A \vdash -\partial^* supp(b)$ so,
using the inference rule for $-\partial^*$ and the rule from part \ref{t1:supp1}
we must have $T(D){+}A \vdash -\partial^* b$. 
$T(D){+}A \vdash {+}\partial^* B_r$
so, using the rules in part \ref{t1:supp1} and part \ref{t1:supp2},
$T(D){+}A \vdash {+}\partial^* supp\_body(r)$.
Hence, for the rules for $\non q$ where $r > s$,
the rules $n_d(s, r)$ can be applied and
$T(D){+}A \vdash -\partial^* \neg comp(s)$.
Consequently, all rules $inf(s)$ for $\non q$ fail.
From this fact and the fact that body of rule $inf(r)$ is proved defeasibly
we conclude 
$T(D){+}A \vdash {+}\partial^* q$.
\finish{clean up ??}

Suppose $-\supp^* q \in \T_{D{+}A} \uparrow (n{+}1)$.
Then, by the $-\supp^*$ inference rule,
$-\Delta q \in \T_{D{+}A} \uparrow n$ and,
for every strict or defeasible rule $r$ in $D$ with head $q$ and body $B_r$, either
$-\supp^* b \in \T_{D{+}A} \uparrow n$ for some $b \in B_r$,
or there is a rule $s$ in $D$ for $\non q$ with body $B_s$
such that ${+}\delta^* B_s \subseteq \T_{D{+}A} \uparrow n$ and $s > r$.
Hence, by the induction hypothesis,
$T(D){+}A \vdash -\Delta q$ and
for every strict or defeasible rule $r$ in $D$ with head $q$ either
$T(D){+}A \vdash -\partial^* supp(b)$  for some $b \in B_r$,
or there is a rule $s$ in $D$ for $\non q$ with $s > r$ and
$T(D){+}A \vdash {+}\partial^* B_s$.
Hence, either
$T(D){+}A \vdash -\partial^* supp\_body(r)$
or 
$T(D){+}A \vdash -\partial^* \neg o(r)$.
In either case, we have 
$T(D){+}A \vdash -\partial^* supp(q)$.

Suppose $-\delta^* q \in \T_{D{+}A} \uparrow (n+1)$.
Then, by the $-\delta^*$ inference rule,
$-\Delta q \in \T_{D{+}A} \uparrow n$ or,
for every strict or defeasible rule $r$ in $D$ with head $q$ and body $B_r$, either
$-\delta^* b \in \T_{D{+}A} \uparrow n$ for some $b \in B_r$,
${+}\Delta \non q \in \T_{D{+}A} \uparrow n$,
or there is a rule $s$ in $D$ for $\non q$ with body $B_s$
such that ${+}\supp^* B_s \subseteq \T_{D{+}A} \uparrow n$ and $r \not> s$.
Hence, by the induction hypothesis,
$T(D){+}A \vdash -\Delta q$ and
for every strict or defeasible rule $r$ in $D$ with head $q$ either
(1) $T(D){+}A \vdash -\partial^* b$  for some $b \in B_r$,
(2) $T(D){+}A \vdash {+}\Delta \non q$,
or (3) there is a rule $s$ in $D$ for $\non q$ with $r \not> s$ and
$T(D){+}A \vdash {+}\partial^* supp(b')$ for every $b'\in B_s$.
We consider these three cases in turn.
In the first case, the rule $inf(r)$ fails.
In the second case, using part \ref{t1:strictq},
we can conclude $T(D){+}A \vdash {+}\Delta strict(\non q)$
and $T(D){+}A \vdash -\partial^* \neg strict(\non q)$,
and hence 
the rule $inf(r)$ fails.
In the third case,
we can conclude 
$T(D){+}A \vdash {+}\partial^* supp\_body(s)$ 
and hence, using part \ref{t1:def},
$T(D){+}A \vdash -\partial^* \neg comp(r)$.
Thus,
the rule $inf(r)$ fails.
In each case,
the rule $inf(r)$ fails.
Thus we can derive $T(D){+}A \vdash -\partial^* q$.


Suppose ${+}\partial^* supp(q) \in \T_{T(D){+}A} \uparrow (n{+}1)$.
Then, by the ${+}\partial^*$ inference rule,
either ${+}\partial^* q  \in \T_{T(D){+}A} \uparrow n$, or
there is a strict or defeasible rule $r$ in $D$ with head $q$ and body $B_r$ such that
${+}\partial^* supp\_body(r) \in \T_{T(D){+}A} \uparrow n$
and 
${+}\partial^* \neg o(r) \in \T_{T(D){+}A} \uparrow n$.
Consequently,
${+}\partial^* supp(b)  \in \T_{T(D){+}A} \uparrow n$, for each $b \in B_r$
for every rule $s$ in $D$ for $\non q$ where $s > r$, 
there is $b$ in the body of $s$ such that $-\partial^* b \in \T_{T(D){+}A} \uparrow n$.
In the first case,
by the induction hypothesis,
$D{+}A \vdash {+}\delta^* q$
and then, by the inclusion theorem,
$D{+}A \vdash {+}\supp^* q$.
In the second case,
by the induction hypothesis,
$D{+}A \vdash {+}\supp^* B_r$
for every rule $s$ in $D$ for $\non q$ where $s > r$, 
there is $b$ in the body of $s$ such that $D{+}A \vdash -\delta^* b$.
Applying the inference rule for ${+}\supp^*$,
$D{+}A \vdash {+}\supp^* q$.

Suppose $-\partial^* supp(q) \in \T_{T(D){+}A} \uparrow (n+1)$.
Then, by the $-\partial^*$ inference rule,
$-\partial^* q  \in \T_{T(D){+}A} \uparrow n$,
and for every strict or defeasible rule $r$ in $D$ for $q$ either
$-\partial^* supp\_body(r)  \in \T_{T(D){+}A} \uparrow n$
or
$-\partial^* \neg o(r)  \in \T_{T(D){+}A} \uparrow n$.
In the former case we must have
$-\partial^* supp(b)  \in \T_{T(D){+}A} \uparrow n$ for some $b$ in the body $B_r$ of $r$.
In the latter case we must have that for some rule $s$ in $D$ with body $B_s$,
$s > r$ and ${+}\partial^* B_s  \subseteq \T_{T(D){+}A} \uparrow n$.
By the induction hypothesis,
we have
$D{+}A \vdash -\delta^* q$ (and hence $D{+}A \vdash -\Delta q$)
and, for each $r$ either
$D{+}A \vdash -\supp^* b$ for some $b \in B_r$
or there is an opposing rule $s$ with $s > r$ and
$D{+}A \vdash {+}\delta^* B_s$.
Applying the inference rule for $-\supp^*$
we conclude
$D{+}A \vdash -\supp^* q$.

Suppose ${+}\partial^* q \in \T_{T(D){+}A} \uparrow (n+1)$.
Then, by the ${+}\partial^*$ inference rule,
there is a strict or defeasible rule $r$ in $D$ with head $q$ and body $B_r$ such that
${+}\partial^* B_r \subseteq \T_{T(D){+}A} \uparrow n$,
${+}\partial^* \neg strict(\non q) \in \T_{T(D){+}A} \uparrow n$,
and
${+}\partial^* \neg comp(r) \in \T_{T(D){+}A} \uparrow n$.
By Lemma \ref{lemma:strictAB}, $D{+}A \vdash -\Delta \non q$.
Using the structure of $T(D)$ and the ${+}\partial^*$ inference rule,
for every rule $s$ in $D$ for $\non q$ where $r \not> s$
we must have $-\partial^* supp\_body(s) \in \T_{T(D){+}A} \uparrow n$,
and hence $-\partial^* supp(b) \in \T_{T(D){+}A} \uparrow n$, for some $b$ in the body of $s$.
By the induction hypothesis,
$D{+}A \vdash  {+}\delta^* B_r$ and
for every rule $s$ in $D$ for $\non q$ where $r \not> s$, there is $b$ in the body of $s$ such that
$D{+}A \vdash -\supp^* b$.
Now, applying the ${+}\delta^*$ inference rule, we have
$D{+}A \vdash {+}\delta^* q$.

Suppose $-\partial^* q \in \T_{T(D){+}A} \uparrow (n+1)$.
Then, by the $-\partial^*$ inference rule,
$-\Delta q \in \T_{T(D){+}A} \uparrow n$ and,
for every strict or defeasible rule $r$ for $q$ in $D$ with body $B_r$,
either
(1) $-\partial^* b \in \T_{T(D){+}A} \uparrow n$ for some $b \in B_r$,
(2) $-\partial^* \neg comp(r) \in \T_{T(D){+}A} \uparrow n$,
(3) $-\partial^* \neg strict(\non q) \in \T_{T(D){+}A} \uparrow n$,
or,
(4) for some rule $s$ for $\non q$ in $D$,
${+}\partial^* B_s \subseteq \T_{T(D){+}A} \uparrow n$,
${+}\partial^* \neg comp(s) \in \T_{T(D){+}A} \uparrow n$, and
${+}\partial^* \neg strict(q) \in \T_{T(D){+}A} \uparrow n$.

Hence, using the structure of $T(D)$ and Lemma \ref{lemma:strictAB},
$-\Delta q \in \T_{T(D){+}A} \uparrow n$ and,
for every rule $r$ for $q$ in $D$ with body $B_r$,
either
(1) $-\partial^* b \in \T_{T(D){+}A} \uparrow n$ for some $b \in B_r$,
(2) for some rule $s'$ for $\non q$ in $D$ we have
${+}\partial^* supp(b) \in \T_{T(D){+}A} \uparrow n$ for each $b \in B_{s'}$,
(3)${+}\Delta \non q  \in \T_{T(D){+}A} \uparrow n$,
or,
(4) for some rule $s$ for $\non q$ in $D$,
${+}\partial^* B_s \subseteq \T_{T(D){+}A} \uparrow n$,
for every rule $r'$ for $q$, there is $b'$ in its body such that
$-\partial^* supp(b') \in \T_{T(D){+}A} \uparrow n$, and
$-\Delta q \in \T_{T(D){+}A} \uparrow n$.

By the induction hypothesis,
$D{+}A \vdash -\Delta q$ and, 
for every strict or defeasible rule $r$ for $q$ in $D$ with body $B_r$,
either
(1) $D{+}A \vdash -\delta b$ for some $b \in B_r$,
(2) for some rule $s$ for $\non q$ in $D$ we have
  $D{+}A \vdash {+}\supp^* b$  for each $b \in B_s$,
(3) $D{+}A \vdash {+}\Delta \non q$,
or
(4) for some rule $s$ for $\non q$ in $D$,
$D{+}A \vdash {+}\delta^* B_s$,
for every rule $r'$ for $q$, there is $b'$ in its body such that
$D{+}A \vdash -\supp^* b'$,
and
$D{+}A \vdash -\Delta q$.
For each disjunct,
applying the inference rule for $-\delta^*$,
we can conclude
$D{+}A \vdash -\delta^* q$.

\end{proof}

This result concerns only addition of facts.
It was established in \cite{Maher12} that
it cannot be extended to addition of rules.

Given that the ambiguity blocking logics can simulate each other, as can the ambiguity propagating logics
(see \cite{Maher12}) we have

\begin{theorem} 
The ambiguity blocking logics ($\DL(\partial)$ and $\DL(\partial^*)$)
can simulate the ambiguity propagating logics ($\DL(\delta)$ and $\DL(\delta^*)$)
with respect to addition of facts.
\end{theorem}

This is Theorem \ref{thm:ABsimAP} from the body of the paper.

\section{Propagated Ambiguity Simulates Blocked Ambiguity}

As with the previous simulation,
the facts and strict rules of $D$ and $T(D)$ are the same, except for rules for $strict(q)$ in $T(D)$.
Thus, again, for any addition $A$, 
$D{+}A$ and $T(D){+}A$ draw the same strict conclusions in $\Sigma(D{+}A)$.
Furthermore, these conclusions are reflected in the defeasible conclusions of $strict(q)$, $true(q)$ and $\neg true(q)$,
and also in support conclusions.

\begin{lemma}  \label{lemma:strictAP}
Let $D$ be a defeasible theory, 
$T(D)$ be the transformed defeasible theory as described in Definition \ref{defn:APsimAB},
and let $A$ be a modular defeasible theory.
Let $\Sigma$ be the language of $D{+}A$ and let $q \in \Sigma$.
Then
\begin{itemize}
\item
$D{+}A \vdash {+}\Delta q$ iff $T(D){+}A \vdash {+}\Delta q$ \hspace{32.5pt} iff $T(D){+}A \vdash {+}\delta^* strict(q)$ 

\hspace{57pt}
iff $T(D){+}A \vdash {+}\delta^* true(q)$  \hspace{5pt} iff $T(D){+}A \vdash {+}\supp^* true(q)$

\hspace{57pt}
iff $T(D){+}A \vdash -\delta^* \neg true(q)$ iff $T(D){+}A \vdash -\supp^* \neg true(q)$ \\

\item
$D{+}A \vdash -\Delta q$ iff $T(D){+}A \vdash -\Delta q$ \hspace{32.5pt} iff $T(D){+}A \vdash -\delta^* strict(q)$  

\hspace{57pt}
iff $T(D){+}A \vdash -\delta^* true(q)$  \hspace{5pt} iff $T(D){+}A \vdash -\supp^* true(q)$

\hspace{57pt}
iff $T(D){+}A \vdash {+}\delta^* \neg true(q)$ iff $T(D){+}A \vdash {+}\supp^* \neg true(q)$

\end{itemize}
\end{lemma}
\begin{proof}
The proof of
$D{+}A \vdash \pm\Delta q$ iff $T(D){+}A \vdash \pm\Delta q$ 
is straightforward, by induction on length of proofs.

In the inference rule for ${+}\delta^* strict(q)$,
clause $.2.3$ must be false, by the structure of the rules in part \ref{t2:strictq} of the transformation.
Consequently, we infer ${+}\delta^* strict(q)$ iff we infer ${+}\Delta strict(q)$,
which happens iff we infer ${+}\Delta q$ since there is only the one rule for $strict(q)$.
Similarly,
clause $.2.3$ of the inference rule for $-\delta^* strict(q)$ is true, so
we infer $-\delta^* strict(q)$ iff we infer $-\Delta strict(q)$,
which happens iff  we infer $-\Delta q$ since there is only the one rule for $strict(q)$.

\ignore{
In the inference rule for $-\delta^* \neg strict(q)$,
clause $.2.1$ is false because the body of $nstr(q)$ is empty,
and clause $.2.3$ is false because $nstr(q) >' str(q)$.
Thus we infer $-\delta^* \neg strict(q)$ iff we infer ${+}\Delta strict(q)$.
Finally,
in the inference rule for ${+}\delta^* \neg strict(q)$,
clause $.1$ is false, because there is no fact or strict rule for $\neg strict(q)$.
and clauses $.2.1$ and $.2.3$ are true (the latter because $nstr(q) >' str(q)$).
Thus, we can infer ${+}\delta^* \neg strict(q)$ iff we can infer $-\Delta strict(q)$.

Similarly,
we can infer ${+}\supp^* strict(q)$ iff we can infer ${+}\Delta strict(q)$,
and infer $-\supp^* strict(q)$ iff we can infer $-\Delta strict(q)$.
}

Note that $-\Delta true(q)$ and $-\Delta \neg true(q)$ are consequences of $T(D){+}A$
because there are no strict rules for such literals in $T(D){+}A$.
Using this fact, the two rules $t(q)$ and $nt(q)$ and the superiority $t(q) > nt(q)$,
using the inference rule for ${+}\delta^*$,
we can infer ${+}\delta^* \neg true(q)$ iff we can infer $-\supp^* strict(q)$, 
because $.1$ of the inference rule is false,
$.2.1$ and $.2.2$ are true, and $.2.3.2$ is false.
Similarly, using the inference rule for ${+}\supp^*$,
we can infer ${+}\supp^* \neg true(q)$ iff we can infer $-\delta^* strict(q)$. 
Using the inference rules for  $-\delta^*$ and  $-\supp^*$,
we can infer $-\delta^* \neg true(q)$ iff we can infer ${+}\supp^* strict(q)$, 
and
we can infer $-\supp^* \neg true(q)$ iff we can infer ${+}\delta^* strict(q)$. 
\end{proof}

We need this more detailed characterization of strict consequence,
compared to Lemma \ref{lemma:strictAB},
because both $\delta^*$ and $\supp^*$ intermediate conclusions influence $\delta^*$ conclusions.

The next lemma is a key part of the proof.
It shows that the structure of $T(D){+}A$ tightly constrains the inferences that can be made
in the sense that, for the literals of interest,
the inference rules $\delta^*$ and $\supp^*$ draw the same conclusions.

\begin{lemma}  \label{lemma:tight}
Let $D$ be a defeasible theory, 
$T(D)$ be the transformed defeasible theory as described in Definition \ref{defn:APsimAB},
and let $A$ be a modular set of facts.
Let $\Sigma$ be the language of $D{+}A$ extended with 
literals of the forms $\f(p)$, $\neg \f(p)$ and $\neg true(p)$, for $p \in \Sigma(D)$.
 
Then, for any $q \in \Sigma$,
\begin{itemize}
\item
$T(D){+}A \vdash {+}\delta^* q$  iff $T(D){+}A \vdash {+}\supp^* q$ 
\item
$T(D){+}A \vdash {-}\delta^* q$  iff $T(D){+}A \vdash {-}\supp^* q$ 
\end{itemize}
\end{lemma}
\begin{proof}
Two parts of the proof follow immediately from the inclusion theorem.
These are the forward direction of the first statement and the backward direction of the second statement.
Furthermore,
it is immediate from  Lemma \ref{lemma:strictAP}
that the result holds for literals involving $true$ and for literals that are proved strictly.
The remaining parts are proved by induction.

Recall that
$T(D){+}A \vdash s$ iff there is an integer $n$ such that $s \in \T_{T(D){+}A} \uparrow n$.
Note that the result holds in $\T_{T(D){+}A} \uparrow 0$, since it is empty.
Suppose the result holds for conclusions $s$ with $s \in \T_{T(D){+}A} \uparrow n$.
We show that it also holds for conclusions in $\T_{T(D){+}A} \uparrow (n+1)$.

(1)
If ${+}\supp^* q \in \T_{T(D){+}A} \uparrow (n+1)$ then ${+}\supp^* \f(q) \in \T_{T(D){+}A} \uparrow n$,
because there is only one rule for $q$ and it cannot be overruled.
Further, if ${+}\supp^* \f(q) \in \T_{T(D){+}A} \uparrow n$ then
for some rule $r$ of $D$ we must have ${+}\supp^* B_r \subseteq \T_{T(D){+}A} \uparrow n$
and ${+}\supp^* \neg true(\non q) \in \T_{T(D){+}A} \uparrow n$ and,
for every rule $s$ in $D$ for $\non q$ with $r \not> s$, 
there is $p \in B_s$ with $-\delta^* p \in \T_{T(D){+}A} \uparrow n$,
because clause $.2.2.2$ must be false, since $n_d(r,s) > p_d(r)$ for every such $s$.
By the induction hypothesis, 
${+}\delta^* B_r \subseteq \T_{T(D){+}A} \uparrow n$,
and for each $s$ there is $p \in B_s$ with $-\supp^* p \in \T_{T(D){+}A} \uparrow n$
and, by Lemma \ref{lemma:strictAP},
${+}\delta^* \neg true(\non q)$ and $-\Delta \non q$ are consequences of $T(D){+}A$.
Applying the ${+}\delta^*$ inference rule, $T(D){+}A \vdash +\delta^* \f(q)$ and,
applying the ${-}\supp^*$ inference rule, $T(D){+}A \vdash -\supp^* \f(\non q)$
since every rule $p_d(s)$ contains a $p$ with $-\supp^* p \in \T_{T(D){+}A} \uparrow n$.
Hence, applying the ${+}\delta^*$ inference rule, $T(D){+}A \vdash +\delta^* q$.

If ${+}\supp^* \neg \f(q)  \in \T_{T(D){+}A} \uparrow (n+1)$
then, for some rule $s$ for $\non q$ in $D$, ${+}\supp^* B_s \subseteq \T_{T(D){+}A} \uparrow n$.
By the induction hypothesis, ${+}\delta^* B_s \subseteq \T_{T(D){+}A} \uparrow n$.
Applying the ${+}\delta^*$ inference rule, 
noting that there is no fact or strict rule for $\f(q)$ and that $n_d(r,s) > p_d(r)$, we have
$T(D){+}A \vdash +\delta^* \neg \f(q)$.

\finish{
Proof does not work for ${-}\delta^* q$
because we cannot be sure that
??j?????????????????????????????????????????
}

(2)
If ${-}\delta^* q \in \T_{T(D){+}A} \uparrow (n+1)$ then,
using the ${-}\delta^*$ inference rule and the structure of $T(D){+}A$,
${-}\Delta q \in \T_{T(D){+}A} \uparrow n$ and either
${-}\delta^* \f(q) \in \T_{T(D){+}A} \uparrow n$ or
${+}\Delta \non q \in \T_{T(D){+}A} \uparrow n$ or
${+}\supp^* \f(\non q) \in \T_{T(D){+}A} \uparrow n$.

If ${-}\delta^* \f(q) \in \T_{T(D){+}A} \uparrow (n+1)$ then
either
for every rule $p_d(r)$, for some $p \in B_r$, ${-}\delta^* p  \in \T_{T(D){+}A} \uparrow n$ or
${-}\delta^* \neg true(\non q) \in \T_{T(D){+}A} \uparrow n$,
or, for some rule $n_d(r,s)$, ${+}\supp^* B_s$.
By the induction hypothesis, either
for each $p_d(r)$ there is a $p$ in its body where $T(D){+}A \vdash -\supp^* p$,
or $T(D){+}A \vdash {+}\delta^* B_s$ for some $s$.
Applying the ${-}\supp^*$, we have $T(D){+}A \vdash -\supp^* \f(q)$.
If ${+}\Delta \non q \in \T_{T(D){+}A} \uparrow n$ then, by Lemma \ref{lemma:strictAP},
$T(D){+}A \vdash -\supp^* \neg true(\non q)$.
Hence we must have $T(D){+}A \vdash -\supp^* \f(q)$, 
since $\neg true(\non q)$ appears in each rule for $\f(q)$.

If ${+}\supp^* \f(\non q) \in \T_{T(D){+}A} \uparrow n$ then
there is a rule $p_d(s)$ for $ \f(\non q)$ where
$T(D){+}A \vdash {+}\supp^* B_s$ and $T(D){+}A \vdash {+}\supp^* \neg true(q)$
and, for every rule $n_d(s,r)$, $T(D){+}A \vdash {-}\delta^* B_r$.
By the induction hypothesis,
$T(D){+}A \vdash {+}\delta^* B_s$ and,
for every rule $n_d(s,r)$ (where we must have $s \not> r$ in $D$), $T(D){+}A \vdash {-}\delta^* B_r$.
Hence, for every $r$ for $q$ in $D$ where $r > s$ we have $T(D){+}A \vdash {-}\supp^* B_r$.
For every other $r$ for $q$ in $D$ there is $n_d(r, s)$ where $T(D){+}A \vdash {+}\delta^* B_s$.
Hence, applying the ${-}\supp^*$ inference rule for $\f(q)$,
we must have $T(D){+}A \vdash {-}\supp^* \f(q)$.

Thus, in every case we have $T(D){+}A \vdash {-}\supp^* \f(q)$ and consequently
$T(D){+}A \vdash {-}\supp^* q$.

If ${-}\delta^* \neg \f(q) \in \T_{T(D){+}A} \uparrow (n+1)$ then
for every rule $p_d(r)$, for some $p$ in its body, ${-}\delta^* p  \in \T_{T(D){+}A} \uparrow n$.
By the induction hypothesis,
for every rule $p_d(r)$, for some $p$ in its body, $T(D){+}A \vdash -\supp^* p$.
Applying the ${-}\supp^*$ inference rule, $T(D){+}A \vdash -\supp^* \neg \f(q)$.

\end{proof}
As a consequence of the inclusion theorem and the previous lemma,
any inference rule between $\supp^*$ and $\delta^*$ 
(that is, any inference rule except for $\Delta$ and $\delta$)
behaves the same way on $\Sigma$-literals
in $T(D){+}A$.
In particular, it applies to $\partial^*$.

\begin{corollary}  \label{cor:tight}
Let $\Sigma$ be the language of $D$, $\Sigma'$ be as defined in the previous lemma.
Let $A$ be any set of facts.
Then if $q \in \Sigma'$

\begin{itemize}
\item
$T(D){+}A \vdash {+}\delta^* q$  iff $T(D){+}A \vdash {+}\partial^* q$ 
\item
$T(D){+}A \vdash {-}\delta^* q$  iff $T(D){+}A \vdash {-}\partial^* q$ 
\end{itemize}
\end{corollary}

Now we show that the transformation preserves the $\partial^*$ consequences of $D{+}A$.

\begin{theorem}   \label{thm:partial*}
Let $D$ be a defeasible theory, 
$T(D)$ be the transformed defeasible theory as described in Definition \ref{defn:APsimAB},
and let $A$ be a modular set of facts.
Let $\Sigma$ be the language of $D{+}A$ and let $q \in \Sigma$.
Then
\begin{itemize}
\item
$D{+}A \vdash +\partial^* q$ iff
$T(D){+}A \vdash +\partial^* q$
\item
$D{+}A \vdash -\partial^* q$ iff
$T(D){+}A \vdash -\partial^* q$
\end{itemize}
\end{theorem}
\begin{proof}

Suppose ${+}\partial^* q \in \T_{D{+}A} \uparrow (n{+}1)$.
Then, by the ${+}\partial^*$ inference rule,
either ${+}\Delta q \in  \T_{D{+}A} \uparrow n$
(in which case, we must have $T(D)+A \vdash +\delta^*$)
or
${+}\Delta q \notin  \T_{D{+}A} \uparrow n$ and
there is a strict or defeasible rule $r$ in $D$ with head $q$ and body $B_r$ such that
${+}\partial^* B_r \subseteq \T_{D{+}A} \uparrow n$,
$-\Delta \non q \in \T_{D{+}A} \uparrow n$,
and for every rule $s$ in $D$ for $\non q$
either there is a literal $b$ in the body of $s$ such that $-\partial^* b \in  \T_{D{+}A} \uparrow n$
or $r > s$.
Hence, in the latter case, by the induction hypothesis, 
there is a strict or defeasible rule $r$ in $D{+}A$ with head $q$ and body $B_r$ such that
$T(D){+}A \vdash {+}\partial^* B_r$,
$T(D){+}A \vdash -\Delta \non q$,
and for every rule $s$ in $D{+}A$ for $\non q$
either $T(D){+}A \vdash -\partial^* B_s$
or $r > s$.

From this statement we derive several facts.
(1) By Lemma \ref{lemma:strictAP} and the inclusion theorem, 
$T(D){+}A \vdash {+}\partial^* \neg true(\non q)$.
(2) Thus, $T(D){+}A \vdash {+}\partial^* (B_r, \neg true(\non q))$ and,
for every rule $n_d(r,s)$ in $T(D)$, $T(D){+}A \vdash --\partial^* B_s$
(since rules $s$ where $r > s$ do not give rise to a rule $n_d(r,s)$).
Hence, $T(D){+}A \vdash {+}\partial^* \f(q)$.
(3) Conversely,
$T(D){+}A \vdash -\partial^* \f(\non q)$
because, for every rule $p_d(s)$ for $\f(\non q)$,
either $T(D){+}A \vdash -\partial^* B_s$ or
there is a rule $n_d(s,r)$ superior to $p_d(s)$ with $T(D){+}A \vdash {+}\partial^* B_r$.
Consequently, since the only rule in $T(D)$ for $q$ has body $\f(q)$ (and similarly for $\non q$),
applying the ${+}\partial^*$ inference rule, we have
$T(D){+}A \vdash {+}\partial^* q$.

Suppose $-\partial^* q \in \T_{D{+}A} \uparrow (n+1)$.
Then, by the $-\partial^*$ inference rule,
$-\Delta q \in \T_{D{+}A} \uparrow n$ and,
for every strict or defeasible rule $r$ in $D$ with head $q$ and body $B_r$, either
$-\partial^* b \subseteq \T_{D{+}A} \uparrow n$ for some $b \in B_r$,
${+}\Delta \non q \in \T_{D{+}A} \uparrow n$,
or there is a rule $s$ in $D$ for $\non q$ with body $B_s$
such that ${+}\partial B_s \subseteq \T_{D{+}A} \uparrow n$ and $r \not> s$.
Hence, by the induction hypothesis,
$T(D){+}A \vdash -\Delta q$ and
for every strict or defeasible rule $r$ in $D$ with head $q$ either
$T(D){+}A \vdash -\partial^* b$  for some $b \in B_r$,
$T(D){+}A \vdash {+}\Delta \non q$,
or there is a rule $s$ in $D$ for $\non q$ where
$T(D){+}A \vdash {+}\partial^* B_s$
and$r \not> s$.
Hence, for every rule $p_d(r)$ in $T(D)$ for $\f(q)$ either 
$T(D){+}A \vdash -\delta^* b$  for some $b \in B_r$, or
$T(D){+}A \vdash -\delta^* \neg strict(\non q)$ (by Lemma \ref{lemma:strictAP}), or
there is a rule $n_d(r,s)$ where $T(D){+}A \vdash {+}\supp^* B_s$.
Applying the inference rule for $-\delta^* \f(q)$, we conclude
$T(D){+}A \vdash -\delta^* \f(q)$
and, hence, $T(D){+}A \vdash -\delta^* q$.

Suppose ${+}\partial^* q \in \T_{T(D){+}A} \uparrow (n{+}1)$.
Then, by the ${+}\partial^*$ inference rule and using the structure of $T(D)$,
either ${+}\Delta q \in \T_{T(D){+}A} \uparrow n$, or
${+}\partial^* \f(q) \in \T_{T(D){+}A} \uparrow n$,
$-\Delta \non q \in \T_{T(D){+}A} \uparrow n$,
and $-\partial^* \f(\non q) \in  \T_{D{+}A} \uparrow n$.
In the first case we have $D{+}A \vdash {+}\Delta q$ and thus $D{+}A \vdash {+}\partial^* q$,
Alternatively,
there is a strict or defeasible rule $r$ in $D$ with head $q$ and body $B_r$ such that
${+}\partial^* B_r \subseteq \T_{T(D){+}A} \uparrow n$,
${+}\partial^* \neg true(\non q) \in \T_{T(D){+}A} \uparrow n$,
and for every rule $s$ in $D$ for $\non q$ where $r \not> s$
there is a literal $b$ in the body $B_s$ of $s$ such that $-\partial^* b \in  \T_{T(D){+}A} \uparrow n$.
By the induction hypothesis and Lemma \ref{lemma:strictAP},
$D{+}A \vdash {+}\partial^* B_r $, 
$D{+}A \vdash {-}\Delta \non q$,
and for every rule $s$ in $D$ for $\non q$ where $r \not> s$
there is $b$ in the body  of $s$ such that  $D{+}A \vdash {-}\partial b$.
Applying the ${+}\partial$ inference rule, we conclude $D{+}A \vdash {+}\partial^* q$.

Suppose ${-}\partial^* q \in \T_{T(D){+}A} \uparrow (n{+}1)$.
Then, by the ${-}\partial^*$ inference rule and using the structure of $T(D)$,
${-}\Delta q \in \T_{T(D){+}A} \uparrow n$ and either
(1) ${-}\partial \f(q) \in \T_{T(D){+}A} \uparrow n$. or
(2) ${+}\Delta \non q \in \T_{T(D){+}A} \uparrow n$, or
(3) ${+}\partial \f(\non q) \in \T_{T(D){+}A} \uparrow n$.
By Lemma \ref{lemma:strictAP} and Corollary \ref {cor:tight} we have $D{+}A \vdash {-}\Delta q$ and 
$D{+}A \vdash {+}\partial^* \neg true(q)$. 

In the first case,
for each rule $r$ for $q$ in $D$ either
there is a literal $p \in B_r$ and $-\partial p \in  \T_{T(D){+}A} \uparrow n$
or ${-}\partial^* \neg true(\non q) \in \T_{T(D){+}A} \uparrow n$
or 
for some rule $s$ for $\non q$ in $D$ where $r \not> s$,
${+}\partial^* B_s \subseteq \T_{T(D){+}A} \uparrow n$.
By the induction hypothesis (and Lemma \ref{lemma:strictAP} and Corollary \ref {cor:tight}),
for each rule $r$ for $q$ in $D$ either
there is a literal $p \in B_r$ and $D{+}A \vdash {-}\partial^* p$,
or $D{+}A \vdash {+}\Delta \non q$,
or 
for some rule $s$ for $\non q$ in $D$ where $r \not> s$,
$D{+}A \vdash {+}\partial^* B_s$.
Applying the ${-}\partial$ inference rule, $D{+}A \vdash {-}\partial^* q$.

In the second case, by Lemma \ref{lemma:strictAP} and Corollary \ref {cor:tight}, 
$D{+}A \vdash {+}\Delta \non q$.
Consequently, applying the ${-}\partial$ inference rule, $D{+}A \vdash {-}\partial^* q$.
In the third case, for some rule $s$ for $\non q$ in $D$,
${+}\partial^* B_s \subseteq \T_{T(D){+}A} \uparrow n$ and, for all rules $r$ for $q$ in $D$ where $s \not> r$,
for some $p \in B_r$, ${-}\partial^* p \in \T_{T(D){+}A} \uparrow n$.
By the induction hypothesis,
for every rule $r$ for $q$ in $D$ where $r > s$, for some $p \in B_r$, $D{+}A \vdash {-}\partial^* p$,
and $D{+}A \vdash {+}\partial^* B_s$.
Applying the ${-}\partial$ inference rule, $D{+}A \vdash {-}\partial^* q$.

\end{proof}

Combining Theorem \ref{thm:partial*} with Lemma \ref{lemma:tight} and the inclusion theorem,
we see that $\DL(\partial^*)$ can be simulated by $\DL(\delta^*)$ and $\DL(\delta)$.

\begin{theorem}   
For $d \in \{\delta, \delta^*, \partial \}$,
$\DL(d)$ can simulate $\DL(\partial^*)$ with respect to addition of facts
\end{theorem}
\begin{proof}
$D{+}A \vdash +\partial^* q$ iff
$T(D){+}A \vdash +\partial^* q$ (by Theorem \ref{thm:partial*})
iff
$T(D){+}A \vdash +\delta^* q$ (by Corollary \ref{cor:tight})
iff
$T(D){+}A \vdash +d q$ (by Lemma \ref{lemma:tight} and the inclusion theorem).
The proof is similar for $-\partial^* q$.
\end{proof}

This is Theorem \ref{thm:APsimAB} from the body of the paper.


\section{Simulation of Individual Defeat wrt Addition of Rules}


Example \ref{ex:failAB} 
does not apply to $\DL(\delta)$ and $\DL(\delta^*)$.
We have $\T(D)+A \vdash +\supp h(r_2)$ and, consequently,
$\T(D)+A \vdash -\delta p$, in agreement with $D$ under $\DL(\delta^*)$.
The weaker inference strength of ambiguity propagation
masks the distinction that is present for blocked ambiguity reasoning.
However, the next example shows that the transformation does not provide
a simulation wrt rules for the propagating ambiguity logics.

\begin{example}   \label{ex:failAP}
Let $D$ consist of the rules

\[
\begin{array}{lrll}
r_1:   &                           & \Rightarrow & \phantom{\neg} p \\ 
r_2:   &                           & \Rightarrow &  \neg p \\
r_3:   &                           & \Rightarrow & \phantom{\neg} p \\ 
r_4:   &                           & \Rightarrow &  \neg p \\
\end{array}
\]
with $r_1 > r_2$ and $r_3 > r_4$.

Then $\T(D)$ consists of the following rules

\[
\begin{array}{lrllrlrll}
p(r_1):          &                           & \Rightarrow & \phantom{\neg} h(r_1) & \phantom{MMMMM}
n(r_1, r_2):   &                           & \Rightarrow & \neg h(r_1) \\ 
& & & & 
n(r_1, r_4):   &                           & \Rightarrow & \neg h(r_1) \\ 
p(r_2):           &                           & \Rightarrow & \phantom{\neg} h(r_2) &
n(r_2, r_1):   &                           & \Rightarrow & \neg h(r_2) \\ 
& & & & 
n(r_2, r_3):   &                           & \Rightarrow & \neg h(r_2) \\ 
p(r_3):          &                           & \Rightarrow & \phantom{\neg} h(r_1) &
n(r_3, r_2):   &                           & \Rightarrow & \neg h(r_3) \\ 
& & & & 
n(r_3, r_4):   &                           & \Rightarrow & \neg h(r_3) \\ 
p(r_4):           &                           & \Rightarrow & \phantom{\neg} h(r_4) &
n(r_4, r_1):   &                           & \Rightarrow & \neg h(r_4) \\ 
& & & & 
n(r_4, r_3):   &                           & \Rightarrow & \neg h(r_4) \\ 
\end{array}
\]

\[
\begin{array}{lrllrlrll}
s(r_1):           &       h(r_1)          & \Rightarrow & \phantom{\neg} p \\ 
s(r_2):            &       h(r_2)        & \Rightarrow &  \neg p \\
s(r_3):           &       h(r_3)          & \Rightarrow & \phantom{\neg} p \\ 
s(r_4):            &       h(r_4)        & \Rightarrow &  \neg p \\
 \\
\end{array}
\]

\noindent
with
$p(r_1) > n(r_1, r_2)$, 
$n(r_2, r_1) > p(r_2)$,
$p(r_3) > n(r_3, r_4)$, 
and
$n(r_4, r_3) > p(r_4)$.

Now, let $A$ be the rule
\[
\begin{array}{lrll}
   &                           & \Rightarrow & p \\ 
\end{array}
\]
Then $D+A \vdash -\delta^* p$,
because for every rule $r$ for $p$, there is a rule for $\neg p$ that is not overruled by $r$
($r_1$ does not overrule $r_4$, $r_3$ does not overrule $r_2$ and $A$ overrules neither). 

However, considering the transformed theory,
$\T(D)+A \vdash -\supp h(r_2)$, because $n(r_2, r_1) > p(r_2)$
and, similarly, $\T(D)+A \vdash -\supp h(r_4)$.
Consequently, both rules for $\neg p$ in $\T(D)+A$ fail.
This leaves the rules for $p$ without competition, and so
$\T(D)+A \vdash +\delta p$,
conflicting with the behaviour of $D{+}A$.

Following essentially the same argument,
this example also applies to $\DL(\partial^*)$ and $\DL(\partial)$.
\end{example}

We show that the transformation defined in Definition \ref{defn:newTDsimID} (and Definition \ref{defn:TDsimID} )
allows the team-defeat logics to simulate their individual-defeat counterparts.
We treat the two cases separately, but first we address the effect of the transformation on strict inference.

\begin{lemma}
Consider the transformation $T$ from Definition \ref{defn:newTDsimID}.
For any $D$ and $A$

\begin{itemize}
\item
$D{+}A \vdash +\Delta q$ iff $T(D){+}A \vdash +\Delta q$ 
\item
$D{+}A \vdash -\Delta q$ iff $T(D){+}A \vdash -\Delta q$ 
\end{itemize}
\end{lemma}
The proof is a straightforward induction.

\finish{check numbers}
\begin{theorem}    \label{thm:TDsimIDAB}
The logic $\DL(\partial^*)$ can be simulated by $\DL(\partial)$
with respect to addition of rules.
\end{theorem}
\begin{proof}
\ignore{
All simulation proofs (of $\DL(d_1)$ by $\DL(d_2)$, say) have two parts:
first we show every consequence of $D{+}A$ in $\DL(d_1)$
has a corresponding consequence of $T(D){+}A$ in $\DL(d_2)$,
and then we show that every consequence of $T(D){+}A$ in $\DL(d_2)$
has a corresponding consequence  of $D{+}A$ in $\DL(d_1)$.
In both cases the proof is by induction on the level $n$ of $\T_{P}{\uparrow} n$
where $\T_{P}$ combines the functions in the inference rules for $\pm d_1$
and $\pm\Delta$ for $D{+}A$ in the first part, and
combines the functions in the inference rules for $\pm d_2$
and $\pm\Delta$ for $T(D){+}A$ in the second part.
The induction hypothesis for the first part is:
for $k \leq n$, if $\alpha \in \T_{D{+}A}{\uparrow} n$ then $T(D){+}A \vdash \alpha'$,
where $\alpha'$ is the counterpart, in $\DL(d_2)$, of $\alpha$.
For the second part it is:
for $k \leq n$, if $\alpha \in \Sigma$ and $\alpha \in \T_{T(D){+}A}{\uparrow} n$ then $D{+}A \vdash \alpha'$,
where $\alpha'$ is the counterpart, in $\DL(d_1)$, of $\alpha$.
Since $\T_{P}{\uparrow} 0 = \emptyset$ the induction hypothesis is always valid for $n=0$.

For the tags $\pm\Delta$ only the strict rules and facts are relevant.
The facts of $T(D){+}A$ are the same as those of $D{+}A$.
Furthermore, by the construction of $T(D)$, a rule $B \rightarrow q$ is in $D$ iff 
$T(D)$ contains the rules $B \rightarrow h(r); h(r) \rightarrow q$.
Now it is clear that if ${+}\Delta q \in \T_{D{+}A}{\uparrow} (n{+}1)$ then, for some strict rule $r$,
${+}\Delta B \subseteq \T_{D{+}A}{\uparrow} n$.  
By the induction hypothesis, $T(D){+}A \vdash {+}\Delta B$,
so we must have $T(D){+}A \vdash {+}\Delta q$.
Thus, by induction, if $D{+}A \vdash {+}\Delta q$ then $T(D){+}A \vdash {+}\Delta q$.
Similarly, if ${-}\Delta q \in \T_{D{+}A}{\uparrow} (n{+}1)$ then $q$ is not a fact in $D{+}A$ and, 
for every strict rule $r$ in $D$,
for some literal $p$ in the body of $r$, ${-}\Delta p \in \T_{D{+}A}{\uparrow} n$.
By the induction hypothesis, $T(D){+}A \vdash {-}\Delta p$.
Since rule $p(r)$ is the only rule for $h(r)$, $T(D){+}A \vdash {-}\Delta h(r)$, for each strict rule $r$ for $q$ in $D$.
Now, each strict rule for $q$ in $T(D){+}A$ contains a body literal $p$ where $T(D){+}A \vdash {-}\Delta p$
and hence $T(D){+}A \vdash {-}\Delta q$.

For the second part, for $\pm\Delta$,
if $q \in \Sigma$ and ${+}\Delta q \in \T_{T(D){+}A}{\uparrow} (n{+}1)$ then either
$q$ is a fact in $T(D){+}A$, or
${+}\Delta h(r) \in \T_{T(D){+}A}{\uparrow} n$, for some strict rule $r$ in $D$
(in which case we must have ${+}\Delta B \subseteq \T_{T(D){+}A}{\uparrow} n$).
By the induction hypothesis,  $q$ is a fact in $D{+}A$ or
$D{+}A \vdash {+}\Delta B$ and then an application of the ${+}\Delta$ inference rule gives us $D{+}A \vdash {+}\Delta q$.
Similarly, if $q \in \Sigma$ and ${-}\Delta q \in \T_{T(D){+}A}{\uparrow} (n{+}1)$ then 
$q$ is not a fact in $T(D){+}A$ and
every strict rule for $q$ in $T(D){+}A$ contains a body literal $p$ where ${-}\Delta p \in \T_{T(D){+}A}{\uparrow} n$.
Specifically, ${-}\Delta h(r)  \T_{T(D){+}A}{\uparrow} n$ for every strict rule $r$ for $q$ in $D$.
Hence, every body $B$ of a strict rule $r$ for $q$ in $D$ contains a literal $p$ where ${-}\Delta p \in \T_{T(D){+}A}{\uparrow} n$.
By the induction hypothesis,
$q$ is not a fact in $D{+}A$ and
every body $B$ of a strict rule $r$ for $q$ in $D{+}A$ contains a literal $p$ where 
$D{+}A \vdash {-}\Delta p$ and, hence, $D{+}A \vdash {-}\Delta q$.

This concludes the verification that the transformation preserves the $\pm\Delta$ consequences in $\Sigma$.
The proofs for the defeasible consequences are substantially more complicated
because of the more intricate inference rules for defeasible conclusions, but the same approach applies.
}

We consider the transformation $T(D)$ of a defeasible theory $D$ as defined in Definition \ref{defn:newTDsimID} (and Definition \ref{defn:TDsimID})
and show that this transformation provides a simulation of each defeasible theory $D$ in $\DL(\partial^*)$
from within $\DL(\partial)$.

Fix any $D$ and any $A$ that satisfies the language separation condition.
Let $\Sigma = \Sigma(D) \cup \Sigma(A)$.

[1]
If ${+}\partial^* q  \in \T_{D{+}A}{\uparrow} (n{+}1)$ then either ${+}\Delta q \in \T_{D{+}A}{\uparrow} n$
(in which case $T(D){+}A \vdash {+}\partial q$)
or ${+}\partial^* B_r  \subseteq \T_{D{+}A}{\uparrow} n$, where $B_r$ is the body of some strict or defeasible rule $r$ in $D{+}A$.
In the latter case, $\T_{D{+}A}{\uparrow} n$ also contains ${-}\Delta \non q$
and for every rule $s$ for $\non q$ in $D{+}A$ either $r > s$ or 
${-}\partial^* p  \in \T_{D{+}A}{\uparrow} n$ for some literal $p$ in the body of $s$.
Then, by the induction hypothesis,
$T(D){+}A \vdash {+}\partial B_r$,
$T(D){+}A \vdash {-}\Delta \non q$ and,
if $r \in D$,
for every rule $n(r, s)$ for $\neg h(r)$ in $T(D)$, either
$p(r) >' n(r, s)$ or $T(D){+}A \vdash {-}\partial p$ where $p$ occurs in the body of $n(r, s)$.
Thus, using the inference rule for ${+}\partial$, if $r \in D$ then $T(D){+}A \vdash {+}\partial h(r)$.
If $r \in A$ then $T(D){+}A \vdash {+}\partial B_r$ so,
whether $r \in D$ or $r \in A$,
there is a rule for $q$ in $T(D){+}A$ with body $B$ and $T(D){+}A \vdash {+}\partial B$.

Applying the inference rule for ${-}\partial$ multiple times,
for each strict or defeasible rule $s$ for $\non q$ in $D$ we have
$T(D){+}A \vdash {-}\partial h(s)$.
Furthermore, as noted above,
for every rule $s \in A$ for $\non q$, since $r \not> s$,
${-}\partial^* p  \in \T_{D{+}A}{\uparrow} n$ for some literal $p$ in the body of $s$.
Thus, every rule for $\non q$ in $T(D){+}A$ fails.
Now, again applying the inference rule for ${+}\partial$,
we have $T(D){+}A \vdash {+}\partial q$.

\finish{DONE}

[2]
If ${-}\partial^* q \in \T_{D{+}A}{\uparrow} (n{+}1)$ then 
${-}\Delta q  \in \T_{D{+}A}{\uparrow} n$ and either
${+}\Delta \non q  \in \T_{D{+}A}{\uparrow} n$ (in which case $T(D){+}A \vdash {-}\partial q$) or,
for every strict or defeasible rule $r$ for $q$ in $D{+}A$, 
either ${-}\partial^* p \in \T_{D{+}A}{\uparrow} n$ for some $p$ in the body of $r$ or
there exists a rule $s$ for $\non q$ with body $B$, ${+}\partial^* B \subseteq \T_{D{+}A}{\uparrow} n$ and $r \not> s$.
Then, for every strict or defeasible rule $p(r)$ in $T(D)$,
either ${-}\partial^* p \in \T_{D{+}A}{\uparrow} n$ for some $p$ in the body of $p(r)$
or there is a rule $n(r, s)$ with body $B$ and $p(r) \not>' n(r, s)$, by the structure of $T(D)$.
By the induction hypothesis,
for every strict or defeasible rule $p(r)$ in $T(D)$,
either $T(D){+}A \vdash {-}\partial p$ for some $p$ in the body of $p(r)$
or there is a rule $n(r, r')$ with body $B$ where $T(D){+}A \vdash {+}\partial B$ and $p(r) \not>' n(r, r')$.
Since there is only one rule for $h(r)$, application of the inference rule for ${-}\partial$ gives us
$T(D){+}A \vdash {-}\partial h(r)$ for each strict or defeasible rule $r \in D$ for $q$.
Also by the induction hypothesis,
for every strict or defeasible rule $r$ for $q$ in $A$,
$T(D){+}A \vdash {-}\partial p$ for some $p$ in the body of $r$.
Hence $T(D){+}A \vdash {-}\partial q$.

\finish{DONE}

[3]
If  $q \in \Sigma$ and ${+}\partial q \in \T_{T(D){+}A}{\uparrow} (n{+}1)$ then either 
(1) ${+}\Delta q \in \T_{T(D){+}A}{\uparrow} n$  (in which case $D{+}A \vdash {+}\partial^* q$),
or else (2) ${+}\partial h(r) \in \T_{T(D){+}A}{\uparrow} n$ for some strict or defeasible rule $r$ for $q$ in $D$,
or else (3) ${+}\partial B_r \subseteq \T_{T(D){+}A}{\uparrow} n$ for some strict or defeasible rule $r$ for $q$ in $A$ with body $B_r$.
In case (3), by the induction hypothesis, $D{+}A \vdash {+}\partial B_r$.
In case (2)
we must also have that every rule for $\non q$ in $T(D){+}A$ fails (except for $o(\non q)$, which is overruled);
that is,
for every rule $s$ for $\non q$ in $D$, 
${-}\partial h(s) \in \T_{T(D){+}A}{\uparrow} n$ 
and, for every rule for $\non q$ in $A$ with body $B$,
for some literal $p$ in $B$ ${-}\partial p \in \T_{T(D){+}A}{\uparrow} n$.
In case (3) we must also have that $one(\non q)$ fails,
so that every rule for $\non q$ in $D$ with body $B$,
for some literal $p$ in $B$ ${-}\partial p \in \T_{T(D){+}A}{\uparrow} n$.
Hence, by the induction hypothesis, in case (3), for every rule for $\non q$ in $D{+}A$ with body $B$,
for some literal $p$ in $B$ $D{+}A \vdash {-}\partial^* p$.
In both cases (2) and (3), 
${-}\Delta \non q \in \T_{T(D){+}A}{\uparrow} n$ and hence $D {+} A \vdash {-}\Delta \non q$.
Applying the $+\partial^*$ inference rule in case (3), $D{+}A \vdash {+}\partial^* q$.

In case (2), if ${+}\partial h(r) \in \T_{T(D){+}A}{\uparrow} n$ then
$p(r)$ is not a defeater,
${+}\partial B_r \subseteq \T_{T(D){+}A}{\uparrow} n$ where $B_r$ is the body of $r$
and for every rule $n(r, s)$ with body $B'$
either for some literal $p$ in $B'$ ${-}\partial p \in \T_{T(D){+}A}{\uparrow} n$
or for some rule $t$ for $h(r)$, its body is proved with respect to $\partial$ and $t > s$.
There is only one rule for $h(r)$, so this last disjunct reduces to $p(r) > n(r, s)$.
Using the construction of $T(D)$,
$r$ is not a defeater,
${+}\partial B_r \subseteq \T_{T(D){+}A}{\uparrow} n$ where $B_r$ is the body of $r$
and for every rule $s$ for $\non q$ in $D$ with body $B'$
either for some literal $p$ in $B'$, ${-}\partial p \in \T_{T(D){+}A}{\uparrow} n$
or $r > s$.
Furthermore, from the previous paragraph,
for every rule for $\non q$ in $A$ with body $B$,
for some literal $p$ in $B$ ${-}\partial p \in \T_{T(D){+}A}{\uparrow} n$.
Using the induction hypothesis,
$D{+}A \vdash {+}\partial^* B_r$,
and for every rule for $\non q$ in $D{+}A$
either for some literal $p$ in the body $D{+}A \vdash {-}\partial^* p$ or $r > s$.
Applying the inference rule for ${+}\partial^*$, 
we obtain $D{+}A \vdash {+}\partial^* q$.

\ignore{ TREATED AERLIER
In case (3),
as observed earlier,
for every rule $s$ for $\non q$ in $D$, 
${-}\partial h(s) \in \T_{T(D){+}A}{\uparrow} n$.
Using the structure of $D$, and following the ${-}\partial$ inference rule,
either
for some literal $p$ in the body of $s$, ${-}\partial p \in \T_{T(D){+}A}{\uparrow} n$,
or, for some rule for $q$ in $D$ with body $B$ that is not inferior to $s$,
${+}\partial B \in \T_{T(D){+}A}{\uparrow} n$.
In addition, 
${+}\partial B_r \subseteq \T_{T(D){+}A}{\uparrow} n$ for some strict or defeasible rule $r$ for $q$ in $A$ with body $B_r$, and,
for every rule for $\non q$ in $A$ with body $B$,
for some literal $p$ in $B$ ${-}\partial p \in \T_{T(D){+}A}{\uparrow} n$.
Applying the induction hypothesis,
for some rule $r$ for $q$ in $A$, $D{+}A \vdash {+}\partial^* B_r$ and,
for every rule for $\non q$ in $A$, there is a $p$ in its body such that $D{+}A \vdash {-}\partial^* p$, and,
for every rule $s$ for $\non q$ in $D$, either there is a $p$ in its body such that $D{+}A \vdash {-}\partial^* p$
or there is a rule for $q$ in $D$ with body $B$ that is not inferior to $s$ and $D{+}A \vdash {+}\partial^* B$.
Applying the inference rule for ${+}\partial^*$,
}

\finish{DONE}

[4]
If $q \in \Sigma$ and ${-}\partial q \in \T_{T(D){+}A}{\uparrow} (n{+}1)$ then,
using the $-\partial$ inference rule and the structure of $T(D)$,
${-}\Delta q  \in \T_{T(D){+}A}{\uparrow} n$ and either
(0) ${+}\Delta \non q  \in \T_{T(D){+}A}{\uparrow} n$ (in which case $D{+}A \vdash {-}\partial^* q$), or else
(1) $-\partial h(r) \in \T_{T(D){+}A}{\uparrow} n$ for every rule $r$ for $q$ in $D$,
while for every rule $r$ in $A$ there is a literal $p$ in the body of $r$ with ${-}\partial p \in \T_{T(D){+}A}{\uparrow} n$,
and $-\partial one(q) \in \T_{T(D){+}A}{\uparrow} n$;
or (2) $+\partial h(s) \in \T_{T(D){+}A}{\uparrow} n$ for some rule $s$ for $\non q$ in $D$;
or (3) $+\partial one(\non q) \in \T_{T(D){+}A}{\uparrow} n$, in which case there is
a rule $s$ for $\non q$ in $D$ where ${+}\partial B_s \subseteq \T_{T(D){+}A}{\uparrow} n$,
and $-\partial h(r) \in \T_{T(D){+}A}{\uparrow} n$ for every rule $r$ for $q$ in $D$
(so that $o(\non q)$ is not overruled);
or (4) there is a rule $s$ for $\non q$ in $A$ where ${+}\partial B_s \subseteq \T_{T(D){+}A}{\uparrow} n$.
In any case, using the induction hypothesis, we have $D{+}A \vdash {-}\Delta q$.

In case (1), since $-\partial one(q) \in \T_{T(D){+}A}{\uparrow} n$,
for every rule $r$ in $D$ for $q$ there is a literal $p$ in $B_r$ with $-\partial p  \in \T_{T(D){+}A}{\uparrow} n$.
Thus all rules for $q$ in $D{+}A$ fail, and hence $D{+}A \vdash {-}\partial^* q$.
In case (2), we must have, for every rule $r$ in $D$ for $q$, either
$s > r$ (so that $p(s) > n(s, r)$ or there is a literal $p$ in $B_r$ with $-\partial p  \in \T_{T(D){+}A}{\uparrow} n$.
By the induction hypothesis, we then have $D{+}A \vdash {-}\partial^* p$, for each such $p$.
Furthermore, no rule in $A$ can overrule $s$.
Hence, applying the $-\partial^*$ inference rule, $D{+}A \vdash {-}\partial^* q$.

In case (3), since $-\partial h(r) \in \T_{T(D){+}A}{\uparrow} n$,
either there is a literal $p$ in $B_r$ with $-\partial p  \in \T_{T(D){+}A}{\uparrow} n$ or
there is a rule $s$ in $D$ for $\non q$ with ${+}\partial B_s \subseteq \T_{T(D){+}A}{\uparrow} n$
and $r \not> s$ (so that $p(r) \not> n(s, r)$).
By the induction hypothesis,
for every rule $r$ for $q$ in $D$ either
there is a literal $p$ in $B_r$ with $D{+}A \vdash {-}\partial^* p$ or
there is a rule $s$ in $D$ for $\non q$ with $D{+}A \vdash {+}\partial^* B_s$ and $r \not> s$.
Furthermore, from $+\partial one(\non q)$ we know there is an $s$ in $D$ with
(using the induction hypothesis) $D{+}A \vdash {+}\partial^* B_s$,
and this $s$ cannot be overruled by any rule $r$ in $A$.
Consequently, applying the $-\partial^*$ inference rule, $D{+}A \vdash {-}\partial^* q$.

In case (4), by the induction hypothesis, we have
there is a rule $s$ for $\non q$ in $A$ where $D{+}A \vdash {+}\partial^* B_s$
and, since $s$ cannot be inferior to any rule,
applying the $-\partial^*$ inference rule we have $D{+}A \vdash {-}\partial^* q$.
\end{proof}

\ignore{
(2) $-\partial one(q) \in \T_{T(D){+}A}{\uparrow} n$, and hence ${-}\partial p  \in \T_{T(D){+}A}{\uparrow} n$ for some $p$ in the body of $r$ for every rule $r$ for $q$ in $D$; 
and for every rule $r$ in $A$ there is a literal $p$ in the body of $r$ with ${-}\partial p \in \T_{T(D){+}A}{\uparrow} n$, or
(3) there is a strict or defeasible rule $s$ for $\non q$ in $D$ where ${+}\partial h(s) \in \T_{T(D){+}A}{\uparrow} n$, or
(4) there is a strict or defeasible rule $s$ for $\non q$ in $A$ where ${+}\partial B_r \subseteq \T_{T(D){+}A}{\uparrow} n$.
Consequently, $D{+}A \vdash {-}\Delta q$.
In the first case, using the induction hypothesis, we have
$D{+}A \vdash {-}\Delta q$ and $D{+}A \vdash {+}\Delta \non q$; hence, $D{+}A \vdash {-}\partial^* q$.
In the second case, for $r \in D$,
either
$r$ is a defeater, or
there is a literal $p$ in the body of $r$ such that ${-}\partial p \in \T_{T(D){+}A}{\uparrow} n$, or
there is a rule $s$ for $\non q$ in $D$ (corresponding to rule $n(r, s)$ in $T(D)$)
with body $B'$ where ${+}\partial B' \subseteq \T_{T(D){+}A}{\uparrow} n$  and $r \not> s$.
By the induction hypothesis,
either 
$r$ is a defeater, or
there is a literal $p$ in the body of $r$ such that $D{+}A \vdash {-}\partial^* p$, or
there is a rule $s$ for $\non q$ in $D$ 
with body $B'$ where $D{+}A \vdash {+}\partial^* B' $  and $r \not> s$.
Further,
for every rule $r$ for $q$ in $A$, $D{+}A \vdash {-}\partial^* p$, for some $p$ in the body of $r$.
Applying the inference rule for ${-}\partial^*$,
we obtain $D{+}A \vdash {-}\partial^* q$.

In the third case, either
${+}\Delta h(s) \in \T_{T(D){+}A}{\uparrow} n$, or
${+}\partial B \subseteq \T_{T(D){+}A}{\uparrow} n$, where $B$ is the body of $s$,
and ${-}\Delta \neg h(s) \in \T_{T(D){+}A}{\uparrow} n$,
and, for every rule $r$ for $q$ in $D$, either
${-}\partial p \in \T_{T(D){+}A}{\uparrow} n$ for some literal $p$ in the body of $s$ or
$s > r$ (using the fact that there is only one rule for $h(s)$ in $T(D){+}A$).
If ${+}\Delta h(s) \in \T_{T(D){+}A}{\uparrow} n$ then ${+}\Delta B \subseteq \T_{T(D){+}A}{\uparrow} n$ and $s$ is strict.
Using the induction hypothesis, $D{+}A \vdash {+}\Delta B$ and, hence, $D{+}A \vdash {+}\Delta \non q$ and,
like case (1) above, $D{+}A \vdash {-}\partial^* q$.
In the other case, by the induction hypothesis,
$D{+}A \vdash {+}\partial^* B$ and, for every rule $r$ for $q$ in $D$, either
$D{+}A \vdash {-}\partial^* p$ for some literal $p$ in the body of $s$ or
$s > r$.
Applying the inference rule for ${-}\partial^*$ we conclude $D{+}A \vdash {-}\partial^* q$.
In the fourth case, by the induction hypothesis,
there is a strict or defeasible rule $s$ in $A$ with body $B_s$ and $D{+}A \vdash {+}\partial^* B_s$.
Since no rule can overrule $s$, we conclude $D{+}A \vdash {-}\partial^* q$.
 }
\finish{DONE}

This concludes the proof that $\DL(\partial)$ can simulate $\DL(\partial^*)$ with respect to addition of rules.
We now turn to the corresponding proof for $\DL(\delta)$ and $\DL(\delta^*)$. 

\begin{theorem}  \label{thm:TDsimIDAP}
The logic  $\DL(\delta^*)$ can be simulated by $\DL(\delta)$
with respect to addition of rules.
\end{theorem}
\begin{proof}




Let $A$ be any set of rules.
Let $\Sigma$ be the language of $D{+}A$ and let $q \in \Sigma$.
Let $T(D)$ be the transformed defeasible theory as described in Definition \ref{defn:newTDsimID}.
Then we claim
\begin{itemize}
\item
$D{+}A \vdash {+}\supp^* q$ iff $T(D){+}A \vdash {+}\supp q$
\item
$D{+}A \vdash {-}\supp^* q$ iff $T(D){+}A \vdash {-}\supp q$
\item
$D{+}A \vdash {+}\delta^* q$ iff $T(D){+}A \vdash {+}\delta q$
\item
$D{+}A \vdash {-}\delta^* q$ iff $T(D){+}A \vdash {-}\delta q$
\end{itemize}

If ${+}\delta^* q  \in \T_{D{+}A}{\uparrow} (n{+}1)$ then either ${+}\Delta q \in \T_{D{+}A}{\uparrow} n$
(in which case $T(D){+}A \vdash {+}\delta q$)
or else ${+}\delta^* B_r  \subseteq \T_{D{+}A}{\uparrow} n$, where $B_r$ is the body of some strict or defeasible rule $r$ in $D{+}A$.
In the latter case, $\T_{D{+}A}{\uparrow} n$ also contains ${-}\Delta \non q$
and for every rule $s$ for $\non q$ in $D{+}A$ either $r > s$ or 
${-}\supp^* p  \in \T_{D{+}A}{\uparrow} n$ for some literal $p$ in the body of $s$.
Then, by the induction hypothesis,
$T(D){+}A \vdash {+}\delta B_r$,
$T(D){+}A \vdash {-}\Delta \non q$ and,
if $r$ and $s$ are in $D$,
for every rule $n(r, s)$ for $\neg h(r)$ in $T(D)$, either
$p(r) >' n(r, s)$ or $T(D){+}A \vdash {-}\supp p$ where $p$ occurs in the body of $n(r, s)$ and, similarly,
the rule $p(s)$ for $h(s)$ in $T(D)$, either
$p(s) <' n(s, r)$ or $T(D){+}A \vdash {-}\supp p$ where $p$ occurs in the body of $p(s)$.
If $s$ is in $A$ then $r > s$ cannot occur (since the rules of $A$ do not participate in the superiority relation)
and $T(D){+}A \vdash {-}\supp p$ where $p$ occurs in the body of $s$.
If $r$ is in $A$ and $s$ is in $D$ then, again, $r > s$ cannot occur
and $T(D){+}A \vdash {-}\supp p$ where $p$ occurs in the body of $p(s)$.
Thus, using the inference rules for ${+}\delta$ and ${-}\supp$, 
if $r$ is in $D$ then $T(D){+}A \vdash {+}\delta h(r)$
and if $s$ is in $D$ then $T(D){+}A \vdash {-}\supp h(s)$.
Now, applying the inference rule for ${+}\delta$,
we conclude $T(D){+}A \vdash {+}\delta q$.

If ${-}\delta^* q \in \T_{D{+}A}{\uparrow} (n{+}1)$ then 
${-}\Delta q  \in \T_{D{+}A}{\uparrow} n$ (and, hence, $T(D){+}A \vdash {-}\Delta q$) and either
${+}\Delta \non q  \in \T_{D{+}A}{\uparrow} n$ (in which case $T(D){+}A \vdash {-}\delta q$) or,
for every strict or defeasible rule $r$ for $q$ in $D{+}A$, 
either ${-}\delta^* p \in \T_{D{+}A}{\uparrow} n$ for some $p$ in the body of $r$ or
there exists a rule $s$ for $\non q$ with body $B_s$, 
where ${+}\supp^* B_s \subseteq \T_{D{+}A}{\uparrow} n$ and $r \not> s$.
Now, if, for some $s$  for $\non q$ in $A$,
${+}\supp^* B_s \subseteq \T_{D{+}A}{\uparrow} n$
then, by the induction hypothesis,
$T(D){+}A \vdash {+}\supp B_s$
and, 
applying the inference rule for ${-}\delta$
(and noting that no rule is superior to $s$),
we have
$T(D){+}A \vdash {-}\delta q$.

Otherwise, for every strict or defeasible rule $p(r)$ in $T(D)$,
either ${-}\delta^* p \in \T_{D{+}A}{\uparrow} n$ for some $p$ in the body of $p(r)$
or there is a rule $n(r, s)$ with body $B_s$ and $p(r) \not>' n(r, s)$, by the structure of $T(D)$.
By the induction hypothesis,
for every strict or defeasible rule $p(r)$ in $T(D)$,
either $T(D){+}A \vdash {-}\delta p$ for some $p$ in the body of $p(r)$
or there is a rule $n(r, s)$ with body $B_s$ where $T(D){+}A \vdash {+}\supp B_s$ and $p(r) \not>' n(r, s)$.
In both cases, since there is only one rule for $h(r)$, application of the inference rule for ${-}\delta$ gives us
$T(D){+}A \vdash {-}\delta h(r)$ for each strict or defeasible rule $r$ for $q$ in $D$.
Hence, no rule $s(r)$ can overrule $o(\non q)$.
Now, if every rule $r$ for $q$ in $A$ has $p \in B_r$ with $-\delta^* p \in \T_{D{+}A}{\uparrow} n$
then, by the induction hypothesis, $T(D){+}A \vdash {-}\delta p$ for every such rule an hence,
by application of the $-\delta$ inference rule, $T(D){+}A \vdash {-}\delta q$. 
Otherwise, there is $s$ for $\non q$ in $D$ with $T(D){+}A \vdash +\supp B_s$.
By the $+\supp$ inference rule $T(D){+}A \vdash +\supp one(\non q)$.
Consequently, since $o(\non q)$ cannot be overruled, $T(D){+}A \vdash {-}\delta q$.

Hence, in every case, $T(D){+}A \vdash {-}\delta q$.

\finish{DONE}

If  $q \in \Sigma$ and ${+}\delta q \in \T_{T(D){+}A}{\uparrow} (n{+}1)$ then either 
${+}\Delta q \in \T_{T(D){+}A}{\uparrow} n$  (in which case $D{+}A \vdash {+}\delta^* q$),
or ${+}\delta h(r) \in \T_{T(D){+}A}{\uparrow} n$ for some strict or defeasible rule $r$ for $q$ in $D$.
or ${+}\delta B \subseteq \T_{T(D){+}A}{\uparrow} n$ for some strict or defeasible rule for $q$ in $A$ with body $B$.
In the latter cases, $\T_{T(D){+}A}{\uparrow} n$ also contains ${-}\Delta \non q$;
hence $D {+} A \vdash {-}\Delta \non q$.
In these cases we must also have, for each rule for $\non q$ in $A$,
for some $p$ in its body $-\supp p \in \T_{T(D){+}A}{\uparrow} n$.
Hence, by the induction hypothesis, for each rule for $\non q$ in $A$,
for some $p$ in its body $D{+}A \vdash -\supp^* p$.
If ${+}\delta h(r) \in \T_{T(D){+}A}{\uparrow} n$ then
$p(r)$ is not a defeater,
${+}\delta B_r \subseteq \T_{T(D){+}A}{\uparrow} n$ where $B_r$ is the body of $r$
and for every rule $n(r, s)$ with body $B_s$
either for some literal $p$ in $B_s$ ${-}\supp p \in \T_{T(D){+}A}{\uparrow} n$
or for some rule $t$ for $h(r)$, its body is proved with respect to $\delta$ and $t > s$.
There is only one rule for $h(r)$, so this last disjunct reduces to $p(r) > n(r, s)$.
Using the construction of $T(D)$,
$r$ is not a defeater,
${+}\delta B_r \subseteq \T_{T(D){+}A}{\uparrow} n$
and for every rule $s$ for $\non q$  in $D$
either for some literal $p$ in $B_s$, ${-}\supp p \in \T_{T(D){+}A}{\uparrow} n$
or $r > s$.
Using the induction hypothesis,
$D{+}A \vdash {+}\delta^* B_r$,
and for every rule for $\non q$ in $D$
either for some literal $p$ in the body $D{+}A \vdash {-}\supp^* p$ or $r > s$.
Applying the inference rule for ${+}\delta^*$ to this statement,
and given we have shown that all rules for $\non q$ in $A$ fail,
we obtain $D{+}A \vdash {+}\delta^* q$.

\finish{DONE}

If $q \in \Sigma$ and ${-}\delta q \in \T_{T(D){+}A}{\uparrow} (n{+}1)$ then
${-}\Delta q  \in \T_{T(D){+}A}{\uparrow} n$ and either
(1) ${+}\Delta \non q  \in \T_{T(D){+}A}{\uparrow} n$ (in which case $D{+}A \vdash {-}\delta^* q$), or
(2) ${-}\delta h(r)  \in \T_{T(D){+}A}{\uparrow} n$ for every rule $r$ for $q$ in $D$
and every rule for $q$ in $A$ has a literal $p$ in its body with ${-}\delta q  \in \T_{T(D){+}A}{\uparrow} n$, or
(3) there is a rule $s$ for $\non q$ in $D$ where ${+}\supp h(s) \in \T_{T(D){+}A}{\uparrow} n$, or 
(4) there is a rule for $\non q$ in $A$ with body $B$ and ${+}\supp B \subseteq \T_{T(D){+}A}{\uparrow} n$.
(Some conditions are simpler than the inference rule for ${-}\delta$ might suggest because the superiority relation
in $T(D){+}A$ does not involve the rules for $q$ and $\non q$.)
Consequently, $D{+}A \vdash {-}\Delta q$.
In the first case, using the induction hypothesis, we have
$D{+}A \vdash {-}\Delta q$ and $D{+}A \vdash {+}\Delta \non q$; hence, $D{+}A \vdash {-}\delta^* q$.
In the second case, for each $r$,
either
$r$ is a defeater, or
there is a literal $p$ in the body of $r$ such that ${-}\delta p \in \T_{T(D){+}A}{\uparrow} n$, or
there is a rule $s$ for $\non q$ in $D$ (corresponding to rule $n(r, s)$ in $T(D)$)
with body $B_s$ where ${+}\supp B_s \subseteq \T_{T(D){+}A}{\uparrow} n$  and $r \not> s$.
By the induction hypothesis,
either 
$r$ is a defeater, or
there is a literal $p$ in the body of $r$ such that $D{+}A \vdash {-}\delta^* p$, or
there is a rule $s$ for $\non q$ in $D$ 
with body $B_s$ where $D{+}A \vdash {+}\supp^* B_s$  and $r \not> s$.
Similarly, using the induction hypothesis,
every rule for $q$ in $A$ has a literal $p$ in its body with $D{+}A \vdash {-}\delta^* p$.
Applying the inference rule for ${-}\delta^*$,
we obtain $D{+}A \vdash {-}\delta^* q$.

In the third case, either
${+}\Delta h(s) \in \T_{T(D){+}A}{\uparrow} n$, or
${+}\supp B \subseteq \T_{T(D){+}A}{\uparrow} n$, where $B_s$ is the body of $s$,
and, for every rule $r$ for $q$ in $D$, either
${-}\delta p \in \T_{T(D){+}A}{\uparrow} n$ for some literal $p$ in the body of $r$ or
$r \not> s$.
If ${+}\Delta h(s) \in \T_{T(D){+}A}{\uparrow} n$ then ${+}\Delta B \subseteq \T_{T(D){+}A}{\uparrow} n$ and $s$ is strict.
Using the induction hypothesis, $D{+}A \vdash {+}\Delta B$ and, hence, $D{+}A \vdash {+}\Delta \non q$ and,
like case (1) above, $D{+}A \vdash {-}\delta^* q$.
In the other case, by the induction hypothesis,
$D{+}A \vdash {+}\supp^* B_s$ and, 
for every rule $r$ for $q$ in $D$, either
$D{+}A \vdash {-}\delta^* p$ for some literal $p$ in the body of $r$ or
$r \not> s$.
Applying the inference rule for ${-}\delta^*$ we conclude $D{+}A \vdash {-}\delta^* q$.

In the fourth case,  using the induction hypothesis,
there is a rule for $\non q$ in $A$ with body $B$ and $D{+}A \vdash {+}\supp^* B$.
Applying the inference rule for ${-}\delta^*$ we conclude $D{+}A \vdash {-}\delta^* q$.


If ${+}\supp^* q  \in \T_{D{+}A}{\uparrow} (n{+}1)$ then either ${+}\Delta q \in \T_{D{+}A}{\uparrow} n$
(in which case $T(D){+}A \vdash {+}\supp q$)
or ${+}\supp^* B_r  \subseteq \T_{D{+}A}{\uparrow} n$, where $B_r$ is the body of some strict or defeasible rule $r$ in $D{+}A$.
In the latter case, 
for every rule $s$ for $\non q$ in $D{+}A$ either $s \not> r$ or 
${-}\delta^* p  \in \T_{D{+}A}{\uparrow} n$ for some literal $p$ in the body of $s$.
If $r \in A$ then $r$ is not inferior to any rule.
So, by the induction hypothesis,
$T(D){+}A \vdash {+}\supp B_r$, and, by the $+\supp$ inference rule $T(D){+}A \vdash {+}\supp q$.
If $r \in D$, by the induction hypothesis,
$T(D){+}A \vdash {+}\supp B_r$,
for every rule $n(r, s)$ for $\neg h(r)$ in $T(D)$, either
$n(r, s) \not>' p(r)$ or $T(D){+}A \vdash {-}\delta p$ where $p$ occurs in the body of $n(r, s)$.
Thus, using the inference rule for ${+}\supp$, $T(D){+}A \vdash {+}\supp h(r)$.
Applying the inference rule for ${-}\delta$ multiple times,
for each strict or defeasible rule $s$ for $\non q$ in $D$ we have
$T(D){+}A \vdash {-}\delta h(s)$.
Now, applying the inference rule for ${+}\supp$,
we have $T(D){+}A \vdash {+}\supp q$.

If ${-}\supp^* q \in \T_{D{+}A}{\uparrow} (n{+}1)$ then 
${-}\Delta q  \in \T_{D{+}A}{\uparrow} n$ and 
for every strict or defeasible rule $r$ for $q$ in $D{+}A$, 
either ${-}\supp^* p \in \T_{D{+}A}{\uparrow} n$ for some $p$ in the body of $r$ or
there exists a rule $s$ for $\non q$ in $D{+}A$ with body $B_s$, 
${+}\delta^* B_s \subseteq \T_{D{+}A}{\uparrow} n$ and $s > r$.
If $r \in A$ then ${-}\supp^* p \in \T_{D{+}A}{\uparrow} n$ for some $p$ in the body of $r$
and hence, by the induction hypothesis, $T(D){+}A \vdash -\supp  p$.
If $r \in D$
then, for every strict or defeasible rule $p(r)$ in $T(D)$,
either ${-}\supp^* p \in \T_{D{+}A}{\uparrow} n$ for some $p$ in the body of $p(r)$
or there is a rule $n(r, s)$ with body $B_s$ with ${+}\delta^* B_s \subseteq \T_{D{+}A}{\uparrow} n$ and $n(r, s) >' p(r)$, by the structure of $T(D)$.
By the induction hypothesis,
for every strict or defeasible rule $p(r)$ in $T(D)$,
either $T(D){+}A \vdash {-}\supp p$ for some $p$ in the body of $p(r)$
or there is a rule $n(r, s)$ with body $B_s$ where $T(D){+}A \vdash {+}\delta B$ and $n(r, s) >' p(r)$.
Application of the inference rule for ${-}\supp$ gives us
$T(D){+}A \vdash {-}\supp h(r)$ for each strict or defeasible rule $r$ for $q$ in $D$.
Rules in $A$ for $q$ also fail, as mentioned above.
Hence $T(D){+}A \vdash {-}\supp q$.


If  $q \in \Sigma$ and ${+}\supp q \in \T_{T(D){+}A}{\uparrow} (n{+}1)$ then either 
${+}\Delta q \in \T_{T(D){+}A}{\uparrow} n$  (in which case $D{+}A \vdash {+}\supp^* q$),
or ${+}\supp  B \subseteq  \T_{T(D){+}A}{\uparrow} n$ for some strict or defeasible rule for $q$ in $A$ with body $B$,
or ${+}\supp h(r) \in \T_{T(D){+}A}{\uparrow} n$ for some strict or defeasible rule $r$ for $q$ in $D$.
In the second case, 
by the induction hypothesis, $D{+}A \vdash {+}\supp^* B$ and, applying the ${+}\supp^*$ inference rule,
$D{+}A \vdash {+}\supp^* q$.
In the third case, if ${+}\supp h(r) \in \T_{T(D){+}A}{\uparrow} n$ then
$p(r)$ is not a defeater,
${+}\supp B_r \subseteq \T_{T(D){+}A}{\uparrow} n$ where $B_r$ is the body of $r$
and for every rule $n(r, s)$ with body $B_s$
either for some literal $p$ in $B_s$, ${-}\delta p \in \T_{T(D){+}A}{\uparrow} n$ or
$n(r,s) \not> p(r)$.
Using the construction of $T(D)$,
$r$ is not a defeater,
${+}\supp B_r \subseteq \T_{T(D){+}A}{\uparrow} n$ where $B_r$ is the body of $r$
and for every rule $s$ for $\non q$ with body $B_s$ in $D$
either for some literal $p$ in $B'$, ${-}\delta p \in \T_{T(D){+}A}{\uparrow} n$
or $s \not> r$.
Using the induction hypothesis,
$D{+}A \vdash {+}\supp^* B$,
and for every rule for $\non q$ in $D$
either for some literal $p$ in the body $D{+}A \vdash {-}\delta^* p$ or $s \not> r$.
Note also that no rule $s$ for $\non q$ in $A$ can be superior to $r$.
Applying the inference rule for ${+}\supp^*$ to this statement,
we obtain $D{+}A \vdash {+}\supp^* q$.

If $q \in \Sigma$ and ${-}\supp q \in \T_{T(D){+}A}{\uparrow} (n{+}1)$ then
${-}\Delta q  \in \T_{T(D){+}A}{\uparrow} n$ (and, consequently, $D{+}A \vdash {-}\Delta q$) and
for every strict or defeasible rule $r$ for $q$ in $A$ with body $B$ 
there is $p$ in $B$ with $-\supp p \in \T_{T(D){+}A}{\uparrow} n$,
and, for every rule $r$ for $q$ in $D$,
${-}\supp h(r)  \in \T_{T(D){+}A}{\uparrow} n$.
From ${-}\supp h(r)$ either
$r$ is a defeater, or
there is a literal $p$ in the body of $r$ such that ${-}\supp p \in \T_{T(D){+}A}{\uparrow} n$, or
there is a rule $s$ for $\non q$ in $D$ (corresponding to rule $n(r, s)$ in $T(D)$)
with body $B_s$ where ${+}\delta B_s \subseteq \T_{T(D){+}A}{\uparrow} n$  and $s > r$.
By the induction hypothesis,
either 
$r$ is a defeater, or
there is a literal $p$ in the body of $r$ such that $D{+}A \vdash {-}\supp^* p$, or
there is a rule $s$ for $\non q$ in $D$ 
with body $B_s$ where $D{+}A \vdash {+}\delta^* B_s$  and $s > r$.
Applying the inference rule for ${-}\supp^*$,
we obtain $D{+}A \vdash {-}\supp^* q$.

\finish{DONE $\supp$}
\end{proof}

\ignore{
In the third case, either
${+}\Delta h(s) \in \T_{T(D){+}A}{\uparrow} n$, or
${+}\supp B \subseteq \T_{T(D){+}A}{\uparrow} n$, where $B$ is the body of $s$,
and ${-}\Delta \neg h(s) \in \T_{T(D){+}A}{\uparrow} n$,
and, for every rule $r$ for $q$ in $D$, either
${-}\delta p \in \T_{T(D){+}A}{\uparrow} n$ for some literal $p$ in the body of $s$ or
$r \not> s$.
If ${+}\Delta h(s) \in \T_{T(D){+}A}{\uparrow} n$ then ${+}\Delta B \subseteq \T_{T(D){+}A}{\uparrow} n$ and $s$ is strict.
Using the induction hypothesis, $D{+}A \vdash {+}\Delta B$ and, hence, $D{+}A \vdash {+}\Delta \non q$ and,
like case (1) above, $D{+}A \vdash {-}\supp^* q$.
In the other case, by the induction hypothesis,
$D{+}A \vdash {+}\supp^* B$ and, for every rule $r$ for $q$ in $D$, either
$D{+}A \vdash {-}\delta^* p$ for some literal $p$ in the body of $s$ or
$r \not> s$.
Applying the inference rule for ${-}\supp^*$ we conclude $D{+}A \vdash {-}\supp^* q$.
}

Combining Theorems \ref{thm:TDsimIDAB} and \ref{thm:TDsimIDAP}
we have Theorem \ref{thm:TDsimAD}.


\section{Simulation of Team Defeat wrt Addition of Rules}

The same theory $D$ and addition $A$ as in Example \ref{ex:failAB} demonstrates
that the simulation of $\DL(\partial)$ by $\DL(\partial^*)$ wrt addition of facts exhibited in \cite{Maher12}
does not extend to addition of rules.

\begin{example}   \label{ex:failAB_2}
Let $D$ consist of the rules
\[
\begin{array}{lrll}
r_1:   &                           & \Rightarrow & \phantom{\neg} p \\ 
r_2:   &                           & \Rightarrow &  \neg p \\
\end{array}
\]
and let $A$ be the rule
\[
\begin{array}{lrll}
   &                           & \Rightarrow & p \\ 
\end{array}
\]
Then $D+A \vdash -\partial p$.

The transformation presented in \cite{Maher12}
simulates $D$ wrt addition of facts with the following theory $D'$:

\[
\begin{array}{lrllrlrll}
R1_{12}:         &                           & \Rightarrow & \neg d(r_1, r_2) \phantom{MM} &
R1_{21}:         &                           & \Rightarrow & \neg d(r_2, r_1) \\
R2_{12}:         &                           & \Rightarrow & \phantom{\neg}  d(r_1, r_2) &
R2_{21}:         &                           & \Rightarrow & \phantom{\neg}  d(r_2, r_1) \\
& & & & \\
NF_1:   &                           & \Rightarrow & \neg fail(r_1) &
NF_2:   &                           & \Rightarrow & \neg fail(r_2) \\
F_1:   &                           & \Rightarrow & \phantom{\neg} fail(r_1) &
F_2:   &                           & \Rightarrow & \phantom{\neg} fail(r_2) \\
& & & & \\
               & d(r_1, r_2)       & \Rightarrow & d(r_1) &
               & d(r_2, r_1)       & \Rightarrow & d(r_2) \\
               & fail(r_1)            & \Rightarrow & d(r_1) &
               & fail(r_2)            & \Rightarrow & d(r_2) \\
& & & & \\               
               &                           & \Rightarrow & one(p) &
               & one(p), d(r_2) & \Rightarrow & \phantom{\neg} p \\
               &                           & \Rightarrow & one(\neg p) &
               & one(\neg p), d(r_1) & \Rightarrow & \neg p \\
\end{array}
\]
with $NF_1 > F_1$ and $NF_2 > F_2$.
(Rules $R3_{ij}$ have been omitted because there are no strict rules in $D$.)

Then consequences of $D'$ (and $D' {+} A$) include
$-\partial^* d(r_1, r_2)$ and $-\partial^* fail(r_1)$,
and hence also $-\partial^* d(r_1)$.
Consequently, the only rule for $\neg p$ in $D' {+} A$ fails and hence,
using the rule in $A$, we can conclude $+\partial^* p$.

Thus $D'$ does not simulate $D$ wrt addition of rules.
The weakness of the transformation in the previous section is also evident here.
\end{example}

\begin{lemma}  \label{lemma:strict_partialstar}
Let $D$ be a defeasible theory, 
$T(D)$ be the transformed defeasible theory as described in Definition \ref{defn:IDsimTD},
and let $A$ be a modular defeasible theory.
Let $\Sigma$ be the language of $D{+}A$ and let $q \in \Sigma$.
Then
\begin{itemize}
\item
$D{+}A \vdash {+}\Delta q$ iff $T(D){+}A \vdash {+}\Delta q$ \hspace{32.5pt} iff $T(D){+}A \vdash {+}\partial^* strict(q)$ 

\hspace{57pt}
iff $T(D){+}A \vdash {+}\partial^* true(q)$  \hspace{3pt} iff $T(D){+}A \vdash -\partial^* \neg true(q)$  \\

\item
$D{+}A \vdash -\Delta q$ iff $T(D){+}A \vdash -\Delta q$ \hspace{32.5pt} iff $T(D){+}A \vdash -\partial^* strict(q)$  

\hspace{57pt}
iff $T(D){+}A \vdash -\partial^* true(q)$  \hspace{3pt} iff $T(D){+}A \vdash {+}\partial^* \neg true(q)$

\end{itemize}
\end{lemma}
\begin{proof}
This result follows immediately from Lemma \ref{lemma:strictAP}
and the inclusion theorem, since $\delta^* \subseteq \partial^* \subseteq \supp^*$.
\end{proof}

We say a rule $r$ fails in $D$ if, for some literal $p$ in the body of $r$, $D \vdash -\partial^* p$.
Similarly, $r$ fails in $\T\uparrow n$ if $-\partial^* p \in \T\uparrow n$ for some literal $p$ in the body of $r$.

\begin{theorem}   \label{thm:IDsimTDpartial}
The logic $\DL(\partial)$ can be simulated by $\DL(\partial^*)$
with respect to addition of rules.
\end{theorem}
\begin{proof}
Let $\Sigma$ be the language of $D{+}A$.
Note, that, employing Lemma \ref{lemma:strictAP},
$T(D){+}A \vdash +\partial^* \neg true(q)$ iff $T(D){+}A \vdash -\Delta q$ iff $D{+}A  \vdash -\Delta q$.
Because $T(D){+}A \vdash -\partial^* g$, we can essentially ignore the rules $supp(q)$,
which are only included for the simulation of $\DL(\delta)$ by $\DL(\delta^*)$.

Suppose $+\partial q \in \T_{D+A}\uparrow (n+1)$.
Then 
either $+\Delta q \in \T_{D+A}\uparrow n$ (in which case $T(D)+A \vdash +\partial^* q$), or
$-\Delta \non q \in \T_{D+A}\uparrow n$ and
there is a non-empty team of strict or defeasible rules for $q$
such that $+\partial B_r \subseteq \T_{D+A}\uparrow n$ for each body $B_r$ of each rule $r$
and
every rule $s$ for $\non q$ either has a body that fails in $\T_{D+A}\uparrow n$
or $s < t$ for some rule $t$ in the team.
$t \notin A$ because rules in $A$ do not participate in the superiority relation.
Then, by the induction hypothesis,
$T(D)+A \vdash -\Delta \non q$,
$T(D)+A \vdash +\partial^* B_r$ for each rule $r$ in the team,
and
for every rule $s$ for $\non q$ either its body fails in $T(D)+A$ or
there is a rule $t$ in the team
and $t > s$.
If $s \in A$ then its body $B_s$ fails in $T(D)+A$.
If $s \in D$ then either $T(D)+A \vdash +\partial^* fail(s)$ or $T(D)+A \vdash +\partial^* d(s,t)$;
in either case, $T(D)+A \vdash +\partial^* d(s)$.
Considering $T(D)$ and the inference rule for $+\partial^*$, we have
$T(D)+A \vdash +\partial^* one(q)$.
By Lemma \ref{lemma:strict_partialstar}, $T(D)+A \vdash +\partial^* \neg true(\non q)$.
Hence, the body of $s(q)$ is proved.
Because $>$ is acyclic,
there is a rule in the team for $q$ that is not inferior to any rule in the team for $\non q$.
Hence this rule $r'$ is not defeated, so $d(r')$ fails, and hence the rule $s(\non q)$ in $T(D)$ (from point 6) fails.
Hence all rules for $\non q$ fail, with the possible exception of $o(\non q)$.
However $s(q) > o(\non q)$ and hence, applying the $+\partial^*$ inference rule,
$T(D)+A \vdash + \partial^* q$.

Suppose $-\partial q \in \T_{D+A}\uparrow (n+1)$.
Then $-\Delta q \in \T_{D+A}\uparrow n$ (and hence $T(D)+A \vdash -\Delta q$) and
either 
(1) $+\Delta \non q \in  \T_{D+A}\uparrow n$ (in which case $T(D)+A \vdash -\partial^* q$), or
(2) every rule $r$ for $q$ fails, or
(3) there is a rule $s$ for $\non q$ with body $B_s$ such that $+\partial B_s \subseteq  \T_{D+A}\uparrow n$ and,
for every strict or defeasible rule $t$ for $q$, 
either
$t$ fails in $\T_{D+A}\uparrow n$, or
$t \not> s$.
In case (2), 
the rules $r$ in $A$ for $q$ fail and,
by the induction hypothesis and the inference rule for $-\partial^*$,
the rules $r$ in $A$ for $q$ fail and,
$T(D)+A \vdash -\partial^* one(q)$
and hence $T(D)+A \vdash -\partial^* q$.
In case (3), by the induction hypothesis,
there is a rule $s$ for $\non q$ with body $B_s$ such that 
$T(D)+A \vdash +\partial^* B_s$ and
for every strict or defeasible rule $t$ for $q$, 
either
$t$ fails in $T(D)+A$, or
$t \not> s$.
If $s \in A$ then $t \not> s$, for every $t$, and hence $T(D)+A \vdash -\partial^* q$.
If $s \in D$ then
$T(D)+A \vdash -\partial^* d(s,t)$ (since, via Lemma \ref{lemma:strict_partialstar}, 
we also have $T(D)+A \vdash -\partial^* true(q)$).
Using the $-\partial^*$ inference rule,
$T(D)+A \vdash -\partial^* d(s)$ and hence
$T(D)+A \vdash -\partial^* q$.


Suppose $q \in \Sigma$ and $+\partial^* q \in \T_{T(D)+A}\uparrow (n+1)$.
Then either
(1) $+\Delta q \in \T_{T(D)+A}\uparrow n$ (in which case $D+A \vdash +\partial q$),
or
$-\Delta \non q \in \T_{T(D)+A}\uparrow n$ (and hence $D+A \vdash -\Delta \non q$) and either
(2) for some $r$ in $A$ for $q$,  $B_r \subseteq \T_{T(D)+A}\uparrow n$,
or (3)
$+\partial^* one(q) \in \T_{T(D)+A}\uparrow n$,
$+\partial^* \neg true(\non q) \in \T_{T(D)+A}\uparrow n$,
and $+\partial^* d(r)$ occurs in $\T_{T(D)+A}\uparrow n$, for each rule $r$ for $\non q$ in $D$.
In both cases (2) and (3) we must have, for any rule $s$ for $\non q$ in $A$,
for some $p$ in the body $B_s$ of $s$, $-\partial^* p  \in \T_{T(D)+A}\uparrow n$.
By the induction hypothesis, $D+A \vdash -\partial p$ for each such $p$.

In case (2), by the induction hypothesis, $D+A \vdash B_r$.
Also, in case (2), the rule $o(\non q)$ must fail.
Consequently, every rule $s$ for $\non q$ in $D$ fails in $\T_{T(D)+A}\uparrow n$.
By the induction hypothesis, every rule $s$ for $\non q$ in $D$ fails in $D+A$.
Now, applying  the inference rule for $+\partial$, 
$D+A \vdash +\partial q$.

In case (3) there must be a strict or defeasible rule $r$ for $q$ in $D$ with body $B_r$ such that
$+\partial^* B_r \subseteq \T_{T(D)+A}\uparrow n$
and, using the rules for $d(s)$ and $d(s,t)$, for every rule $s$ for $\non q$,
either the body $B_s$ of $s$ fails
or there is a strict or defeasible rule $t$ for $q$ with body $B_t$ such that $+\partial^* B_t \subseteq \T_{T(D)+A}\uparrow n$
and $t > s$.
By the induction hypothesis,
$D+A \vdash +\partial B_r$, 
and, for every rule $s$ for $\non q$ in $D$,
either $D+A \vdash -\partial B_s$
or
there is a strict or defeasible rule $t$ for $q$ with body $B_t$ such that $D+A \vdash +\partial B_t$
and $t > s$.
As noted above, for any rule $s$ for $\non q$ in $A$, $B_s$ fails in $D+A$.
Hence,
by the inference rule for $+\partial$, 
$D+A \vdash +\partial q$.


If $q \in \Sigma$ and $-\partial^* q \in \T_{T(D)+A}\uparrow (n+1)$ then, 
using the inference rule for $-\partial^*$ and the structure of $T(D)$,
$-\Delta q  \in \T_{T(D)+A}\uparrow n$ and either
(a) $+\Delta \non q  \in \T_{T(D)+A}\uparrow n$ (in which case $D+A \vdash -\partial^* q$), or
(b) $-\partial^* one(q) \in \T_{T(D)+A}\uparrow n$, or
(c) $-\partial^* d(s) \in \T_{T(D)+A}\uparrow n$ for some rule $s$ for $\non q$ in $D$, or
(d) $+\partial^* one(\non q) \in \T_{T(D)+A}\uparrow n$ and
$+\partial^* d(r) \in \T_{T(D)+A}\uparrow n$ for every rule $r$ for $q$ in $D$,
or (e) for some $s$ for $\non q$ in $A$, $+\partial^* B_s \subseteq  \T_{T(D)+A}\uparrow n$.

If (b) $-\partial^* one(q) \in \T_{T(D)+A}\uparrow n$ then
for every strict or defeasible rule for $q$ in $D$ fails.
Applying the induction hypothesis and the inference rule for $-\partial$,
we have 
$D+A \vdash -\partial q$.
If c) $-\partial^* d(s) \in \T_{T(D)+A}\uparrow n$ for some rule $s$ for $\non q$ in $D$,
then there is no strict or defeasible rule $r_2$ for $q$ that defeats $s$.
If (d) then there is a rule $s$ for $\non q$ with body $B_s$ such that 
$+\partial^* B_s \subseteq \T_{T(D)+A}\uparrow n$ and
every rule $r$ for $q$ in $D$ is defeated by a strict or defeasible rule for $\non q$.
In both cases (c) and (d), applying the induction hypothesis and the inference rule for $-\partial$,
we have 
$D+A \vdash -\partial q$.
If (e) then, by the induction hypothesis, $D+A \vdash +\partial B_s$ and hence,
applying the inference rule for $-\partial$,
$D+A \vdash -\partial q$.

\end{proof}

\finish{
Interestingly, this transformation uses the superiority relation only when there are two single rules opposed.
In this case team defeat and non-team defeat produce the same conclusions.
Thus this transformation is also correct for transforming a theory within the logic $\DL(\partial)$.
}

\begin{theorem}   \label{thm:IDsimTDdelta}
The logic $\DL(\delta)$ can be simulated by $\DL(\delta^*)$
with respect to addition of rules.
\end{theorem}
\begin{proof}
Let $\Sigma$ be the language of $D{+}A$.
Note that, for any $q \in \Sigma(D)$,  $T(D){+}A \vdash +\supp^* \neg true(q)$ and,
employing Lemma \ref{lemma:strictAP},
$T(D){+}A \vdash +\delta^* \neg true(q)$ iff $T(D){+}A \vdash -\Delta q$ iff $D{+}A  \vdash -\Delta q$.
Also note that
$T(D){+}A \vdash -\delta^* g$, $T(D){+}A \vdash +\supp^* g$, and $T(D){+}A \vdash +\supp^* \neg g$,
where $g$ is the proposition used in part \ref{t4:support} of Definition \ref{defn:IDsimTD}.


\finish{ Lemma \ref{lemma:strictAP} handles strict consequences 

$+d B$ means $+d p$ for all $p \in B$
$-d B$ means $-d p$ for some $p \in B$

}

Suppose $+\delta q \in \T_{D+A}\uparrow (n+1)$.
Then 
either $+\Delta q \in \T_{D+A}\uparrow n$, or
$-\Delta \non q \in \T_{D+A}\uparrow n$ and
there is a non-empty team of strict or defeasible rules for $q$
such that $+\delta B_r \subseteq \T_{D+A}\uparrow n$ for each body $B_r$ of each rule $r$
and
every rule $s$ for $\non q$ either has a body that fails in $\T_{D+A}\uparrow n$
or $s < t$ for some rule $t$ in the team.
Then, by the induction hypothesis,
either $T(D)+A \vdash +\Delta q$ (in which case $T(D)+A \vdash +\delta^* q$), or
$T(D)+A \vdash -\Delta \non q$,
$T(D)+A \vdash +\delta^* B_r$ for each rule $r$ in the team,
and
for every rule $s$ for $\non q$ with body $B_s$ either $T(D)+A \vdash -\sigma B_s$ or
there is a rule $t$ in the team
and $t > s$.
If $s \in A$ then $T(D)+A \vdash -\sigma B_s$.
If $s \in D$ then either $T(D)+A \vdash +\delta^* fail(s)$  or $T(D)+A \vdash +\delta^* d(s,t)$;
in either case, $T(D)+A \vdash +\delta^* d(s)$.
Considering $T(D)$, and the inference rule for $+\delta^*$ we have
$T(D)+A \vdash +\delta^* one(q)$.
By Lemma \ref{lemma:strictAP}, $T(D)+A \vdash +\\delta^* \neg true(\non q)$.
Hence, the body of $s(q)$ is proved.
Because $>$ is acyclic,
there is a rule in the team for $q$ that is not inferior to any rule in the team for $\non q$.
Hence this rule $r'$ is not defeated, so $d(r')$ fails, and hence the rule for $\non q$ in $T(D)$ from point 6 fails.
Similarly, $d_\supp(r',s)$ fails, and hence the rules for $\non q$ in $T(D)$ from point 7 fail.
Hence all rules for $\non q$ fail, with the possible exception of $o(\non q)$.
However $s(q) > o(\non q)$ and hence, applying the $+\delta^*$ inference rule,
$T(D)+A \vdash + \delta^* q$.


Suppose $-\delta q \in \T_{D+A}\uparrow (n+1)$.
Then $-\Delta q \in \T_{D+A}\uparrow n$ (and hence $T(D)+A \vdash -\Delta q$) and
either 
(1) $+\Delta \non q \in  \T_{D+A}\uparrow n$ (in which case $T(D)+A \vdash -\delta^* q$), or
(2) every rule $r$ for $q$ contains a body literal $p$ with $-\delta p \in \T_{D+A}\uparrow n$, or
(3) there is a rule $s$ for $\non q$ with body $B_s$ such that $+\supp B_s \subseteq  \T_{D+A}\uparrow n$ and,
for every strict or defeasible rule $t$ for $q$, 
either
$t$ fails in $\T_{D+A}\uparrow n$, or
$t \not> s$.
In case (2), 
the rules $r$ in $A$ for $q$ fail and,
by the induction hypothesis and the inference rule for $-\delta^*$,
the rules $r$ in $A$ for $q$ fail in $T(D)+A$, so
$T(D)+A \vdash -\delta^* one(q)$
and hence $T(D)+A \vdash -\delta^* q$.
In case (3), by the induction hypothesis,
there is a rule $s$ for $\non q$ with body $B_s$ such that 
$T(D)+A \vdash +\supp^* B_s$ and
for every strict or defeasible rule $t$ for $q$, 
either
$t$ fails in $T(D)+A$, or
$t \not> s$.
If $s \in A$ then $t \not> s$, for every $t$, and hence $T(D)+A \vdash -\delta^* q$.
If $s \in D$ then
$T(D)+A \vdash -\delta^* d(s,t)$ (since, via Lemma \ref{lemma:strictAP}, 
we also have $T(D)+A \vdash -\delta^* true(q)$).
Using the $-\delta^*$ inference rule,
$T(D)+A \vdash -\delta^* d(s)$.
The bodies of rules from point 7 of the transformation also fail (wrt $\delta^*$),
because of the presence of $g$.
Hence
$T(D)+A \vdash -\delta^* q$.


Suppose $q \in \Sigma$ and $+\delta^* q \in \T_{T(D)+A}\uparrow (n+1)$.
Then either
(1) $+\Delta q \in \T_{T(D)+A}\uparrow n$ (in which case $D+A \vdash +\delta q$),
or else $-\Delta \non q \in \T_{T(D)+A}\uparrow n$ and either
(2) there is a strict or defeasible rule $r$ for $q$ in $A$ where
$+\delta^* B_r \subseteq \T_{T(D)+A}\uparrow n$
and for all rules for $\non q$ in $T(D){+}A$,
the body of the rule contains a literal $p$ with $-\sigma^* p  \in \T_{T(D)+A}\uparrow n$,
or (3) each of
$+\delta^* one(q)$,
$+\delta^* \neg true(\non q)$,
and $+\delta^* d(s)$ occurs in $\T_{T(D)+A}\uparrow n$, for each rule $s$ for $\non q$ in $D$.

Hence, in case (3), there is a strict or defeasible rule $r$ for $q$ with body $B_r$ such that
$+\delta^* B_r \subseteq \T_{T(D)+A}\uparrow n$
and, for every rule $s$ for $\non q$,
either $-\sigma^* p  \in \T_{T(D)+A}\uparrow n$, for some $p$ in the body $B_s$ of $s$, or
there exists $t$ in $D$ for $q$ with $+\delta^* B_t \subseteq \T_{T(D)+A}\uparrow n$
and $t>s$.
By the induction hypothesis,
$D+A \vdash -\Delta \non q$,
$D+A \vdash +\delta B_r$, 
and, for every rule $s$ for $\non q$,
either $D+A \vdash -\sigma B_s$
or
$D+A \vdash +\delta B_t$ and $t > s$.
By the inference rule for $+\delta$, 
$D+A \vdash +\delta q$.

In case (2), using the structure of $T(D)$,
for the rules $supp(\non q)$, originating from some rule $s$ for $\non q$ in $D$,
either for some $p$ in $B_s$, $-\sigma^* p  \in \T_{T(D)+A}\uparrow n$ or,
for some $t$, $-\sigma^* d_\supp(t, s)  \in \T_{T(D)+A}\uparrow n$
(and, hence, $+\delta^* B_t \subseteq \T_{T(D)+A}\uparrow n$ and $t > s$).
Now, by the induction hypothesis,
$D+A \vdash +\delta B_r$;
$D+A \vdash -\Delta \non q$;
for all rules for $\non q$ in $A$, the body of the rule contains a literal $p$ with $D+A \vdash -\sigma^* p$; and
for all rules for $\non q$ in $D$, either the body of the rule contains a literal $p$ with $D+A \vdash -\sigma^* p$
or there is a rule $t$ for $q$ in $D$ with $t > s$ and $D+A \vdash +\delta B_t$.
Applying the inference rule for $+\delta$, 
$D+A \vdash +\delta q$.


If $q \in \Sigma$ and $-\delta^* q \in \T_{T(D)+A}\uparrow (n+1)$ then, 
using the inference rule for $-\delta^*$ and the structure of $T(D){+}A$,
$-\Delta q  \in \T_{T(D)+A}\uparrow n$ (and, hence, $D+A \vdash -\Delta q$) and either
$+\Delta \non q  \in \T_{T(D)+A}\uparrow n$ (in which case $D+A \vdash -\delta q$), or else
for every rule $r$ for $q$ in $A$, there is a literal $p$ in $B_r$ such that $-\delta^* p \in \T_{T(D)+A}\uparrow n$
and either
(1) $-\delta^* \neg true(\non q) \in \T_{T(D)+A}\uparrow n$
(in which case $D+A \vdash +\Delta \non q$ and hence $D+A \vdash -\delta q$), or
(2) $-\delta^* one(q) \in \T_{T(D)+A}\uparrow n$, or
(3) $-\delta^* d(s) \in \T_{T(D)+A}\uparrow n$ for some rule $s$ for $\non q$ in $D$.
Or
(4) $+\supp^* one(\non q) \in \T_{T(D)+A}\uparrow n$ and
$+\supp^* d(r) \in \T_{T(D)+A}\uparrow n$ for every rule $r$ for $q$ in $D$, or
(5) there is a rule $s$ for $\non q$ in $A$ and $+\supp^* B_s \subseteq  \T_{T(D)+A}\uparrow n$.
Or (6) the body of a rule $supp(\non q, s)$ is supported for some rule $s$ for $\non q$
(that is, $+\supp^* B_s \subseteq  \T_{T(D)+A}\uparrow n$ and, for each rule $r$ for $q$,
$+\supp^* d_\supp(r, s) \in  \T_{T(D)+A}\uparrow n$).

For (2) and (3), by the induction hypothesis,
for every rule $r$ for $q$ in $A$, there is a literal $p$ in $B_r$ such that $D+A \vdash -\delta p$.
If (2) $-\delta^* one(q) \in \T_{T(D)+A}\uparrow n$ then
every strict or defeasible rule for $q$ in $D$ fails.
Applying the induction hypothesis and the inference rule for $-\delta$,
we have 
$D+A \vdash -\delta q$.
If (3) $-\delta^* d(s) \in \T_{T(D)+A}\uparrow n$ for some rule $s$ for $\non q$ in $D$,
then $+\sigma^* B_s \subseteq \T_{T(D)+A}\uparrow n$, where $B_s$ is the body of $s$,
and
for every strict or defeasible rule $r$ for $q$ with body $B$ either 
$-\delta^* B_r \in \T_{T(D)+A}\uparrow n$ or  
$+\sigma^* B_s \subseteq \T_{T(D)+A}\uparrow n$, where $B_s$ is the body of $s$,
and $r \not> s$.
Applying the induction hypothesis, 
$D+A \vdash +\sigma B_s$ and,
for every $r$ for $q$,
$D+A \vdash -\delta B_r$ or
$D+A \vdash +\sigma B_s$ and $r \not> s$.
Hence, by the inference rule for $-\delta$,
$D+A \vdash -\delta q$.

If (4) then there is a rule $s$ for $\non q$ with body $B_s$ such that 
$+\supp^* B_s \subseteq \T_{T(D)+A}\uparrow n$ and
for every rule $r$ for $q$ in $D$ either
there is a literal $p$ in the body of $r$ such that $-\delta^* p \in \T_{T(D)+A}\uparrow n$ or
there is a rule $s'$ for $\non q$ with body $B'$ such that $+\supp^* B' \subseteq \T_{T(D)+A}\uparrow n$
and $s' > r$.
Applying the induction hypothesis, 
for every rule $r$ for $q$ in $D$ either
there is a literal $p$ in the body of $r$ such that $D+A \vdash -\delta p$ or
there is a rule $s'$ for $\non q$ with body $B'$ such that $D+A \vdash +\supp B'$
and $s' > r$.
t follows, by the inference rule for $-\delta$,
that $D+A \vdash -\delta q$.

If (5) then, by the induction hypothesis, $D+A \vdash +\sigma B_s$ and, since $s$ is not inferior to any rule,
the inference rule for $-\delta$ gives us
$D+A \vdash -\delta q$.

In case (6), since $+\supp^* d_\supp(r, s) \in  \T_{T(D)+A}\uparrow n$, we must have
$+\supp^* B_s \subseteq \T_{T(D)+A}\uparrow n$ and either there is a literal $p$ in $B_r$ such that 
$-\delta^* p \in \T_{T(D)+A}\uparrow n$
or $r \not> s$.
By the induction hypothesis,
$D+A \vdash +\supp^* B_s$ and, for every rule $r$ for $q$ in $D$ either
there is a literal $p$ in $B_r$ such that 
$D+A \vdash -\delta^* p$
or $r \not> s$.
By the $-\delta$ inference rule, $D+A \vdash -\delta q$.


Suppose $+\sigma q \in \T_{D+A}\uparrow (n+1)$.
Then 
either $+\Delta q \in \T_{D+A}\uparrow n$, or
there is a strict or defeasible rule $r$ for $q$
such that $+\sigma B_r \subseteq \T_{D+A}\uparrow n$ where $B_r$ is the body of $r$
and
every rule $s$ for $\non q$ has a body  with a literal $p$ such that  $-\delta p \in \T_{D+A}\uparrow n$
or $s \not> r$.
Then, by the induction hypothesis,
either $T(D)+A \vdash +\Delta q$ (in which case $T(D)+A \vdash +\sigma^* q$), or
$T(D)+A \vdash +\sigma^* B_r$, 
and
every rule $s$ for $\non q$ has a body  with a literal $p$ such that  $-\delta^* p \in \T_{D+A}\uparrow n$
or $s \not> r$.
If $r \in A$ then $r$ is not inferior to any rule and, by the inference rule for $+\sigma^*$,
$T(D)+A \vdash + \sigma^* q$.
If $r \in D$ then, by the $+\sigma^*$ inference rule, $T(D)+A \vdash +\sigma^* B_r$.
Furthermore, again by the $+\sigma^*$ inference rule, for every $s$ for $\non q$ in $D$,
$T(D)+A \vdash +\sigma^* d_\supp(s, r)$,
since $a(s,r) \not> b(s,r)$ iff $s \not> r$.
Note that there is no superiority relation between the rules in $T(D)$ for $q$ and $\non q$.
Hence, applying the inference rule for $+\sigma^*$,  $T(D)+A \vdash + \sigma^* q$.


Suppose $-\sigma q \in \T_{D+A}\uparrow (n+1)$.
Then $-\Delta q \in \T_{D+A}\uparrow n$ and
either 
every rule $r$ for $q$ contains a body literal $p$ and $-\sigma p \in \T_{D+A}\uparrow n$, or
there is a rule $s$ for $\non q$ with body $B_s$ such that $+\delta B_s \subseteq  \T_{D+A}\uparrow n$ and $s > r$.
(Note that, for $r \in A$, only the first possibility can apply.)
Then, by the induction hypothesis,
$T(D)+A \vdash -\Delta q$ and either 
every rule $r$ for $q$ contains a body literal $p$ such that $T(D)+A \vdash -\sigma^* p$, or
there is a rule $s$ for $\non q$ with body $B_s$ such that 
$T(D)+A \vdash +\delta^* B_s$ and $s > r$.
(In particular, every rule for $q$ in $A$ contains a body literal $p$ with $T(D)+A \vdash -\sigma^* p$.)
If all rules for $q$ in $D$ fall in the former case, we have $-\sigma^* one(q)$,
and all rules $supp(q)$ fail.
Otherwise, there is an $s$ that is not inferior to any rule for $q$ and hence
$T(D)+A \vdash -\sigma^* d(s,r)$ and
$T(D)+A \vdash -\sigma^* d(s)$.
Similarly, $T(D)+A \vdash -\sigma^* d_\supp(s,r)$.
In either case, all rules for $q$ fail, and hence
$T(D)+A \vdash -\sigma^* q$.


Suppose $q \in \Sigma$ and $+\supp^* q \in \T_{T(D)+A}\uparrow (n+1)$.
Then either
(1) $+\Delta q \in \T_{T(D)+A}\uparrow n$ (in which case $D+A \vdash +\supp q$), or else either
(2) for some rule $r$ for $q$ in $A$, $+\supp^* B_r \subseteq  \T_{T(D)+A}\uparrow n$, or
(3) $+\supp^* one(q) \in \T_{T(D)+A}\uparrow n$,
and $+\supp^* d(s)$ occurs in $\T_{T(D)+A}\uparrow n$, for each rule $s$ for $\non q$ in $D$, or
(4) for some strict or defeasible rule $r$ for $q$ in $D$, $+\supp^* B_r \subseteq  \T_{T(D)+A}\uparrow n$ and
$+\supp^* d_\supp(s,r)$ occurs in $\T_{T(D)+A}\uparrow n$, for each rule $s$ for $\non q$ in $D$.

In case (2), by the induction hypothesis, $D{+}A \vdash +\supp B_r$ and hence,
by the inference rule for $\supp$, $D{+}A \vdash +\supp q$.

In case (3), there is a strict or defeasible rule $r$ for $q$ with body $B_r$ such that
$+\supp^* B_r \subseteq \T_{T(D)+A}\uparrow n$
and, for every rule $s$ for $\non q$ in $D$,
either $-\delta^* p  \in \T_{T(D)+A}\uparrow n$, for some $p$ in the body $B_s$ of $s$
or there is a rule $t$ for $q$ with body $B_t$ such that $+\supp^* B_t  \in \T_{T(D)+A}\uparrow n$.
and $s \not> t$.
By the induction hypothesis,
$D+A \vdash +\supp B_r$, 
and, for every rule $s$ for $\non q$,
either $D+A \vdash -\delta B_s$
or there is a rule $t$ for $q$ with body $B_t$ such that $D+A \vdash +\supp^* B_t $
and $s \not> t$.
Because > is acyclic, there is a rule $t$ for $q$ such that $D+A \vdash +\supp B_t $
and, for every rule $s$ for $\non q$ either
$D+A \vdash -\delta B_s $ or $s \not> t$.
By the inference rule for $+\supp$, 
$D+A \vdash +\supp q$.

In case (4), 
there is a strict or defeasible rule $r$ for $q$ with body $B_r$ such that
$+\supp^* B_r \subseteq \T_{T(D)+A}\uparrow n$ and,
for each rule $s$ for $\non q$ in $D$,
either $-\delta^* p  \in \T_{T(D)+A}\uparrow n$, for some $p$ in the body $B_s$ of $s$,
or $s \not> r$.
By the induction hypothesis,
$D+A \vdash +\supp B_r$,
and, for each $s$, either $D+A \vdash -\delta p$ or $s \not> r$.
By the $+\supp$ inference rule, 
$D+A \vdash +\supp q$.


If $q \in \Sigma$ and $-\supp^* q \in \T_{T(D)+A}\uparrow (n+1)$ then, 
using the inference rule for $-\supp^*$ and the structure of $T(D)$,
$-\Delta q  \in \T_{T(D)+A}\uparrow n$ (and hence $D+A \vdash -\Delta q$), and either
(1) $+\Delta \non q  \in \T_{T(D)+A}\uparrow n$ (in which case $D+A \vdash -\supp^* q$), or
(2) for each strict or defeasible rule $r$ for $q$ in $A$, 
there is a literal $p$ in $B_r$ such that $-\supp^* p \in \T_{T(D)+A}\uparrow n$ and
for each strict or defeasible rule $r$ for $q$ in $D$, either
there is a literal $p$ in $B_r$ such that $-\supp^* p \in \T_{T(D)+A}\uparrow n$, or
there is a rule $s$ for $\non q$ in $D$ such that 
$+\delta^* B_s \subseteq \T_{T(D)+A}\uparrow n$ and $s > r$.
By the induction hypothesis, in case (2),
$D+A \vdash -\supp B_r$ for the rules $r$ in $A$ and,
for rules $r$ in $D$, either
$D+A \vdash -\supp B_r$ or there is a rule $s$ for $\non q$ in $D$ such that
$D+A \vdash +\delta^* B_s$  and $s > r$.
Applying the $-\supp$ inference rule, 
$D+A \vdash -\supp q$.
\end{proof}

Combining Theorems \ref{thm:IDsimTDpartial} and \ref{thm:IDsimTDdelta},
we have Theorem \ref{thm:IDsimTD}.

\end{appendix}